\documentclass[12pt]{article}
\title{\bf Complexity of representations of coefficients of power 
series in classical statistical mechanics. Their classification and 
complexity criteria}
\author{\bf G.I.~Kalmykov\\
\bigskip}
\date{}
\usepackage{amsthm,latexsym}
\usepackage{amsfonts}
\usepackage{amssymb}
\usepackage{amsmath}
\usepackage{bm}
\usepackage[cp1251]{inputenc}
\usepackage[T1,T2A]{fontenc}
\usepackage[russian,english]{babel}
\usepackage{indentfirst}
\usepackage{enumerate}
\usepackage{verbatim}
\newcommand{\MO}{\mathop}
\newcommand{\LL}{\Lambda}
\newcommand{\LT}{\lefteqn}
\newcommand{\R}{{\bf R}^{\nu}}
\newcommand{\XX}{\widetilde X}

\newcommand{\mf}{\mathfrak}
\renewcommand{\refname}{References}
\begin{document}

\maketitle

\begin{abstract}
It is declared that the aim of simplifying representations 
of coefficients of power series of classical statistical mechanics 
is to simplify a process of obtaining estimates of the coefficients 
using their simplified representations.

The aim of the article is: to formulate criteria for the complexity 
(from the above point of view) of these representations and 
to demonstrate their application by examples of comparing Ree-Hoover 
representations of virial coefficients (briefly --- Ree-Hoover 
representations) and such representations of power series coefficients
that are based on the conception of the frame classification of 
labeled graphs.

To solve these problems, mathematical notions were introduced 
(such as a base product, a base integral, a base linear combination 
of integrals, a base linear combination of integrals with coefficients 
of negligible complexity, a base set of base linear combinations 
of integrals with coefficients of negligible complexity); and 
a classification of representations of coefficients of power series of 
classical statistical mechanics is proposed. In this classification 
the class of base linear combinations of integrals with coefficients 
of negligible complexity is the most important class. It includes 
the most well-known representations of the coefficients of power 
series of classical statistical mechanics.

Three criteria are formulated to estimate the comparative complexity 
of base linear combinations of integrals with coefficients 
of negligible complexity and their extensions to the totality 
of base sets of base linear combinations of integrals 
with coefficients of negligible complexity are constructed. 
The application of all the constructed criteria is demonstrated 
by examples of comparing with each other of Ree-Hoover representations 
and of such power series coefficients representations, which are 
constructed on the basis of the concept of frame classification 
of labeled graphs. The obtained results are presented in the tables 
and commented.
\end{abstract}

\pagebreak

{\bf 1. Introduction} 

The article discusses thermodynamic equilibrium one-component 
systems of classical particles, both enclosed in a bounded set 
$\LL$ of $\nu$-dimensional real Euclidean space ${\bf R}^{\nu}$ and 
enclosed in $\nu$-dimensional real Euclidean space ${\bf R}^{\nu}$.
It is assumed that these particles interact through
central forces, characterized by the potential of pairwise interaction 
$\Phi({\bf r})$, where 
${\bf r} = (r^{(1)}, r^{(2)}, \ldots, r^{(\nu)}) \in {\bf R}^{\nu}$.
It is also assumed that the potential of pairwise interaction
$\Phi({\bf r}) $ is a measurable function, and the interaction
(pairwise interaction) satisfies the stability condition
[24, 17, 49] and regularity condition [24, 17, 49].

As usual, we denote Mayer function
\begin{equation}
f_{ij} = \exp\{-\beta\Phi({\bf r}_i-{\bf r}_j)\} - 1,
\end{equation}
where $i \ne j$, ${\bf r}_i,{\bf r}_j \in {\bf R}^{\nu}$,
$\beta = 1 / kT$ is inverse temperature, $k$ is the Boltzmann 
constant, $T$~is absolute temperature.
By $\widetilde f_{ij}$ we denote {\bf Boltzmann function}
[24, 49], assuming
\begin{equation}
\widetilde f_{ij} = 1 + f_{ij} = 
\exp\{-\beta\Phi({\bf r}_i-{\bf r}_j)\}.
\end{equation}
In the case, when such a system of particles is enclosed in a limited
set $\LL$, the dependence of the pressure $p(\LL)$ on the density 
$\varrho$ in such a system can be presented in two forms: 
in the form of virial expansion of pressure $p(\LL)$ in powers 
of density $\varrho$ and in
parametric form, i.e. as two
equations expressing the dependence of the pressure $p(\LL)$ and
the density $\varrho(\LL)$ on the parameter $z$, called activity
[23, 24, 44, 49].

The virial expansion is:
\begin{equation}
p(\beta, \LL) = 
\beta^{-1}\sum_{n = 1}^{\infty}B_n(\beta, \LL)\varrho^n.
\end{equation}
Below we will omit the argument $\beta$ of the coefficients $B_n$
for simplicity. In this expansion, the coefficients
$B_n(\LL)$ are called virial coefficients. The virial coefficient 
$B_1(\LL)$ is $1$, and for $n > 1$ virial coefficients are defined
by the formula:
\begin{equation}
B_n(\LL) =-\frac{n-1}{|\LL|n!}\sum_{B \in {\mathfrak B}_n}
\int_{(\LL^{\nu})^n}
\prod_{\{u, v\} \in X(B)}f_{uv}(d{\bf r})_n,
\end{equation}
where $|\LL|$ is the measure of the set $\LL$, ${\mathfrak B}_n$ is 
the totality of all doubly connected labeled graphs (blocks) 
with the set of vertices $V_n = \{1, 2, \ldots,n\}$, 
$X(B)$ is the set of all  edges of block
$B$; $(d{\bf r})_n = d{\bf r}_1d{\bf r}_2\ldots d{\bf r}_n$, \quad
$d{\bf r}_i=dr_i^{(1)}dr_i^{(2)}\cdot\ldots\cdot dr_i^{(\nu)}$.

Here and in what follows, following [25, 28], we assume that
every graph $G$, by definition, has neither multiple edges nor loops.

Hereinafter in the text, we assume that the vertices of edges and of
graphs are labeled with natural numbers. Therefore, throughout
the article, we identify vertices of graphs with their labels.
In the same way, we identify the vertices,
incident to edges, with their labels.

These representations of virial coefficients were obtained by J. Mayer.
He also noticed that for $n \ge 2$ the virial coefficients $B_n(\LL)$
quickly tend to their limit $B_n$ as $\LL$ grows. This makes 
it possible as an estimate of the limit of the coefficient $B_n(\LL)$ 
to take the value of the virial coefficient $B_n(\LL)$ where 
the set $\LL$ is not very large.

He also found a parametric representation of the pressure dependence
$p(\beta, \LL)$ on the density $\varrho(\beta, \LL)$:
\begin{equation}
p(\beta, \LL) = \beta^{-1}\sum_{n = 1}^{\infty}b_n(\beta, \LL)z^n;
\end{equation}
\begin{equation}
\varrho(\beta, \LL) = \sum_{n = 1}^{\infty}nb_n(\beta, \LL)z^n.
\end{equation}
Below we will omit the argument $\beta$ of the coefficients 
$b_n(\beta, \LL)$ for simplicity. In expansions (5) and (6) in degrees 
of activity $z$ the coefficients $b_n(\LL)$ are called, like 
virial coefficients, Mayer coefficients. Unlike virial coefficients,
we will call them Mayer coefficients in the degrees of activity 
$z$. And in those cases where their meaning is uniquely determined 
by the context, we will briefly call them Mayer coefficients.

Mayer coefficient $b_1(\LL)$ is $1$, and for $n > 1$ the Mayer 
coefficients $b_n(\LL)$ are defined by the formula:
\begin{equation}
b_n(\LL) = 
\frac{1}{|\LL|n!}\sum_{G \in {\bf G_n}}\int_{({\LL}^{\nu})^{n}}
\prod_{\{u, v\} \in X(G)}f_{uv}d{\bf r}_n,
\end{equation}
where $\bf G_n$ is the totality of all connected labeled graphs
with the set of vertices $V_n = \{1, 2, \ldots, n\}$,
$X(G)$ is the set of all edges of the graph $G$.

However, it was subsequently noticed that these representations have
very unpleasant property, thanks to which they are practically
unsuitable both for the calculation of virial coefficients (except
for the first three) and for the theoretical analysis of the behavior
of the higher coefficients. For the first time this property
of Mayer representations of the coefficients of power series
of classical statistical mechanics was pointed out by I.I. Ivanchik.
In his works [1, 30], he was the first to qualitatively describe this
property and called it an {\bf asymptotic catastrophe}. What is
the manifestation of an asymptotic catastrophe? The fact is that
Mayer representation of the $n$-th coefficient of the power series
contains a factor that is the sum of integrals. Such sums of integrals
have the following feature: even with not very large values of $n$
a significant part of the integrals of such a sum with large
accuracy mutually cancel out as values of opposite signs.

Relatively small the remainder remaining after such a mutual
annihilation is, for $n \to \infty$, an infinitesimal quantity
compared to with the number of terms in the sum traditionally
determining this coefficient. This "remainder" of primary interest
becomes inaccessible for direct research even for small $n$.

Further, the author of this article in the book [17]
gave a rigorous mathematical definition of the asymptotic catastrophe.
For the convenience of the reader, we present this definition here.

D\,e\,f\,i\,n\,i\,t\,i\,o\,n 1. In representations of power series
coefficients there is the asymptotic catastrophe phenomenon if
for any $B > 0$ the number of terms in the sum, representing
the coefficient of the variable to the power of $n$,
for $n \to \infty$ grows faster than the value
$(n!)^2B^n$. $\blacksquare$

The meaning of this definition is that it enables
to separate those representations of the coefficients of the power
series, where already for relatively small $n$ the number of terms is
too large, from representations, in which the number of terms grows
significantly slower.

When trying to estimate the coefficients of Mayer expansions,
based on those representations where the phenomenon of an asymptotic
catastrophe is present, it is almost inevitable that with an increase
in $n$, a catastrophically rapid increase in the estimation errors
of these coefficients takes place.

Over the past few decades, the efforts of a number of scientists have
been directed towards to simplify representations of
coefficients of power series of classical statistical
mechanics and their estimation.

The aim of simplifying the representations of the coefficients
of these power series was to simplify the process of obtaining
estimates of these coefficients using their simplified
representations.
For brevity, a complexity of the process of obtaining an estimate
of a given coefficient by means of this representation, we will call
the {\bf complexity of the given representation of this coefficient}.

The most famous results in simplifying the representations of the
virial coefficients are apparently Ree-Hoover representations [46],
[47], [48]. In these representations, for each $n \ge 4$, the virial
coefficient $B_n(\LL)$ is represented as a linear combination of 
integrals, the integrands of which are labeled with complete labeled 
graphs. In every integral, 
which is a term of such a linear combination, the integrand
is the product of Mayer and Boltzmann functions. And the set
of all Mayer and Boltzmann functions included in this product,
is in one-to-one correspondence with the set of edges of the graph,
labeling the integrand of this integral. Moreover, each edge of this 
graph labeled with Mayer function corresponds to Mayer function that 
is a label of this edge. And each edge labeled with Boltzmann function
corresponds to Boltzmann function that is a label of this edge. So 
the virial coefficient $B_n(\LL)$ is represented as a linear 
combination of integrals, in each of which the integrand is the 
product of Mayer and Boltzmann functions, total number of which is 
${n(n - 1) / 2}$. These representations are called {\bf Ree-Hoover 
representations}.

Using Ree-Hoover representations of virial coefficients,
a number of scientists have calculated [50] estimates of the virial
coefficients $B_n(\LL)$ (for $n = \overline{4,8})$ for a number
of different values temperatures. Later, on a graphical computer,
the estimates of the virial coefficients $B_n(\LL)$ were calculated 
[51] for $n = \overline{6.9}$ for the Lennard-Jones potential for 
different temperatures. At that the previously calculated estimates
of the values of these coefficients were made precise.
Moreover, estimates of the values of these coefficients were calculated
for $n = \overline{10,16}$ for several (from one to four) temperatures.
By the way, the fact that for $n = \overline{10,16}$
it was possible to find estimates for the value of the virial
coefficient $B_n(\LL)$ at no more than four different temperatures,
indicates that for $n > 9$ the calculations volume required
to estimate one of the values of the virial coefficient $B_n(\LL)$
by Ree-Hoover method is so large that these calculations require
a very considerable time even when working on a modern computer
with high performance. However, the question remains: are the
Ree-Hoover representations free from the asymptotic catastrophe?

A different approach to simplifying the representations of
the coefficients of power series of classical statistical mechanics
is developed by the author of this article. It is based on a concept
of classification of labeled graphs. This concept is developing
by the author \mbox {[2--9, 13--20, 31--34, 37--39]}.
We will call it {\bf the frame sum method}.

Within the bounds of this method, he obtained the avoiding 
the asymptotic catastrophe representations:
of Mayer coefficients of expansions of pressure and density
in powers of activity, of coefficients of expansion of $m$-partial
distribution function in powers of activity, of coefficients
of expansion of the ratio of activity to density in powers
of activity and  of virial coefficients
[3, 4, 6--9, 15, 17, 31--34, 37, 39].

The advantage of these representations is that they are free from
asymptotic catastrophe [9, 11, 15, 17, 36, 37, 39]. Using these
representations, it was possible to obtain [9, 10, 12, 17, 35, 39] 
an upper bound for the radius of convergence of Mayer expansions 
in degrees of activity (for non-negative potential).
And also it was possible, using these representations, on a personal
computer calculate, fairly accurately, the estimates 
of the thermodynamic limits of the 4th, 5th and 6th virial 
coefficients at one of the temperature values.

2.  {\bf The aim of the article and the results obtained}

The aim of the article is: to define criteria for estimation of
a complexity of representations of coefficients of power series
of the classical statistical mechanics;
to demonstrate application of these criteria with examples
of comparison of Ree-Hoover representations of virial coefficients
and such power series coefficients representations that are based
on the concept of frame classification of labeled graphs.

It is obvious that even for comparison in the complexity of two
different representations of a given coefficient of a certain power
series you must have a criterion. This kind of criterion is all the
more necessary if the task is set to compare the complexity of given
representations of given coefficients of a variable in a power $n$
of two different power series.

The creation of such criteria facilitates the fact that many 
well-known representations of the coefficients of power series 
of classical statistical mechanics are linear combinations of 
multidimensional integrals, the integrands of which are labeled with 
labeled graphs, in which each edge is labeled with either Mayer or 
Boltzmann functions. In every integral that is a term of such a 
linear combination, the integrand is the product of Mayer and 
Boltzmann functions (such are, for example, proposed by Ree and 
Hoover [46, 47, 48] representations of virial coefficients).

In the article [39], a classification of the representations 
of the coefficients of power series of classical statistical mechanics 
is made. The most important class of this classification contains 
obtained by the frame sums method the virial coefficients 
representations in the thermodynamic limit and the representations of 
the thermodynamic limits of  Mayer coefficients 
of the pressure and density expansions in the degrees of activity .
These representations are linear combinations of multidimensional 
integrals described in the previous parbox.

To estimate the comparative complexity of the included in this class 
representations of the coefficients of power series, in [39], 
for the first time, three criteria were constructed, ordered by their 
accuracy. Also, in [39], three criteria  were constructed, ordered 
by their accuracy, for a comparative estimation of the complexity 
of polynomials in linear combinations included in the above mentioned 
class of representations of the coefficients of power series 
of classical statistical mechanics.

In the given article, this class is 
extended so that this extension includes many well-known 
representations of the coefficients of power series arising 
in the investigations of thermodynamic equilibrium one-component 
systems of classical particles as enclosed in $\nu$-dimensional real 
Euclidean space ${\bf R}^{\nu}$, and those enclosed in bounded
the set $\LL$ contained in the space ${\bf R}^{\nu}$.
This article introduces the concept of {\bf comparable} linear 
combinations belonging to this extension and constructs criteria 
for a comparative estimation of the complexity of comparable 
linear combinations.
Also proposed criteria for comparative estimation of complexity of
polynomials in linear combinations included in this extension.

To describe these criteria, the mathematical concepts introduced 
in [39] and some properties of these concepts are used.
For the convenience of readers, all these mathematical concepts and 
their properties are given in this article. In those cases when 
the proofs of theorems and lemmas taken from [39] were
not clear enough, or not detailed enough, they were replaced by clear 
and detailed proofs with references to sources and used formulas.

The application of these criteria is demonstrated by examples 
of the estimates of the comparative complexity of Ree-Hoover 
representations of the virial coefficients and of the power series
coefficients representations based on the concept of frame 
classification of labeled graphs.

3. {\bf Some mathematical concepts and their properties}

Before proceeding to the description of the proposed classification
and the proposed criteria of the complexity of representations
of the coefficients of power series, we will give definitions
of the mathematical concepts necessary for their descriptions,
and dwell on some properties of these concepts.

First of all, we will slightly expand the concept of an edge
of a labeled graph, introducing the following

D\,e\,f\,i\,n\,i\,t\,i\,o\,n 2 [39]. An unordered pair $\{i, j\}$ of
different natural numbers is called an {\bf edge}. $\blacksquare$

In this article, we will consider only the sets of pairwise distinct
edges without mention this circumstance. $\blacksquare$

D\,e\,f\,i\,n\,i\,t\,i\,o\,n 3 [39]. We will say that a set of
edges $X_f = \{\{i, j\}\}$ {\bf defines} the set
$F = \{f_{i j}\}$ of Mayer functions, if any Mayer function
$f_{i j}$ belongs to the set $F$ if and only if
the edge $\{i, j\}$ belongs to the set $X_f$. At that, the set of
edges $X_f$ will be called a {\bf set of Mayer edges with respect
to this set $F$ of Mayer functions}. $\blacksquare$

D\,e\,f\,i\,n\,i\,t\,i\,o\,n 4 [39]. We will also say that a set of
edges $X_{\widetilde f} = \{\{i', j'\}\}$ {\bf defines} the set
\mbox{$\widetilde F = \{\widetilde f_{i' j'}\}$} of Boltzmann
functions if any Boltzmann function
$\widetilde f_{i' j'} = f_{i' j'} + 1$ is contained in the set
$\widetilde F$ if and only if the edge
$\{i', j'\}$ belongs to the set $X_{\widetilde f}$. At that
the set $X_{\widetilde f}$ will be called a {\bf set of Boltzmann
edges with respect to this set $\widetilde F$ of Boltzmann
functions}. $\blacksquare$

Let's introduce the notations:
\begin{equation}
P(F, \widetilde F) = \prod_{f_{i j} \in F}
\prod_{\widetilde f_{i' j'} \in \widetilde F}
f_{i j}\widetilde f_{i' j'}
\label{p}
\end{equation}
is the product of all Mayer functions belonging to a set of
Mayer functions $F$, and all Boltzmann functions belonging to
a set of Boltzmann functions $\widetilde F$. It is obvious that
the product $P(F, \widetilde F)$ is a function of sets $F$ and
$\widetilde F$. For brevity, we will omit the arguments $F$ and
$\widetilde F$ of the product $P$. The product $P$ will be called
a {\bf product of Mayer and Boltzmann functions}.

${\bf X} = \{X_f, X_{\widetilde f}\}$ is an ordered pair of disjoint
sets: a set of edges $X_f = \{\{i, j\}\}$ and a set of edges
$X_{\widetilde f} = \{\{i', j'\}\}$.

$V(X_f)$ is the set of ends (vertices) of all edges
from the set $X_f$.

$V(X_{\widetilde f})$ is the set of ends (vertices) of all edges from
the set $X_{\widetilde f}$.

$\left|V(X_f)\bigcup V(X_{\widetilde f})\right|$ is the cardinality
of the sum of sets $V(X_f)$ and $V(X_{\widetilde f})$.

we will also consider such ordered pairs
${\bf X} = \{X_f, X_{\widetilde f}\}$ of disjoint sets,
in which the second set is empty, that is
pairs of the form ${\bf X} = \{X_f, \emptyset \}$.

D\,e\,f\,i\,n\,i\,t\,i\,o\,n 5 [39]. If disjoint sets of edges $X_f$ 
and $X_{\widetilde f}$ satisfy the condition
\begin{equation}
V(X_f)\bigcup V(X_{\widetilde f}) = V_n = \{1, 2, \ldots, n\},
\end{equation}
where
\begin{equation}
n = \left|V(X_f)\bigcup V(X_{\widetilde f})\right|,
\end{equation}
then the ordered pair ${\bf X} = \{X_f, X_{\widetilde f}\}$ of these
sets will be called a {\bf canonical pair of sets}, and
the number $n$ will be called the {\bf order} of this canonical pair 
of sets.
In a canonical pair of sets ${\bf X} = \{X_f, X_{\widetilde f}\}$,
the first set $X_f$ will be called a {\bf set of Mayer edges},
and the second set $X_{\widetilde f}$ will be called a {\bf set of 
Boltzmann edges}. $\blacksquare$

By ${\mathfrak X}_n = \{{\bf X}=(X_f,\, X_{\widetilde f})\}$
we denote the totality of all canonical pairs of sets of order $n$.
Note that in a pair ${\bf X} = (X_f, X_{\widetilde f})$, included
in the totality ${\mathfrak X}_n$, the set of Boltzmann edges
$X_{\widetilde f}$ can be empty.

To each canonical pair of sets ${\bf X} = (X_f, X_{\widetilde f})$
of order $n$ we assign the product of Mayer and
Boltzmann functions $P_n({\bf X})$ defined by the formula
\begin{equation}
 P_n({\bf X}) = \prod_{\{i,j\} \in X_f({\bf X})}
\prod_{\{i',j'\} \in X_{\widetilde f}
({\bf X})}f_{ij}\widetilde f_{i'j'}.
\label{8}
\end{equation}
Obviously, the product of Mayer and Boltzmann functions
$P_n({\bf X})$
is the restriction to the set ${\mathfrak X}_n$ of the function
$P(F, \widetilde F)$, defined by formula (8).

D\,e\,f\,i\,n\,i\,t\,i\,o\,n 6 [39]. We will say that a canonical
pair of sets ${\bf X} = (X_f, X_{\widetilde f})$ of order $n$
{\bf defines} the product of functions $P_n({\bf X})$ and call this
product of functions a {\bf canonical product}, and
number $n$ is {\bf order} of this product. $\blacksquare$

By ${\mathfrak P}_n = \{P \colon P =
P_n({\bf X}), \; {\bf X} \in {\mathfrak X}_n\}$ denote
the set of all canonical products defined by
canonical pairs of sets from the totality ${\mathfrak X}_n$.

From the definitions of the totality ${\mathfrak X}_n$, of
the set ${\mathfrak P}_n$ and of the product $P_n({\bf X})$
by formula (11) it follows that the correlation
\begin{equation}
P = P_n({\bf X})
\label{12}
\end{equation}
between the elements ${\bf X} \in {\mathfrak X}_n$ and
$P \in {\mathfrak P}_n$ is a mapping of the totality
${\mathfrak X}_n = \{{\bf X}\}$ onto the set ${\mathfrak P}_n = \{P\}$.

Note that the mapping $P_n \colon {\mathfrak X}_n \to {\mathfrak P}_n$
is a one-to-one mapping of the totality ${\mathfrak X}_n$
onto the set ${\mathfrak P}_n$.
Since each functions product $P$ from the set
${\mathfrak P}_n $ under the mapping $P_n$ has, and, moreover,
the only one, preimage ${\bf X} = (X_f, X_{\widetilde f})$
in the totality ${\mathfrak X}_n$, then this preimage can be taken as
the label of this product and this product can be considered labeled
with the canonical pair of sets ${\bf X} = (X_f, X_{\widetilde f})$.
At that, any canonical pair of sets
${\bf X} = (X_f, X_{\widetilde f})$ from the totality
${\mathfrak X}_n$ turns out to be the label of the canonical product
of functions, which is included in the set ${\mathfrak P}_n$ and
is uniquely defined by this pair of sets by formulas (12) and (11).
Other methods of labeling the canonical products of functions will be
described below. All these methods have found their application
in this article.

Let us denote by
${\mathfrak G}_n = \{G(V_n;\, X_f,\, X_{\widetilde f})\}$ a set of all
labeled graphs with the vertex set $V_n = \{1, 2, \ldots, n\}$ and
an edges set $X$, which is the union of two disjoint
sets: a set $X_f = \{\{i, j\}\}$ and a set
$X_{\widetilde f} = \{\{i', j'\}\}$, is forming a canonical pair of
sets $(X_f, X_{\widetilde f}) \in {\mathfrak X}_n$.

For graphs belonging to the set
${\mathfrak G}_n = \{G(V_n;\, X_f,\, X_{\widetilde f})\}$,
we introduce the notation: 
$X_f(G) = X_f, \; X_{\widetilde f}(G) = X_{\widetilde f}$ where 
$G = G(V_n;\, X_f,\, X_{\widetilde f}) \in {\mathfrak G}_n$. 
The edges set $ X_f (G) $ will be called {\bf the set of Mayer edges 
of the graph} $G \in {\mathfrak G}_n$, and the set 
$X_{\widetilde f}(G)$ will be called {\bf the set of Boltzmann edges 
of the graph} $G \in {\mathfrak G}_n$.

We define a mapping $A_n$ of the set ${\mathfrak G}_n$ onto the set
${\mathfrak X}_n$, setting
\begin{equation}
A_n(G) = (X_f(G), X_{\widetilde f}(G)),
\label{13}
\end{equation}
where $G \in {\mathfrak G}_n$. The mapping $A_n$ defined 
by formula (13) is a one-to-one mapping
of the set ${\mathfrak G}_n $ onto the set ${\mathfrak X}_n$.

Recall that the mapping $P_n$, defined by the formulas (11) and (12),
is a mapping of the set ${\mathfrak X}_n$ onto the set
${\mathfrak P}_n$. Hence, there is the mappings composition
$P_n \circ A_n$, which is a map of the set ${\mathfrak G}_n$
onto the set ${\mathfrak P}_n$. Since the mappings $A_n$ and $P_n$ are
one-to-one, their composition $P_n \circ A_n$ is also [22, 40]
one-to-one.

{\bf Remark 1} [39]. Each product of functions $P$ from the set
${\mathfrak P}_n$ under the mapping $P_n \circ A_n$ has, and moreover
unique, preimage in the set ${\mathfrak G}_n$. This means that
this preimage can be taken as a graph-label of this product and this
product can be considered labeled.
Moreover, any graph $G(V_n; X_f, X_{\widetilde f})$ from the set
${\mathfrak G}_n$ turns out to be a label of a functions product,
which we will denote $P_{1n}(G)$. This product is included
in the set ${\mathfrak P}_n$ and is uniquely defined by this graph
according to the formula
\begin{multline}
P_{1n}(G) = (P_n\circ A_n)(G) = P_n(A_n(G)) =
P_n((X_f(G), X_{\widetilde f}(G))) = \\
\prod_{\{i,j\} \in X_f(G)}\prod_{\{i',j'\} \in X_{\widetilde f}(G)}
f_{ij}\widetilde f_{i'j'}.
\label{14}
\end{multline}
\hbox to \textwidth{\hfil \raisebox{5pt}[0pt][0pt]{$\blacksquare$}}
Since the product $P_{1n}(G)$ is included in the set
${\mathfrak P}_n$, then the definition of this set implies that
the product $P_{1n}(G)$ is canonical.

Based on Remark 1, we formulate the following

D\,e\,f\,i\,n\,i\,t\,i\,o\,n 7 [39]. If a graph
$G(V_n;\, X_f,\, X_{\widetilde f})$ belongs
to the set ${\mathfrak G}_n$, then the canonical functions product
$P_{1n}(G)$ defined by formula (14) will be called
{\bf the product labeled with the graph}
$G = G(V_n; X_f, X_{\widetilde f})$, and the graph
$G = G(V_n; X_f, X_{\widetilde f})$ will be called 
{\bf the graph-label} of this product of functions. $\blacksquare$

Let us consider a graph $G = G(V_n; X_f, X_{\widetilde f})$, belonging
to the set of graphs ${\mathfrak G}_n$.
We denote by $R(G) = (V_n; X_f)$ the graph with the set of vertices
$V_n$ and the set of edges $X_f$. The graph $R(G)$ is a subgraph
of the graph $G$. By definition, the set of edges of the graph $R(G)$
is the set $X_f(G)$ of Mayer edges of the graph $G$. This set of edges
defines the set of Mayer functions included in the functions 
product $P_{1n}(G)$. But the graph $R(G)$, by definition, does not
contain, unlike the graph $G$, the set $X_{\widetilde f}(G)$
of Boltzmann edges. By Definition 4 this set $X_{\widetilde f}(G)$
of Boltzmann edges defines the set of Boltzmann functions included
in the functions product $P_{1n}(G)$. Therefore, we will call
subgraph $R(G)$ of graph $G$ {\bf insufficient label}
of the functions product $P_{1n}(G)$ labeled with the graph $G$.

D\,e\,f\,i\,n\,i\,t\,i\,o\,n 8 [39]. Product of functions
$P \in {\mathfrak P}_n$ will be called {\bf base product of order $n$},
if its graph-label $G \in {\mathfrak G}_n$ satisfies the condition:
the subgraph $R(G)$ of the graph $G$ is a connected graph.
If the subgraph $R(G)$ of the graph-label $G \in {\mathfrak G}_n$
is not connected, then the product of functions $P$ labeled
with the graph $G$ will be called {\bf pseudobase product}.
$\blacksquare$

Let's introduce the notation:
${\mathfrak P}_{bn} = \{P\}$ is the set of all base products,
belonging to the set ${\mathfrak P}_n$;
${\mathfrak G}_{bn}$ is the set of all graphs that are graphs-label
of base products belonging to the set ${\mathfrak P}_{bn}$.

Definitions 7 and 8 and Remark 1 imply

{\bf Corollary 1.} {\it The sets ${\mathfrak P}_{bn}$ and
${\mathfrak G}_{bn}$ are in one-to-one correspondence.}

{\bf Lemma 1} [39]. {\it If the subgraph $R(G)$ of a graph-label
$G \in {\mathfrak G}_n$ is connected, then, firstly, each edge
from the set $X_{\widetilde f}(G)$ connects two non-adjacent vertices
of the graph $R(G)$ and, secondly, the canonical product $P_{1n}(G)$,
which is labeled with graph $G$, is a function of $n$ variables
${\bf r}_1, {\bf r}_2, \ldots, {\bf r}_n$.}

{\bf Proof.} Since any edge from the set $X_{\widetilde f}(G)$
belongs to the graph $G$ by the definition of this graph, then
both vertices incident to this edge belong to the set $V_n$.
Therefore, these vertices belong to the graph $R(G)$ by its definition.
From the conditions of the lemma by Definition 8 it follows that
the graph $G$ belongs to the set
${\mathfrak G}_n$. From here by the definition of this set it follows
that the sets $X_{\widetilde f}$ and $X_f$ have no common edges and
form a canonical pair of order $n$. This means that the set $X_f$
does not contain an edge connecting two vertices incident to some edge
from the set $X_{\widetilde f}(G)$. Hence, each edge from the set
$X_{\widetilde f}$ connects two non-adjacent vertices of the graph
$R(G)$. The first assertion of the lemma is proved. 

Let us now prove the second assertion of the lemma.
Let $i$ be a vertex belonging to the set $V_n$. As
the subgraph $R(G) = (V_n; X_f)$ of the graph $G$ is connected,
then in the set of edges $X_f(G)$ there exists an edge connecting
the vertex $i$ with some vertex $j \in V_n$. Hence, by the definition
of the product $P_{1n}(G)$ by formula (14), it follows that
the Mayer function $f_{i j}$ is included in this product. And since
the Mayer function $f_{i j}$ by the definition is a function
of the variables ${\bf r}_i$ and ${\bf r}_j$, then these
variables are included in the set of variables of the functions
product $P_{1n}(G)$. Thus, for any $i \in V_n$ the variable
${\bf r}_i$ is a variable of the function that is the functions
product $P_{1n}(G)$.

On the other hand, if $i \notin V_n$, then $i$ is not a vertex
of the graph $G$ and cannot be a vertex incident to any edge
of this graph. Therefore, it follows from the definition
of the product $P_{1n}(G)$ that the variable ${\bf r}_i$ is not
a variable of any of the functions, included in this product.
The results obtained imply the second assertion of the lemma.
$\blacktriangleright$

Lemma 1 implies the following.

{\bf Corollary 2} [39]. {\it A base product $P \in {\mathfrak P}_{bn}$
is a function of $n$ variables
${\bf r}_1, {\bf r}_2, \ldots, {\bf r}_n$,
where $n$ is the number of vertices of the graph-label $G$.}

D\,e\,f\,i\,n\,i\,t\,i\,o\,n 9. If the integrand of an integral 
is a base product $P \in {\mathfrak P}_{bn}$ of order $n$, and 
the integration domain of this integral is either real space 
$({\bf R}^{\nu})^{n - 1}$, or a connected bounded Lebesgue measurable 
set contained in the space $({\bf R}^{\nu})^n$, then this integral 
will be called a {\bf base integral}, and the number $n$ will be 
called its {\bf order}. $\blacksquare$

Let $G \in {\mathfrak G}_{bn}$, and $U$ be a connected bounded
Lebesgue measurable set contained in the space $({\bf R}^{\nu})^n$.
Let's introduce the notation:
\begin{equation}
I(G, U) = 
\int_U P_{1n}(G)(d{\bf r})_n
\label{15"}
\end{equation}
\begin{equation}
I(G) = I(P_{1n}(G)) = 
\int_{({\bf R}^{\nu})^{n-1}} P_{1n}(G)(d{\bf r})_{1, n-1},
\label{15'}
\end{equation}
where 
$(d{\bf r})_{1, n-1} = d{\bf r}_2d{\bf r}_3 \ldots d{\bf r}_n$.

{\bf Theorem 1.} {\it If the potential of the pairwise 
interaction $\Phi({\bf r})$ is a measurable function, the pairwise 
interaction satisfies the conditions of stability 
and regularity, and the graph $G$ belongs 
to the set ${\mathfrak G}_ {bn}$, then the following statements are 
true:

$A_1)$ the function $P_{1n}(G)$ is integrable over the space 
$({\bf R}^{\nu})^{n - 1}$, and the integral $I(G)$ converges and 
does not depend on the value of the variable $\bf r_1$;

$A_2)$ the function $P_{1n}(G)$ is integrable on any connected bounded
Lebesgue measurable set $U$ contained in the space
$({\bf R}^{\nu})^n$, and the integral $I(G, U)$ converges.}

{\bf Proof.} First of all, note that the regularity of the pairwise
interaction means that the Mayer function $f({\bf r})$ at some $C > 0$
satisfies the inequality
\begin{equation}
\int_{{\bf R}^{\nu}}|f({\bf r})|d{\bf r} < C.
\label{a18}
\end{equation}

Recall that this article considers only systems of particles with 
a pairwise interaction. In such systems, the interaction is stable 
in if and only if there is a number $B \ge 0$ such that for all 
$n > 1$, the inequality
\begin{equation}
\sum_{1 \le i < j \le n }\Phi({\bf r}_i - {\bf r}_j) > -nB.
\label{a20}
\end{equation}
takes place. In particular, for $n = 2$, the inequality
\begin{equation}
\Phi({\bf r}_1 - {\bf r}_2) > -2B.
\label{a21}
\end{equation}
takes place. Therefore, the Boltzmann function 
$\widetilde f({\bf r})$ satisfies the inequality
\begin{equation}
\widetilde f({\bf r}) < \exp(2\beta B).
\label{a22}
\end{equation}
It follows that the Mayer function $f({\bf r})$ for some $D \ge 1$ 
satisfies the inequality
\begin{equation}
|f({\bf r})| < D.
\label{a23}
\end{equation}

From the definition of the function $P_{1n}(G)$ 
by the formula (14) and from the inequalities (20) 
and (21) it follows that the function $P_{1n}(G)$
for some $E > 0$ satisfies the inequality
\begin{equation}
|P_{1n}(G)| < E.
\label{a25}
\end{equation}
for all $({\bf r})_n \in ({\bf R}^{\nu})^n$.

Since the potential of pairwise interaction $\Phi({\bf r})$ is
measurable function, and Boltzmann function $\widetilde f$ by its
definition is a continuous function of this potential $\Phi$,
then, by the properties of measurable functions [21], Boltzmann 
function $\widetilde f$ is also measurable. Hence, by the properties 
of measurable functions [21] it follows that the Mayer function 
$f({\bf r})$ is measurable.

By Lemma 1, the function $P_{1n}(G)$ is a function of the $n$ 
variables ${\bf r}_1, {\bf r}_2, \ldots, {\bf r}_n$. And according to
its definition by formula (14), this function is the product 
of a finite number of functions, which, as we have already established, 
are measurable.

So, the function $P_{1n}(G)$ is a product of a finite number of
measurable functions and is defined in real space
$({\bf R}^{\nu})^n$. Hence, by the properties of measurable functions,
it follows that the function $P_{1n}(G)$ is a measurable function 
in the space $({\bf R}^{\nu})^n$. From this and from the inequality 
(22) by the properties of integrable functions it follows 
that the function $P_{1n}(G)$ is integrable on any connected bounded 
Lebesgue measurable set $U$, contained in the space 
$({\bf R}^{\nu})^n$ and the integral $I(G, U)$ converges.

It follows from the conditions of the theorem that the graph $R(G)$ 
is connected. Therefore, there is a tree $t(G)$, which is a subgraph 
of the graph $R(G)$. Therefore, the integrand $P_{1n}(G)$ 
of the integral $I(G)$ can be present as follows
\begin{equation}
P_{1n}(G) = \Omega({\bf r})_n\prod_{\{i,j\} \in 
X(t(G))}y({\bf r}_i - {\bf r}_j),
\label{a15'}
\end{equation}
where
\begin{equation}
\Omega({\bf r})_n = 
\prod_{\{ij\} \in [X_f(G) \setminus X(t(G))]}
f_{ij}({\bf r}_i - {\bf r}_j)
\prod_{\{i' j'\} \in X_{\widetilde f}(t(G))}
\widetilde f_{i' j'}({\bf r}_{i'} - {\bf r}_{j'}),
\label{a16}
\end{equation}
\begin{equation}
  y({\bf r}) = f({\bf r}).
\label{a17}
\end{equation}
From the inequalities (17) and (21) and from the definition (25) 
of the function $y({\bf r})$ it follows that the function $y({\bf r})$
also satisfies inequalities
\begin{equation}
\int_{{\bf R}^{\nu}}|y({\bf r})|d{\bf r} < C. 
\label{a19}
\end{equation}
and
\begin{equation}
|y({\bf r})| < D.
\label{a24}
\end{equation}

From the definition of the function $\Omega$ by the formula 
(24) and from the inequalities (20) and (21)
it follows that the function $\Omega({\bf r})_n$ for some $E'> 0$ 
the inequality
\begin{equation}
|\Omega({\bf r})_n| < E'
\label{a27}
\end{equation}
satisfies.

Since Mayer function $f({\bf r})$ is measurable, then 
by the properties of measurable functions [21] it follows that 
the function $y({\bf r})$, defined by the formula (25) 
is also measurable in the space ${\bf R}^{\nu}$.

The function $\Omega({\bf r})_n$, defined by the formula (24),
is a product of a finite number of functions, which, as we have 
already established, are measurable in their definition domain. 
Hence, by the properties of measurable functions [21], it follows 
that the function $\Omega({\bf r})_n$ is measurable in the space 
$({\bf R}^{\nu})^n$. From the definition of the function 
$\Omega({\bf r})_n$ by the formula (24) it follows that this 
function is a translationally invariant function [17], 
[24], [49].

So, the integrand $P_{1n}(G)$ of the integral $I(G)$ is represented 
by the formula (23), where the measurable function 
$y({\bf r})$ satisfies the inequalities (26) and (27), 
and the measurable function $\Omega({\bf r})_n$ satisfies 
the inequality (28) and is a translationally invariant 
function. Hence, by Theorem 3 from Chapter III of [17], it follows 
that the function $P_{1n}(G)$ represented by the formula (23) 
is a function integrable over the space $({\bf R}^{\nu})^{n -1}$, 
and the improper integral $I(G)$ converges and does not depend 
on the value of the variable $\bf r_1$. Theorem 1 is proved.
$\blacktriangleright$

{\bf Remark 2.} Since the article deals only with particles systems
 satisfying the conditions of Theorem 1, then every improper integral 
$I(G)$, taken over the space $({\bf R}^{\nu})^{n-1}$ and labeled 
by a graph $G \in {\mathfrak G}_{bn}$, and every integral 
of the form $I(G, U)$, labeled by a graph 
$G \in {\mathfrak G}_{bn}$ and taken over any connected bounded 
Lebesgue measurable set $U$ contained in the space 
$({\bf R}^{\nu})^n$, satisfy conditions of Theorem 1 and 
are convergent by Theorem 1. $\blacksquare$

D\,e\,f\,i\,n\,i\,t\,i\,o\,n 10 [39]. An Integral of a pseudobase 
product of functions will be called a {\bf pseudobase integral}. 
$\blacksquare$

D\,e\,f\,i\,n\,i\,t\,i\,o\,n 11. If in a linear combination $L$ 
of convergent base integrals of order $n$ all integrals have one 
and the same integration domain $U(L)$, and the coefficient 
for each of the integrals included in it is a real number and 
is defined by the graph labeling the integrand of this integral,  
then the linear combination $L$ is called a {\bf base linear 
combination}, the number $n$ is called its {\bf order}, and 
the integration domain $U(L)$ is called  a {\bf set, associated with 
the given linear combination} $L$.  $\blacksquare$

{\bf Remark 3.} Definition 11 implies that any base integral 
of a given base linear combination is completely defined by
the set, which is associated with a given linear combination, and 
by its integrand, which, being the base product 
$P \in {\mathfrak P}_{bn}$, is defined by the graph-label
$G \in {\mathfrak G}_{bn}$ of this base product. Hence,
any base integral of a given base linear combination is completely
defined by the set associated with the given linear combination and by
the graph-label $G$ of the base product, which is its integrand.
 $\blacksquare$

D\,e\,f\,i\,n\,i\,t\,i\,o\,n 12. If in a linear combination
of integrals of products of Mayer and Boltzmann functions at least
one integral is not convergent base integral, then this linear
combination of integrals is called a {\bf pseudo-base linear
combination}. $\blacksquare$
                   
{\bf Example 1} Consider Ree-Hoover representation [48]
of a virial coefficient $B_n(\LL)$ for $n \ge 2$. It was stated 
above that this representation is a linear combination of integrals. 
In each of these integrals, the integrand is a product of Mayer and
Boltzmann functions. The definition of Ree-Hoover representation
of the virial coefficient $B_n(\LL)$ implies that in this linear 
combination each integral is labeled (in sense Ree-Hoover [48]) 
with some complete graph $G(V_n;\, X_f,\, X_{\widetilde f})$.
Moreover, the edges set $X_f$ by Definition 3 defines the set
$F = \{f_{i j}\}$ of Mayer functions included as factors
in the integrand of the integral, labeled with the graph
$G(V_n;\, X_f,\, X_{\widetilde f})$; and the edges set
$X_{\widetilde f}$ by Definition 4 defines the set
\mbox{$\widetilde F = \{\widetilde f_{i 'j'}\}$} of Boltzmann
functions included as factors in this integrand.

From the definition of the Ree-Hoover representation of the virial 
coefficient $B_n(\LL)$ it follows that the sets $X_f$ and 
$X_{\widetilde f}$ of the graph $G$ are disjoint and form 
a sets canonical pair ${\bf X} = (X_f, X_{\widetilde f})$ 
of order $n$. Two conclusions follow from this:
1) by Definition 6, the integrand of the integral labeled 
(in sense Ree-Hoover) with the graph $G$, is the canonical product 
$P_n({\bf X})$ of order $n$, defined by  the sets canonical pair 
$({\bf X}) = ((X_f, \, X_{\widetilde f}))$ according to formula (11);
2) the graph $G(V_n; \, X_f, \, X_{\widetilde f})$ belongs to the set
${\mathfrak G}_n$ by the definition of this set.

From conclusion 2) by Definition 7, it follows that the graph 
$G(V_n; \, X_f, \, X_{\widetilde f})$ is the graph-label 
of the functions product $P_{1n}(G)$, which is the product labeled 
by this graph, is uniquely defined by this graph according 
to formula (14) and, by Remark 1,  belongs to the set 
${\mathfrak P}_n$.

From the definition of the product of functions $P_{1n}(G)$ 
by formula (14) it follows that this product is 
the canonical product $P(X_f, \, X_{\widetilde f})$ of order $n$,
which is the integrand of the integral included in considered 
Ree-Hoover representation and labeled (in sense Ree-Hoover [48]) 
by the graph $G$.
Since in this case the subgraph $R(G)$ of the graph
$G(V_n; \, X_f, \, X_{\widetilde f})$ is, as is known [48],
doubly connected graph, then by Definition 8 this integrand is 
a base product of order $n$. This base product belongs to the set 
${\mathfrak P}_{bn}$ by the definition of this sets. And 
the graph-label $G$ of this base product belongs to the set 
${\mathfrak G}_{bn}$ by the definition of this set.

So, the integrand of any integral, that is included in Ree-Hoover 
representation of the virial coefficient $B_n(\LL)$, is a base 
product labeled by the complete graph  belonging to the set 
${\mathfrak G}_{bn}$ and labeling (in sense Ree-Hoover)  
this integral. This integrand is defined by the formula (14), 
where $G$ is the above graph.
From the formula (14) it follows that the number of Mayer and 
Boltzmann functions included in the canonical product labeled 
by a complete graph with $n$ vertices is equal to the number 
$n (n - 1) / 2$ of edges of this graph.

In [48], Ree and Hoover considered systems of particles enclosed 
in a bounded volume $\LL$ and obtained representations of the virial 
coefficients $B_n(\LL)$ for a case of a bounded volume $\LL$ 
as  integrals linear combination in which all integrals have 
the same domain of integration $\LL^n$. We can hold that 
Ree-Hoover representations are integrals linear combinations, 
in each of which all integrals have the same integration domain, 
completely defined by this linear combination.

In what follows, we will assume that the set $\LL^n$ is connected, 
bounded, and Lebesgue measurable. Since in this case the integrand 
of each integral of this linear combination is a base product
of order $n$, then, by Definition 9, each integral in 
the linear combination that is Ree-Hoover representation 
of a virial coefficient $B_n(\LL)$ is a base integral of order $n$.

So, under the above conditions, the Ree-Hoover representation 
of the virial coefficient $B_n(\LL)$ has the following properties:
1) this representation is a linear combination of the
integrals whose domain of integration is the connected
bounded and Lebesgue measurable set contained in
space $({\bf R}^{\nu})^n$;
2) the integrand of each integral of this linear combination
is a base product whose graph-label belongs to the set 
${\mathfrak G}_{bn}$.

This article deals only with thermodynamic equilibrium one-component 
systems of classical particles with pair interaction [24, 49]. 
In this case, it is assumed that the pair interaction satisfies 
the conditions of stability and regularity, and the pair potential 
$\Phi({\bf r})$ is a measurable function.
Under these restrictions and for $n \ge 2$, the integrands
of all integrals included in the Ree-Hoover representation 
of the virial coefficient $B_n(\LL)$, by Theorem 1, are integrable 
on any connected, bounded and Lebesgue measurable set $U$,
contained in the space $({\bf R}^{\nu})^n$, and all these integrals 
converge.

So, in the case when systems of particles enclosed in a bounded 
volume satisfies the conditions listed above in this example, 
for $n \ge 2$ the Ree-Hoover representation of the virial coefficient 
$B_n(\LL)$ is a linear combination of converging base integrals.

As is known [48], the integrals linear combination, which is 
Ree-Hoover representation of the virial coefficient $B_n(\LL)$, 
satisfies the condition: the coefficient of each integral included 
in this linear combination is a real number and is defined by 
the graph labeling (in sense Ree-Hoover) this integral. Based on this 
fact and the fact that everyone included in this linear combination 
integrals are convergent base integrals of order $n$, having 
the same domain of integration, 
we come to the conclusion: by Definition 11, this linear combination 
is a base one of order $n$. So, in the cases considered in this 
example, Ree-Hoover representation of the virial coefficient 
$B_n(\LL)$ for $n \ge 2$ is a base linear combination of order $n$. 

According to Remark 3, each integral in this linear combination 
is completely defined by its integrand and the set, associated 
with this linear combination. It has been established above that this 
integrand is a base product of order $n$ belonging to the set 
${\mathfrak P}_{bn} \subset {\mathfrak P}_n$ and labeled with 
the labeled graph $G$ belonging to the set ${\mathfrak G}_{bn}$.
By Corollary 1, this base product is uniquely determined by its 
graph-label $G \in {\mathfrak G}_{bn}$. Therefore, each 
integral in this linear combination is completely defined by the set,
associated with the given linear combination, and by the graph-label 
of the base product, which is the integrand of this integral.
 $\blacktriangleright$

Let's introduce the notation:

${\mathfrak G}(L)$ is the set of all graphs serving as graphs-labels
of such the base products that are the integrands of the integrals
included in the base linear combination $L$;
\begin{equation}
R({\mathfrak G}(L)) = \{R(G): \quad G \in {\mathfrak G}(L)\}.
\end{equation}

D\,e\,f\,i\,n\,i\,t\,i\,o\,n 13. If $L$ is a base linear combination, 
then the set of graphs ${\mathfrak G}(L)$ will be called {\bf the set 
of graphs-labels} of this base linear combination, and the number 
of integrals included in it will be called the {\bf length} of this 
linear combination and denote by $q(L)$. $\blacksquare$ 

There are often cases when for labeling a canonical product
of functions $P \in {\mathfrak P}_n$ it is easier to use other graphs
rather than the graph-label of such a product of functions. For example,
to use the graph $\widetilde G(V_n, X_f)$, where $X_f$
is the set of Mayer edges with respect to the set $F$ of all Mayer
functions, included in this canonical product of functions
$P \in {\mathfrak P}_n$.

The graph $\widetilde G(V_n, X_f)$ makes it possible directly to
define only Mayer functions included in the functions product
$P(X_f, X_{\widetilde f})$. To define the Boltzmann functions
included in such a product, in some cases it is preferable,
bypassing the definition of the graph-label of such a product,
directly to specify the set $X_{\widetilde f}$ of Boltzmann edges
with respect to the set $\widetilde F$ of all Boltzmann functions,
included in this canonical product $P \in {\mathfrak P}_n$,
or to specify a constructive method for constructing this set. This 
gives the ability to directly define the Boltzmann functions included
into the functions product labeled with the graph $\widetilde G$.
The set $X_{\widetilde f}$ complements the set of edges of the graph
$\widetilde G$ to the set of edges of the graph-label of this product.
Let's call this set {\bf complementary} and denote by
$X_{\rm ad}(\widetilde G)$, setting
$X_{\rm ad}(\widetilde G) = X_{\widetilde f}$.

We denote by $\widetilde{\mathfrak G}_n = \{\widetilde G\}$, where
$n \ge 3$, a finite set of pairwise distinct connected labeled
graphs that has the set $V_n$ as their set of vertices and satisfies
the condition: for each graph from this set it  is defined
the complementary set $X_{\rm ad}(\widetilde G)$, that is put in 
correspondence to this graph, and does not intersect with Mayer edges 
set $X_f(\widetilde G)$ and forms with it a canonical pair
$(X_f(\widetilde G), X_{\rm ad}(\widetilde G)) \in {\mathfrak X}_n$.

D\,e\,f\,i\,n\,i\,t\,i\,o\,n 14 [39]. Graphs from a set
$\widetilde{\mathfrak G}_n$ will be called {\bf completed}.
$\blacksquare$

Let's introduce the notation:

${\mathfrak X}({\widetilde{\mathfrak G}_n}) =
\{(X_f(\widetilde G), X_{\rm ad}(\widetilde G)) \colon
\widetilde G \in \widetilde{\mathfrak G}_n\}$

${\mathfrak P}({\widetilde{\mathfrak G}_n})=
P_n({\mathfrak X}({\widetilde{\mathfrak G}_n}))$ is the image
of the set of canonical pairs
${\mathfrak X}({\widetilde{\mathfrak G}_n}) \subset {\mathfrak X}_n$
under the map \mbox{$P_n\colon {\mathfrak X}_n \to {\mathfrak P}_n$};

${\mathfrak P}_{\widetilde{\mathfrak G}_n} = P_n
\mid_{{\mathfrak X}({\widetilde{\mathfrak G}_n})}$ is
 the restriction of mapping $P_n$ on the subset
${\mathfrak X}({\widetilde{\mathfrak G}_n}) \subset {\mathfrak X}_n$.

By definition, the mapping ${\mathfrak P}_{\widetilde{\mathfrak G}_n}$
is the one-to-one mapping the set
${\mathfrak X}({\widetilde{\mathfrak G}_n})$ on the set
${\mathfrak P}({\widetilde{\mathfrak G}_n})$.

We define a mapping $A_{\widetilde{\mathfrak G}_n}$ of the set
$\widetilde {\mathfrak G}_n$ to the set ${\mathfrak X}({\widetilde{\mathfrak G}_n})$,
letting that
\begin{equation}
A_{\widetilde{\mathfrak G}_n}(\widetilde G) =
(X_f(\widetilde G), X_{\rm ad}(\widetilde G)), \quad
\widetilde G \in \widetilde{\mathfrak G}_n.
\label{13'}
\end{equation}
The mapping $A_{\widetilde{\mathfrak G}_n}$ defined by formula
(30) is the one-to-one mapping of the set
$\widetilde {\mathfrak G}_n$
on the set ${\mathfrak X}({\widetilde{\mathfrak G}_n})$.

{\bf Remark 4}. Since the definition domain of the mapping
$P_{\widetilde{\mathfrak G}_n}$ is the same as the values domain 
of the mapping $A_{\widetilde{\mathfrak G}_n}$, then the composition
of the mappings
$P_{\widetilde{\mathfrak G}_n} \circ A_{\widetilde{\mathfrak G}_n}$
exists and is the mapping of the set $\widetilde{\mathfrak G}_n$
on the set ${\mathfrak P}({\widetilde{\mathfrak G}_n})$.

Since the mappings $A_{\widetilde{\mathfrak G}_n} \colon
\widetilde{\mathfrak G}_n \to
{\mathfrak X}({\widetilde{\mathfrak G}_n})$ and
$P_{\widetilde{\mathfrak G}_n}\colon
{\mathfrak X}({\widetilde{\mathfrak G}_n}) \to
{\mathfrak P}({\widetilde{\mathfrak G}_n})$ are the
one-to-one mappings, then their composition
$P_{\widetilde{\mathfrak G}_n}\circ A_{\widetilde{\mathfrak G}_n}
\colon \widetilde{\mathfrak G}_n \to {\mathfrak P}({\widetilde{\mathfrak G}_n})$
is [22, 40] the one-to-one mapping of the set
$\widetilde{\mathfrak G}_n$ to the set
${\mathfrak P}({\widetilde{\mathfrak G}_n})$.
$\blacksquare$

Remark 4 implies

{\bf Corollary 3} [39]. {\it When mapping 
$P_{\widetilde{\mathfrak G}_n} \circ A_{\widetilde{\mathfrak G}_n}$, 
each functions product $\widetilde P$ from the set
${\mathfrak P}({\widetilde{\mathfrak G}_n})$  has, and at that
the only, preimage in the set $\widetilde{\mathfrak G}_n$.
This means that this preimage is a graph,
which can be taken as a label of this product, and this product can
be considered labeled with this graph. At that, every graph
$\widetilde G$ from the set $\widetilde{\mathfrak G}_n$ turns out
to be the label of the functions  product, which is the image of this
graph when mapping
$P_{\widetilde{\mathfrak G}_n} \circ A_{\widetilde{\mathfrak G}_n} 
\colon \widetilde{\mathfrak G}_n \to 
{\mathfrak P}({\widetilde{\mathfrak G}_n})$.}

Image of the graph $\widetilde {G} \in \widetilde{\mathfrak G}_n$
under the mapping $P_{\widetilde{\mathfrak G}_n} \circ
A_{\widetilde{\mathfrak G}_n} \colon
\widetilde{\mathfrak G}_n \to
{\mathfrak P}({\widetilde{\mathfrak G}_n})$ denote
$\widetilde P_{\widetilde{\mathfrak G}_n}(\widetilde G)$.

Based on Remark 4 and Corollary 3, we formulate the following

D\,e\,f\,i\,n\,i\,t\,i\,o\,n 15 [39]. The functions product
$\widetilde P_{\widetilde{\mathfrak G}_n}(\widetilde G)$, which is
the image of a graph
$\widetilde {G}(V_n, X_f) \in \widetilde{\mathfrak G}_n$ under
the mapping 
\mbox{$P_{\widetilde{\mathfrak G}_n} \circ
A_{\widetilde{\mathfrak G}_n} \colon
\widetilde{\mathfrak G}_n \to 
{\mathfrak P}({\widetilde{\mathfrak G}_n})$}, we will
call the {\bf product labeled with the graph}
$\widetilde G = \widetilde G(V_n, X_f)$, and the graph
$\widetilde{G}(V_n, X_f)$ is the {\bf completed graph-label} of this
product. $\blacksquare$

{\bf Lemma 2} [39]. {\it If a graph $\widetilde{G}(V_n, X_f)$
belongs to the set $\widetilde{\mathfrak G}_n$, then the functions
product $\widetilde P_{\widetilde{\mathfrak G}_n}(\widetilde G)$
labeled with this graph is a canonical product of order $n$.
In this case this product is represented by the formula }
\begin{equation}
\widetilde P_{\widetilde{\mathfrak G}_n}(\widetilde G) =
\prod_{\{i,j\} \in X_f(\widetilde G)}\prod_{\{i',j'\} \in X_{\rm ad}(\widetilde G)}
f_{ij}\widetilde f_{i'j'}.
\label{14'}
\end{equation}

{\bf Proof.}  Let us first prove that the functions product
$\widetilde P_{\widetilde{\mathfrak G}_n}(\widetilde G)$ is
a canonical one of order $n$. From the definition of the set
${\mathfrak P}({\widetilde{\mathfrak G}_n})$ it follows that this set
is a subset of the set ${\mathfrak P}_n$ of canonical products
of the order $n$. From this and Remark 4 it follows that the set
 of values of the mapping
$P_{\widetilde{\mathfrak G}_n} \circ A_{\widetilde{\mathfrak G}_n}
\colon \widetilde{\mathfrak G}_n \to
 {\mathfrak P}({\widetilde{\mathfrak G}_n})$ is a set of
canonical products of order $n$. Therefore, whatever a graph
$\widetilde{G}(V_n, X_f) \in \widetilde {\mathfrak G}_n$, its image
$\widetilde P_{\widetilde{\mathfrak G}_n}(\widetilde G)$ under the mapping
$P_{\widetilde{\mathfrak G}_n} \circ A_{\widetilde{\mathfrak G}_n}
\colon \widetilde{\mathfrak G}_n \to
{\mathfrak P}({\widetilde{\mathfrak G}_n})$ is a canonical
product of order $n$. By Definition 15, the product
$\widetilde P_{\widetilde{\mathfrak G}_n}(\widetilde G)$ is a product labeled
with the graph $\widetilde G$. So, it is proved that the functions
product $\widetilde P_{\widetilde{\mathfrak G}_n}(\widetilde G)$ labeled
with the graph $\widetilde {G} \in \widetilde {\mathfrak G}_n$  is
a canonical product of order $n$.

Let us now prove that the functions product
$\widetilde P_{\widetilde{\mathfrak G}_n}(\widetilde G)$, which is
labeled with the  graph
$\widetilde {G} \in \widetilde{\mathfrak G}_n$,
is represented by formula (31).
From the definition of the functions product
$\widetilde P_{\widetilde{\mathfrak G}_n}(\widetilde G)$,
the definitions of the mapping
$P_{\widetilde{\mathfrak G}_n} \colon
{\mathfrak X}({\widetilde{\mathfrak G}_n}) \to {\mathfrak P}({\widetilde{\mathfrak G}_n})$,
the definitions of the mapping $P_n \colon{\mathfrak X}_n \to
{\mathfrak P}_n$ by formulas (11) and (12) and
the definitions of the mapping $A_{\widetilde{\mathfrak G}_n} \colon
\widetilde{\mathfrak G}_n \to {\mathfrak X}({\widetilde{\mathfrak G}_n})$
by the formula (30) it follow that
\begin{multline}
\widetilde P_{\widetilde{\mathfrak G}_n}(\widetilde G) =
P_{\widetilde{\mathfrak G}_n}\circ A_{\widetilde{\mathfrak G}_n}(\widetilde G) =
P_{\widetilde{\mathfrak G}_n}(A_{\widetilde{\mathfrak G}_n}(\widetilde G)) =
P_{\widetilde{\mathfrak G}_n}((X_f(\widetilde G), X_{\rm ad}(\widetilde G) = \\
P_n((X_f(\widetilde G), X_{\rm ad}(\widetilde G))) =
\prod_{\{i,j\} \in X_f(\widetilde G)}
\prod_{\{i',j'\} \in X_{\rm ad}(\widetilde G)}
f_{ij}\widetilde f_{i'j'}.
\label{14'.1}
\end{multline}
Hence formula (31) follows. Lemma 2 is completely proved.
$\blacktriangleright$

{\bf Theorem 2.} {\it If the graph $\widetilde G(V_n, X_f)$ belongs
to the set $\widetilde{\mathfrak G}_n$ and to it has assigned 
the complementary set $X_{\rm ad}(\widetilde G)$, then the following
assertions are true:

$A_1$. The graph
$G(V_n; \, X_f(\widetilde G), \, X_{\rm ad}(\widetilde G))$
belongs to the set ${\mathfrak G}_{bn}$ and is the graph-label
of the product
${\widetilde P}_{\widetilde{\mathfrak G}_n}(\widetilde G)$.

$A_2$. The graph $\widetilde G$ is the image of the graph-label
$G(V_n; \, X_f(\widetilde G), \, X_{\rm ad}(\widetilde G))$ under
the mapping $R$.

$A_3$. The product
$\widetilde P_{\widetilde{\mathfrak G}_n}(\widetilde G)$ of Mayer
and Boltzmann functions is a base product of order $n$,
and the graph $\widetilde G$ is its completed graph-label.}

{\bf Proof.} By the definition of the set
$\widetilde{\mathfrak G}_n$, the complementary set
$X_{\rm ad}(\widetilde G)$ forms with the edges set
$X_f(\widetilde G)$ a canonical pair
$(X_f(\widetilde G), X_{\rm ad}(\widetilde G)) \in {\mathfrak X}_n$.

Hence it follows that the graph
$G(V_n; \, X_f(\widetilde G), \, X_{\rm ad}(\widetilde G))$
belongs to the  graphs set ${\mathfrak G}_n$ by the definition
of this set. By Remark 1, the functions product $P_{1n}(G)$, which is
labeled with this graph $G$, belongs to the set ${\mathfrak P}_n$ and
is canonical by the definition of this set. By Definition 7,
the functions product $P_{1n}(G)$ is defined by formula (14),
which in this case has the form
\begin{multline}
P_{1n}(G) = (P_n\circ A_n)(G) = P_n(A_n(G)) =
P_n((X_f(\widetilde G), X_{\rm ad}(\widetilde G))) = \\
\prod_{\{i,j\} \in X_f(\widetilde G)}\prod_{\{i',j'\} \in X_{\rm ad}(\widetilde G)}
f_{ij}\widetilde f_{i'j'}.
\label{14"}
\end{multline}

By Lemma 2, the functions product
$\widetilde P_{\widetilde{\mathfrak G}_n}(\widetilde G)$ is canonical
and is defined by
formula (31). From formulas (33) and (31) it 
follows that
\begin{equation} 
P_{1n}(G) = \widetilde P_{\widetilde{\mathfrak G}_n}(\widetilde G).
\label{122'} 
\end{equation} 

Hence, by Definition 7 it follows that the graph
$G(V_n; \, X_f(\widetilde G), \, X_{\rm ad}(\widetilde G))$, is the
graph-label of the product
${\widetilde P}_{\widetilde{\mathfrak G}_n}(\widetilde G)$.

Since the graph 
$G(V_n; \, X_f(\widetilde G), \, X_{\rm ad}(\widetilde G))$
belongs to the graphs set ${\mathfrak G}_n$, then it belongs 
to the definition domain of the mapping $R$ by the definition of this 
mapping. Assertion $A_2$ follows from the definitions of the graphs 
$\widetilde G$ and $G$ by the conditions of Theorem 2 and from 
the definition of the mapping $R$.

By the conditions of Theorem 2, the graph $\widetilde G$ belongs 
to the graphs set $\widetilde{\mathfrak G}_n$ and, therefore, 
is a connected graph by the definition of this set. Since in this case
the graph $G(V_n; \, X_f(\widetilde G), \, X_{\rm ad}(\widetilde G))$ 
is the graph-label of the product
$\widetilde P_{\widetilde{\mathfrak G}_n}(\widetilde G)$, then
Assertion $A_2$ by Definition 8 implies that this product
is the base one of order $n$. Hence it follows that its graph-label
$G(V_n; \, X_f(\widetilde G), \, X_{\rm ad}(\widetilde G))$ belongs 
to the graphs set ${\mathfrak G}_{bn}$ by the definition of this set.
Statement $ A_1 $ is completely proved.

 From the conditions of Theorem 2 it follows that by Definition 15
the product $\widetilde P_{\widetilde{\mathfrak G}_n}(\widetilde G)$
is the product labeled with the graph
$\widetilde G = \widetilde G(V_n, X_f)$ and the graph
$\widetilde G$ is the completed graph-label of this product.
The Assertion $A_3$ is proved. Theorem 2 is completely proved.
$\blacktriangleright$

For each graph ${\widetilde G} \in \widetilde{\mathfrak G}_n$ let's 
define the integrals $\widetilde I(\widetilde G)$ and 
$\widetilde I(\widetilde G, U)$, setting
\begin{equation}
\widetilde I(\widetilde G) = \int_{({\bf R}^{\nu})^{n-1}}
{\widetilde P}_{\widetilde{\mathfrak G}_n}(\widetilde G)
(d{\bf r})_{1, n-1};
\label{22} 
\end{equation} 
\begin{equation}
\widetilde I(\widetilde G, U) = 
\int_U{\widetilde P}_{\widetilde{\mathfrak G}_n}(\widetilde G)
(d{\bf r})_n,
\label{22'} 
\end{equation} 
where $U$ is a connected, bounded and Lebesgue measurable set,
contained in the space $({\bf R}^{\nu})^n$.

{\bf Remark 5.} If the graph  $\widetilde G(V_n, X_f)$,  to which
the complementary set $X_{\rm ad}(\widetilde G)$ has been assigned, 
belongs to the set $\widetilde{\mathfrak G}_n$, then by Theorem 2, 
the functions product 
$\widetilde P_{\widetilde{\mathfrak G}_n}(\widetilde G)$, defined 
by the formula (31), is a base one of order $n$, and the graph
$G(V_n; \, X_f(\widetilde G), \, X_{\rm ad}(\widetilde G))$,
belongs to the set ${\mathfrak G}_{bn}$ and is the label of this 
product.

Hence, it follows that, by Definition 9, the integral 
$\widetilde I(\widetilde G)$ and integrals of the form 
$\widetilde I(\widetilde G, U)$, defined by the formulas (35) 
and (36), respectively, are base integrals of order $n$. 
Their integrand is the base functions product 
$\widetilde P_{\widetilde{\mathfrak G}_n}(\widetilde G)$ of order $n$. 
 $\blacksquare$

{\bf Theorem 3.} {\it Let us the potential of a pairwise 
interaction $\Phi({\bf r})$ be a measurable function, the pairwise
interaction satisfies the conditions of stability and regularity, and
the graph $\widetilde G(V_n, X_f)$, to which the complementary set
$X_{\rm ad}(\widetilde G)$ is putted in correspondence, belongs
to the set $\widetilde{\mathfrak G}_n$. Then the product of functions
$\widetilde P_{\widetilde{\mathfrak G}_n}(\widetilde G)$, defined by
formula {\rm (31)} has the following properties:

$A_1)$ it is integrable over the space $({\bf R}^{\nu})^{n - 1}$, and 
its integral $\widetilde I(\widetilde G)$ is a base convergent 
integral of order $n$ that does not depend on the value 
of the variable $\bf r_1$;

$A_2)$ it is integrable on any connected bounded
Lebesgue measurable set $U$ contained in the space
$({\bf R}^{\nu})^n$, and the integral $\widetilde I(\widetilde G, U)$
is a base convergent integral of order $n$.}

{\bf Proof.} By Theorem 2, the product of functions
$\widetilde P_{\widetilde{\mathfrak G}_n}(\widetilde G)$, defined 
by the formula (31), is a functions base product of order $n$, 
the graph $G(V_n;\, X_f(\widetilde G),\, X_{\rm ad}(\widetilde G))$
belongs to the set ${\mathfrak G}_{bn}$ and is the label 
of the product 
${\widetilde P}_{\widetilde{\mathfrak G}_n}(\widetilde G)$, that is
equality (34) holds. 

By Remark 5, the integral $\widetilde I(\widetilde G)$ is a base 
integral of order $n$. By Remark 5, the integral 
$\widetilde I(\widetilde G, U)$ is also a base integral of order $n$ 
for any connected bounded Lebesgue measurable  set $U$ contained 
in the space $({\bf R}^{\nu})^n$.
This and the conditions of Theorem 3 by Theorem 1 imply 
Assertions $A_1$) and $A_2$) of Theorem 3. Theorem 3 is completely 
proved. $\blacktriangleright$

{\bf Theorem 4.} {\it Let the potential $\Phi({\bf r})$ of a pairwise 
 interaction  be a measurable function, the pairwise 
interaction satisfies the conditions of stability and regularity,
and a non-empty subset $\widetilde{\mathfrak G}_n^{(0)}$ of 
the graphs set $\widetilde {\mathfrak G}_n$ satisfies the condition:
for each graph
$\widetilde G(V_n; X_f) \in \widetilde {\mathfrak G}_n^{(0)}$
a coefficient $c(\widetilde G)$, which corresponds to this graph
and is a real number, is been defined.

Then the following statements are true:

$A_1$. The linear combination
\begin{equation}
L = \sum_{\widetilde G \in \widetilde{\mathfrak G}_n^{(0)}}
c(\widetilde G)\widetilde I(\widetilde G),
\label{23}
\end{equation}
of  the integrals over the space $({\bf R}^{\nu})^{n - 1}$,
where every integral $\widetilde I(\widetilde G)$ is defined by 
the formula {\rm (35)}, is a base linear combination 
of order $n$.

$A_2$. For any connected bounded Lebesgue measurable set $U$
contained in the space $({\bf R}^{\nu})^n$, the linear combination
\begin{equation}
\widetilde L =
\sum_{\widetilde G \in \widetilde{\mathfrak G}_n^{(0)}}
c(\widetilde G)\widetilde I(\widetilde G, U),
\label{23'}
\end{equation} 
of the integrals of the form {\rm (36)} over the set $U$ 
is a base linear combination of order $n$.}

{\bf Proof.} It follows from the conditions of Theorem 4 that every 
integral in the linear combination $L$, and every 
integral included in the linear combination $\widetilde L$, 
by Theorem 3 are converging base integrals of order $n$. Hence,  
from this and the conditions of Theorem 4 by Definition 11 it follows 
both statements of Theorem 4.  $\blacktriangleright$

Let's denote by $\widetilde{\mathfrak G}(\widetilde L)$ the set 
of all graphs serving as completed graphs-labels of such base
products that are integrands of integrals,
included in the base linear combination $\widetilde L$.

D\,e\,f\,i\,n\,i\,t\,i\,o\,n 16. If $\widetilde L$ is a base linear 
combination, then the set of graphs 
$\widetilde{\mathfrak G}(\widetilde L)$
we will call {\bf the set of the completed graphs-labels} of this
base linear combination. $\blacksquare$

{\bf Remark 6} [39]. For the purpose stated in the article, we have 
enough to establish a criterion for the comparative complexity
of representations of the coefficients of a power series only
for the case when such representations are base linear combinations,
and the complexity of the estimation of the coefficient of any
of the integrals included in such a linear combination is negligible.
In what follows, such base linear combinations will be called
{\bf base linear combinations with coefficients of the negligible
complexity.} $\blacksquare$

4. {\bf Comparative complexity criteria of base linear combinations with 
coefficients of negligible complexities}

The article proposes criteria for comparing the complexity of such
base linear combinations with coefficients of negligible
complexities that satisfy the condition: their associated sets
coincide with each other.

First, let's give the following

D\,e\,f\,i\,n\,i\,t\,i\,o\,n 17. Two base linear combinations
$L$ and $L_1$ with negligible complexity coefficients are called
{\bf comparable} if their orders are equal and $U(L) = U(L_1)$.
$\blacksquare$

Let $U \subset (\R)^n$ be a connected bounded measurable set.

Let's introduce the notation:

${\mf L}(n, U)$ is the set of all linear combinations that are 
base linear combinations of order $n$ with coefficients of negligible 
complexity and have as an associated set the set $U$;

${\mf L}(n)$ is  the set of all  base linear combinations 
of order $n$ with coefficients of negligible complexities and 
with associated sets that are connected bounded measurable sets 
contained in the space $(\R)^n$;

${\mf L}(n, (\R)^{n-1})$ is the set of all base linear combinations 
of convergent improper base integrals of order $n$ over space
$(\R)^{n - 1}$ with coefficients of negligible complexity.

Obviously, the set ${\mf L}(n, (\R)^{n-1})$ consists of pairwise 
comparable base linear combinations of order $n$ with coefficients 
of negligible complexity.

{\bf Remark 7} [39]. Of all the computer time spent on calculations 
performed to estimate the base integral, the overwhelming majority 
are the time spent on calculating the values of Mayer and Boltzmann 
functions included in the representation of the integrand of this
integral. Remaining within the framework of the roughest comparison 
(so to speak, "in the first approximation"), we can hold that 
of the two basic converging integrals whose integration domains 
coincide, more complicated is the estimate of the integral, of which
the integrand representation includes a greater number of Mayer and 
Boltzmann functions. If the representations of the integrands 
of both integrals include equal number of Mayer and Boltzmann 
functions, then we will hold that the estimates of these integrals 
in complexity {\bf are negligibly differ} from each other, and we say 
that the complexity of these estimates {\bf approximately are equal}. 
$\blacksquare$

Thus, remark 7 contains the criterion of the complexity of estimating
of base integrals. All criteria proposed in the article are
based on just this criterion.

The simplest such criterion is length $q(L)$ of a base linear
combination $L$. We denote this criterion $Cr_1$ by setting
$Cr_1(L) = q(L)$. Its definitional domain is denoted by $D(Cr_1)$.
This domain is defined by the formula
\begin{equation}
D(Cr_1) = \left[\bigcup_{n \ge 2}{\mf L}(n)\right] \bigcup 
\left[\bigcup_{n \ge 2}{\mf L}(n, (\R)^{n-1})\right].
\label{Cr_1}
\end{equation}
This criterion is applicable in cases where the compared base linear 
combinations differ from each other in length, while integrals 
included in them 
and their coefficients differ negligibly from each other in their 
complexity. It follows from the definition of the criterion $Cr_1$ 
that its value depends only on the length of a linear combination 
and does not depend on set associated with this linear combination .

As another criterion, it is proposed the sum of all edges of all
graph-labels from the set ${\mathfrak G}(L)$, where $L$ is a given
base linear combination. This criterion will be denoted
by $Cr_2(L)$. It is defined by the formula
\begin{equation}
Cr_2(L) = \sum_{G \in {\mathfrak G}(L)} (\left|X_f(G)\right| +
\left|X_{\widetilde f}(G)\right|),
\label{Cr_2}
\end{equation}
where $\left|X_f(G) \right|$ is the cardinality of the set $X_f(G)$ 
of Mayer functions;
$\left|X_{\widetilde f}(G) \right|$ is the cardinality of the set
$X_{\widetilde f}(G)$ of Boltzmann functions. Its domain of definition
  coincides with the set $D(Cr_1)$.

From the definition of the criterion $Cr_2$ by formula (40) 
it follows that its value on a linear combination included 
in its domain of definition depends only on the set ${\mathfrak G}(L)$
of the graphs serving as labels for the integrands of integrals 
included in this linear combination, and does not depend from 
the set associated with this linear combination.
                                   
One more, more precise, criterion can be proposed. It can be applied 
in the case when an equivalent probabilistic model is used to estimate
each integral from the estimated linear combination.

In this probabilistic model, the estimated integral is
a mathematical expectation of a product of Mayer and Boltzmann
functions of linear combinations of independent
random variables taking values in the $\nu$-dimensional real
Euclidean space ${\bf R}^{\nu}$.

Moreover, each of these random variables is distributed with a density,
equal to the normalized modulus of Mayer function. And the number
of such random values is equal to the number $n - 1$. Thus,
the problem of estimating the base integral, whose integrand is 
labeled with the graph-label $G \in {\mathfrak G}_{bn}$ is reduced 
to the estimation the mathematical expectation of the product of 
Mayer and Boltzmann functions  of the linear combinations of 
independent continuous random variables. This product includes 
$\left|X_f(G)\right| - n + 1$ Mayer
 and $\left|X_{\widetilde f}(G)\right|$ of Boltzmann functions.

The only known way to estimate the mathematical expectation of this
product is the construction of an approximating discrete
stochastic model, which is obtained from the above probabilistic model
by substitution in place of all continuous random variables
by discrete random variables approximating them. As a result,
the problem of an estimation the base integral is reduced
to an estimation mathematical expectation of the product of Mayer and
Boltzmann functions of linear combinations of discrete random
variables.

Of all the computer time spent on calculations performed to estimate 
this mathematical expectation, the overwhelming majority is the time 
spent on calculating the values of Mayer and Boltzmann functions 
whose number $N_1(G)$ is determined by the formula
\begin{equation}
N_1(G) =
\left|X_f(G)\right| - n + 1 + \left|X_{\widetilde f}(G)\right|.
\label{25}
\end{equation}
Therefore, the value $N_1(G)$ defined by the formula (41) can
serve as a {\bf modernized criterion of the complexity of estimation
the improper base integral}, whose integrand is labeled 
with the graph $G$, where $G \in {\mathfrak G}_{bn}$.

D\,e\,f\,i\,n\,i\,t\,i\,o\,n 18. In the case $N_1(G) = 0$, we 
will say that the complexity of the estimation the improper 
convergent base integral, whose integrand is labeled with 
the graph $G$, according to the modernized criterion 
for the complexity of estimating an improper convergent base integral 
is {\bf negligible}. Otherwise, we  will say that the complexity
of estimating the integral, whose integrand is labeled with 
the graph $G$, is {\bf considerable} according to the modernized 
criterion for the complexity of estimation an improper convergent 
base integral. $\blacksquare$

{\bf Example 2.} Consider the graph
$G = G(V_3; X_f, X_{\widetilde f})$, where
$X_f = \{\{1,2\}, \{2, 3\}\}$, \mbox{$X_{\widetilde f} = \varnothing$}.
The graph $G$ belongs to the set ${\mathfrak G}_3$ by the definition 
of the set ${\mathfrak G}_n$.
As its subgraph $R(G) = G$ is connected, then the canonical product
$P_{1n}(G)$ labeled with the graph $G$, where $n = 3$, is a base one
by Definition 8 and belongs to set ${\mathfrak P}_{b3}$ 
by the definition of this set. And the graph $G$ belongs 
to set ${\mathfrak G}_{b3}$ by the definition of this set. Hence, 
by Definition 9, it follows that the defined by formula (16) 
integral $I(G)$, whose integrand is labeled with the graph $G$,
is an improper  base integral of order $3$. In the case when particles 
systems satisfy conditions of Theorem 1, this integral is, 
by Remark 2, a convergent one.

Using the criterion $N_1$ for the complexity of estimating 
an improper convergent base integral, we estimate the complexity 
of this improper integral $I(G)$. From the definition of the sets 
$X_f$ and $X_{\widetilde f}$ it follows:
$\left|X_f \right| = 2$, $\left|X_{\widetilde f} \right| = 0$.
From here by formula (41) we obtain
\begin{equation}
N_1(G) = 0.
\label{26}
\end{equation}
From (42), by Definition 18, it follows that the complexity
of the estimation of the integral $I(G)$ is negligible according 
to the modernized complexity criterion $N_1(G)$. 
$\blacktriangleright$

The proposed third, more precise, criterion for the complexity of base
linear combinations of improper convergent base integrals is denoted 
by $Cr_3(L)$, and its definitional domain is $D(Cr_3)$. This domain 
is defined by the formula
\begin{equation}
D(Cr_3) = \bigcup_{n \ge 2} {\mf L}(n, (\R)^{n-1}).
\label{D3}
\end{equation}

The criterion $Cr_3$, is based on the complexity criterion $N_1(G)$
of the estimation improper convergent base integrals. As such 
a criterion there is proposed the sum over all the integrals, which 
are included in a given base linear combination, of the complexity 
estimates of these integrals. This sum is defined by the formula
\begin{equation}
Cr_3(L) = \sum_{G \in {\mathfrak G}(L)}N_1(G),
\label{27}
\end{equation}
where $N_1(G)$ is defined by formula (41).

From the definition of the criterion $Cr_3$ by the formulas 
(41) and (44) it follows that its value on a linear 
combination included in its definition domain depends only on the set 
of the graphs-labels of the integrands of the integrals included 
in this linear combination, and does not depend from the set associated 
with this linear combination.

D\,e\,f\,i\,n\,i\,t\,i\,o\,n 19. Let $L$ and $L_1$ be two
comparable base linear combinations of integrals 
with the negligible complexity coefficients.  And let
these two linear combinations belong to the domain of definition 
of a criterion $Cr_i$, $i = 1, 2, 3$. We will hold that 
{\bf by the criterion $Cr_i$, the base linear combination $L_1$ is 
considerably more complicated than the base linear combination 
$L$}, if $Cr_i(L_1)> Cr_i (L)$.
If \mbox{$Cr_i(L_1) = Cr_i(L)$}, then we will hold that
by criterion $Cr_i$ the complexity of one of these two base linear
combinations is {\bf equal or negligibly different} from complexity
another of them, and say that according to the criterion $Cr_i$ 
the complexity of one of them {\bf is approximately equal} to
another's complexity. 

If it is known that the base linear combination $L_1$ is more
complicated than the base linear combination $L$, and
$Cr_i(L_1) = Cr_i(L)$, then we will suppose that
according to the criterion $Cr_i$, the linear combination
$L_1$ {\bf is negligibly more complicated} than the linear
combination $L$. $\blacksquare$

The proposed criteria of the complexity of base linear combinations
with coefficients of the negligible complexity are constructed so,
that they, with some exceptions, satisfy the principle:
if, according to this criterion, one of the two base linear
combinations is considerably more complicated than the other one, then
in fact the estimation of the value represented by this base linear
combination is considerably more complicated than estimation
of the value represented by the other base linear combination. 
And in the case when, according to this criterion, the complexity of 
one of the two base linear combinations is negligibly different from 
the complexity of the other of them, then in fact the estimation 
complexity of the value represented by one of these two base linear 
combinations, negligibly differs from the estimation complexity 
of the value represented by the other base linear combination.

In the case when the conclusions drawn on
the values of one of the~criteria are in conflict with the conclusions,
based on values of another, more precise, criterion, preference should
be given to conclusions drawn on the basis of the values
of a more precise criterion.

{\bf Example 3.} Let $L$ and $L_1$ be two linear combinations 
belonging to the set ${\mf L}(3, (\R)^2)$. 
In this case, the linear combination $L_1$ includes two improper 
convergent integrals $I(G)$ and $I(G_1)$, whose integrands are labeled 
by the graphs $G$ and $G_1$ respectively; these integrals are defined 
by the formulas (23) and (14). Here $G$ is the graph considered 
in Example 2, 
and the graph $G_1 = G_1(V_3; X_{f, 1}, X_{{\widetilde f}, 1})$
has a set $X_{f, 1} = \{\{1,2\}, \{ 1, 3\}\}$ of Mayer edges
and the set $X_{{\widetilde f}, 1} = \{\{2, 3 \}\}$ of Boltzmann
edges. The linear combination $L$ contains only one integral $I(G_1)$,
whose integrand is labeled with the graph-label $G_1$. Moreover, in
both linear combinations, the coefficients of the base integrals
$I(G)$ and $I(G_1)$ are defined and equal to $1$.

Graph $G_1 = G_1(V_3; X_{f, 1}, X_{{\widetilde f}, 1})$ belongs
to the set ${\mathfrak G}_3$ by definition of the set
${\mathfrak G}_n$.  Since its subgraph
$R(G_1)$ is connected, then the canonical product $P_{13}(G_1)$
labeled with the graph $G_1$ is a base product by Definition 8 and
belongs to set ${\mathfrak P}_{b3}$. And the graph $G$ belongs
to the set ${\mathfrak G}_{b3}$ by its definition. From this, 
by Definition~9, it follows that the integral $I(G_1)$, 
the integrand of which is labeled with the graph-label $G_1$, 
is an improper base integral of order 3 over the space 
$({\bf R}^{\nu})^2$. In the case when particle systems satisfy
conditions of Theorem 1, this integral is, by Remark 2, converging
and belongs to the set ${\mf L}(3, (\R)^2)$ by its definition.

The linear combination $L$ contains only one integral $I(G_1)$. In the
above case this integral is a convergent base one,
and its coefficient is given and therefore no effort is required
at all to calculate this coefficient. Hence, by Definition 11 and
Remark 7 follows that the linear combination $L$ is a base linear 
combination of order $3$ with the coefficient of the negligible 
complexity and belongs to the set ${\mf L}(3, (\R)^2)$ 
by its definition.

In Example 2, it was proved that the integral $I(G)$, whose integrand 
is labeled with the graph $G$, is a convergent base integral. 
Thus, both the integrals included in
the linear combination $L_1$ are convergent base ones, and
the coefficients of these integrals are given and therefore no
effort is required at all to calculate these coefficients.
This implies that, by Definition 11 and Remark 7, the linear
combination $L_1$ is also a base linear combination of order $3$
with coefficients of the negligible complexity and belongs to the set
${\mf L}(3, (\R)^2)$ by its definition. 

Using the criterion $Cr_3$, we estimate the complexity of linear 
combinations $L$ and $L_1$. Note, that the base linear
combination $L_1$, besides the integral labeled with the graph $G_1$,
also contains one base integral, whose integrand is labeled 
with the graph~$G$. Therefore, it is natural to be of opinion that 
base linear combination~$L_1$ is more complIcated than the base linear 
combination~$L$.

Using the definition of the complexity criterion of
the estimation an improper convergent base integral
by formula (41), let us find the value of this criterion
for the integral labeled with the Graph $G_1$:
\begin{equation}
N_1(G_1) = \left|X_f(G_1)\right| - 3 + 1 +
\left|X_{\widetilde f}(G_1)\right| = 1.
\label{28}
\end{equation}
The value of this criterion for the integral labeled with
graph $G$, was found in example  2 (see formula (42)).

Based on the definition of the criterion $Cr_3$ by formula (44)
and using formulas (42) and (45), we find the values
of this criteria for the base linear combinations $L$ and $L_1$ of
improper integrals:
\begin{equation}
Cr_3(L) = Cr_3(L_1) = 1.
\label{29}
\end{equation}

From formula (46) by Definition 19 it follows that according
to the criterion $Cr_3$ the base linear combination $L_1$ is
negligibly more complicated than the base linear combination $L$.
$\blacktriangleright$

From the definition of the criterion $Cr_3$ and Definition 19 it follows

{\bf Corollary 4.} {\it Let $L$ and $L_1$ be two base linear
combinations of improper integrals with coefficients
of the negligible complexity that belong to the set 
${\mf L}(n, (\R)^{n - 1})$ and satisfy the conditions:

{\rm 1.} The length of the linear combination $L_1$ is greater than
the length of the linear combination $L$.

{\rm 2.} Each integral included in the linear combination $L$ is
included and also into the linear combination $L_1$.

Suppose that among the improper base integrals included 
in the linear combination $L_1$ and not included in the linear 
combination $L$, there is at least one integral such that  
graph-label $G$ of its integrand satisfies the inequality 
$N_1(G) > 0$.
Then, according to the criterion $Cr_3$, the base linear combination 
$L_1$ is considerably more complicated than the base linear 
combination $L$. Otherwise, the base linear combination $L_1$ is 
negligibly more complicated than the base linear combination $L$.}

{\bf Proof.}  Suppose that among the improper base integrals, which are 
included in the linear combination $L_1$ and are not included 
in the linear combination $L$, there is at least one integral having 
a nonzero value of the complexity criterion $Cr_3$ of its estimation.
Then by the definition of the criterion $Cr_3$ by the formula (44) 
from the conditions of Corollary 4 the inequality 
$Cr_3(L_1) > Cr_3(L)$ follows. From this, by Definition 19, it 
follows that by the criterion $Cr_3$ the base linear combination $L_1$
is considerably more complicated than the base linear combination $L$.
In other words, the base linear combination $L$ is considerably 
simpler than the base linear combination $L_1$.

Let us now consider the opposite case, when every integral included
in the linear combination $L_1$ and not included in the linear
combination $L$ is such that the graph-label $G$ of its integrand 
satisfies the equality $N_1(G) = 0$. In this case, by the definition 
of the criterion $Cr_3$ by the formula (44) from the conditions
Corollary 4 the equality $Cr_3(L_1) = Cr_3(L)$ follows. Hence, 
by Definition 19, it follows that the base linear combination $L_1$
is negligibly more complicated than the base linear combination $L$.  
$\blacktriangleright$

D\,e\,f\,i\,n\,i\,t\,i\,o\,n 20 [39]. A base product $P(G)$ is called
{\bf complete} if its graph-label $G$ is complete.
Otherwise, the base product is called {\bf incomplete}. $\blacksquare$

D\,e\,f\,i\,n\,i\,t\,i\,o\,n 21. The base integral is called
{\bf complete} if its integrand is a complete base product.
The base integral is called {\bf incomplete} if its integrand
 is an incomplete base product. $\blacksquare$

D\,e\,f\,i\,n\,i\,t\,i\,o\,n 22. A base linear combination
is called {\bf complete} if all the integrals included in it are
complete. Otherwise, the base linear combination is called
{\bf incomplete}.
$\blacksquare$

From the definition of the Ree-Hoover representations [48], Example 1
and Definitions 20, 21 and 22 follow

{\bf Corollary 5.} {\it For any $n > 1$, Ree-Hoover representation of
of the virial coefficient $B_n$ is a complete base linear
combination of order $n$ with the negligible  complexity coefficients}.

Definitions 20 and 21 and Remark 7 imply the following

{\bf Remark 8.} Let one of the two convergent base integrals be
complete, and the other incomplete, and let the integrands of 
both of these integrals are labeled with graphs with the same set 
of vertices, and let their integration domains of both of these 
integrals coincide with each other. Then the estimate of
the complete integral is considerably more complicated than estimate 
of the incomplete integral. $\blacksquare$

Remark 8 implies

{\bf Corollary 6} {\it Let $L_1$ be an incomplete base linear
combination with the negligible complexity coefficients,
and $L_2$ be a complete base linear combination 
with the negligible complexity coefficients. And let these two linear 
combinations be comparable. And let the number of the integrals 
in the linear combination $L_1$ be at most the number 
of the integrals in the linear combination $L_2$.

If all the integrals included in these base linear combinations are
improper integrals, then, according to Remark {\rm 8}, the linear
combination $L_2$ is considerably more complicated than the linear
combinations $L_1$ by the criteria $Cr_2$ and $Cr_3$. If all
the integrals included in these base
linear combinations are proper, then, according to Remark {\rm 8},
the linear combination $L_2$ is considerably more complicated than
linear combination $L_1$ by the criterion $Cr_2$}.

5. {\bf Tree sum as a special case of the base linear combination}

Within the framework of the frame sums method, two approaches
can be distinguished.

For the exposition of the first of them, we need to introduce
the definition of a tree sum. In order to simplify the exposition
and without striving for maximal generality,
we will give this definition in the sense, although not the most
general, but sufficient for the purposes that set out in this article.

For this, we introduce the following definitions:

$T_n = \{t\}$ is a set of all labeled trees with the set of vertices
$V_n$, where $n > 1$, and with the root $1$;

$ X_f(t) = \{\{u, v\}\}$ is the set of edges of a tree $t \in T_n$;

$\XX_{ad}(t) = \{\{u, v\}\}$ is the set of admissible edges 
[9, 13, 17] of a tree $t \in T_n$;
\begin{equation}
I(t) = 
\int_{(\R)^{n-1}}\prod\limits_{\{u,v\}\in X_f(t)}f_{uv}
\prod\limits_{\{\widetilde u,\widetilde v\}\in \XX_{ad}(t)}
(1+f_{\widetilde u\widetilde v})(d{\bf r})_{1,n-1},
\label{30}
\end{equation}
\begin{equation}
I(t, \LL) = \frac{1}{|\LL|}
\int_{{\LL}^n}\prod\limits_{\{u,v\}\in X_f(t)}f_{uv}
\prod\limits_{\{\widetilde u,\widetilde v\}\in \XX_{ad}(t)}
(1+f_{\widetilde u\widetilde v})(d{\bf r})_n,
\label{30'}
\end{equation}
where $t \in T_n$ and $\LL$ is a connected, bounded and 
Lebesgue measurable set contained in the space $\R$.

Let $T'$ be a non-empty subset of the trees set $T_n$,
where $n > 1$; and to each  tree $t \in T'$ is assigned  the set
$\XX_{ad}(t)$ of admissible edges.  

Let us introduce the notation:

$c(t \mid T')$, $c_1(t  \mid T')$ is real functions defined 
on the trees set $T'$.
\begin{equation}
L(T') = \sum_{t\in T'}c(t  \mid T')I(t), 
\label{31}
\end{equation}
where for each $t \in T'$ the integral $I(t)$ is defined by 
the formula (47). 
\begin{equation}
L(T', \LL) = \sum_{t\in T'}c_1(t  \mid T')I(t, \LL), 
\label{31a}
\end{equation}
where for each $t \in T'$ the integral $I(t, \LL)$ is defined by 
the formula (48). 

D\,e\,f\,i\,n\,i\,t\,i\,o\,n 23. Linear combinations $L(T')$ and 
$L(T', \LL)$, where $T' \subset T_n$ and $n \ge 2$ is called 
 {\bf tree sums}. $\blacksquare$

{\bf Remark 9.} From the definition of the set of admissible edges
$\XX_{ad}(t)$ it follows that this set does not intersect with
the edges set $X_f(t)$ of the tree $t \in T_n$ and consists
of pairwise distinct edges, each of which connects two non-adjacent
vertices of the tree $t$. $\blacksquare$

{\bf Theorem 5.} {\it Let the potential $\Phi({\bf r})$ of a pairwise 
 interaction  be a measurable function, the pairwise 
interaction satisfies the conditions of stability and regularity. 
Then the tree sums $L(T')$ and $L(T', \LL)$, defined by the formulas 
{\rm (49)} and {\rm (50)}, where $T' \subset T_n$ 
and \mbox {$n \ge 2$}, are base linear combinations of the order $n$, 
and each tree $t \in T'$ is the completed graph-label 
of the integrand of the integral $I(t)$ in the tree sum $L(T')$, 
and this tree is the completed graph-label of the integrand 
of the integral $I(t, \LL)$ in the tree sum $L(T', \LL)$.
Moreover, to each such tree $t$ is assigned, as a complementary set, 
the set of admissible edges $\XX_{\rm ad}(t) = \{\{u, v\}\}$.}

{\bf Proof.} Definitions of integrals $I(t)$ and $I(t, \LL)$ by the
formulas (47) and (48) respectively mean that for each tree 
$t \in T'$ is defined the finite set $\XX_{\rm ad}(t)$ of admissible 
edges that is put in correspondebce to this tree. By Remark 9, 
this set does not intersect with the set $X_f(t)$ and consists 
of pairwise distinct edges, each of which connects two non-adjacent 
vertices of the tree $t$.  From the definition of the integrals 
$I(t)$ and $I(t, \LL)$ by the formulas (47) and (48) it follows that 
for each tree $t \in T'$ these integrals have the same integrand, 
which is a product of Mayer and Boltzmann functions.

 Moreover, the set of edges $X_f(t)$ of the tree $t$ 
labeling the integrand of integrals $I(t)$ and $I(t, \LL)$, defines 
the set $F$ of all Mayer functions of this product and is, 
by Definition 3, the set of Mayer edges with respect to the set $F$ 
of Mayer functions. And by Definition 4, the set of admissible edges 
$\XX_{\rm ad}(t)$ defines the set $\widetilde F$ of all Boltzmann 
functions of this product and is the set of Boltzmann edges with 
respect to the set $\widetilde F$ of Boltzmann functions.
Thus, by the definition of a complementary set, the set 
$\XX_{\rm ad}(t)$ is a complementary set put in correspondebce 
to the tree $t$. The sets $X_f(t)$ and $\XX_{\rm ad}(t)$ form 
an ordered pair ${\bf X} = (X_f, X_{\rm ad}(t))$.

By the definition of the trees set $T_n$, every tree $t \in T_n$
is a connected graph with vertex set $V_n$ and, hence, the equality 
$V(X_f(t)) = V_n$ holds. This and Remark 9 imply the equality
$V(X_f) \bigcup V(X_{\widetilde f}) = V_n$. From this equality, 
by Definition 5, it follows that an ordered pair of sets 
${\bf X} = (X_f, X_{\widetilde f})$ is canonical. It follows 
from the results obtained that any tree $t \in T_n$ belongs to 
the set $\widetilde {\mathfrak G}_n$ by its definition.

Hence, by Definition 15 and Lemma 2, it follows that each tree
$t \in T'$ is the completed graph-label of the canonical product 
of functions $\widetilde P_{\widetilde{\mathfrak G}_n}(t)$,
which is labeled by this tree, is of order $n$ and is represented by
formula
\begin{equation}
\widetilde P_{\widetilde{\mathfrak G}_n}(t) =
\prod_{\{i,j\} \in X_f(t)}
\prod_{\{i',j'\} \in X_{\rm ad}(t)}
f_{ij}\widetilde f_{i'j'}.
\label{31'}
\end{equation}
The right-hand side of the formula (51) coincides with both 
the integrand of the integral $I(t)$ and the integrand 
of the integral $I(t, \LL)$. Therefore, the functions product
$\widetilde P_{\widetilde{\mathfrak G}_n}(t)$ labeled by the tree $t$
is the integrand of the integrals $I(t)$ and $I(t, \LL)$; and 
the tree $t$ is the completed graph-label of the integrand 
of the integrals $I(t)$ and $I(t, \LL)$.

Hence, by Remark 5, it follows that the integrand of the integrals 
$I(t)$ and $I(t, \LL)$ is a functions base product of order $n$. 
And the integrals $I(t)$ and $I (t, \LL)$ by Definition 9 are base 
integrals of order $n$. By Theorem 3, for any connected bounded 
Lebesgue measurable set $\LL$ contained in the space 
$({\bf R}^{\nu})$, this functions  product is an integrable function 
on the set $\LL^n$, and the integral $I(t, \LL)$ of this functions 
product converges; moreover, this functions product is an integrable 
function over the space $({\bf R}^{\nu})^{n - 1}$, and the integral 
$I(t)$ of this product converges and does not depend on the value 
of the variable $\bf r_1$.

Recall that functions $c_(t  \mid T')$ and $c_1(t  \mid T')$ are 
defined on the trees set $T'$, and take real values on the trees 
of this set. For each $t \in T'$, the value $c(t  \mid T')$ is 
the coefficient of the integral $I(t)$ belonging to the tree sum 
$L(T')$. In exactly the same way, for each $t \in T'$, the quantity 
$c_1(t  \mid T')$ is the coefficient of the integral $I(t, \LL)$ 
belonging to the tree sum $L(T', \LL)$.

From the results obtained, it follows by Theorem 4 that the tree 
sums $L(T')$ and $L(T', \LL)$ defined by the formulas (49) and (50), 
where $T' \subset T_n$ and \mbox{$ n > 1$}, are base linear 
combinations of order $n$. Theorem 5 is completely proved.
 $\blacktriangleright$

If the tree sum is a base linear combination of order $n$,
then we will call the number $n$ {\bf order} of this tree sum.

6. {\bf Representations of Mayer coefficients $b_n$ by tree sums. Their 
complexity compared to the Ree-Hoover representations}

As an example of representing the coefficients of power series 
by tree sums, one can cite the representations of Mayer 
coefficients $b_n(\LL)$ obtained by the author [3, 9, 17], free of
asymptotic catastrophe. These representations were obtained for
 the case when the thermodynamic equilibrium one-component
 system of classical particles with pairwise interaction [24, 49] 
is enclosed in a bounded volume $\LL$, which is connected, bounded 
and Lebesgue measurable set contained in the space $\R$. It was 
assumed that the pairwise interaction satisfies the conditions 
of stability and regularity, and the pair potential $\Phi({\bf r})$ 
is measurable function. For all $n \ge 2$, each of the representations 
of Mayer coefficient $b_n(\LL)$ obtained by the author under these 
conditions is a tree sum, which is a base linear combination 
of order $n$ with coefficients of insignificant complexity and with 
an associated set ${\LL}^n \subset (\R)^n$.

Initially,  were obtained such representations, in which 
the coefficient $b_n(\LL)$ was expressed as the product of the number 
$1 / n!$ by the sum of all integrals, whose integrands are labeled 
with labeled trees with $n$ verteces [25, 28, 9, 17] and 
with the root vertex labeled with $1$ [3]. Moreover, to each labeling 
tree $t$ was assigned the set of admissible edges 
${\XX_{ad}(t) = \{\{u, v\}\}}$. By Definition 23, such a sum is 
a tree sum. In this sum coefficient of each integral included 
in this sum is equal to unity. Therefore, no calculations are required
to determine the values of the coefficients of the integrals included 
in this sum. Hence, by Theorem 5 and Remark 6, it follows that this 
tree sum is a base linear combination of order $n$ with 
the coefficients of the negligible complexity.

Subsequently, these representations were simplified 
[9, 17].
For this purpose, a binary relation of maximal isomorphism
of labeled rooted trees was introduced. This relation has
the properties of reflexivity,
symmetry and transitivity, that is, it is a relation of
equivalence [21] and decomposes the set $\{T_n\}$, consisting of all
labeled trees with the verteces set $V_n = \{1, 2, \ldots, n\}$
and rooted vertex $1$ into classes of maximally isomorphic trees.
These classes have a very useful property: in the above representation
of Mayer coefficient $b_n(\LL)$ by the tree sum are equal all 
integrals whose integrands are labeled with maximally isomorphic 
trees. In the works [9, 17] is introduced a constructive definition 
of such the subset $TR(n) \subset T_n$ that satisfies the condition: 
a) no two trees belonging to the set $TR(n)$ are maximal isomorphic, 
b) the cardinality of the set $TR(n)$ is equal to the number of 
classes of maximal isomorphic trees, belonging to the set $T_n$.

Using the representation of coefficients $b_n(\LL)$ as the sum 
of all integrals, whose integrands are labeled with the labeled trees 
with the verteces set $V_n = \{1, 2, \ldots, n\}$ and with the rooted 
vertex $1$, decomposition of the set of rooted labeled trees 
with the verteces set $V_n$ and with the rooted vertex labeled $1$ 
into classes of maximally isomorphic trees, and the above property 
of maximally isomorphic trees, the representation 
of Mayer coefficient $b_n(\LL)$ by the tree sum was obtained 
in the form:
\begin{equation}
b_n(\LL) =
\frac{1}{n!|\LL|}\sum_{t\in T\!R(n)}\left|TI(t)\right|I(t, \LL).
\label{32}
\end{equation}
Here $TI(t)$ is the set of the trees belonging to the set $T_n$
and maximally isomorphic to the tree $t$; $\left|TI(t)\right|$ is
cardinality of the set $TI(t)$; $I(t, \LL)$ is the integral defined
by formula (48). 

Passing in the representations of Mayer coefficients $b_n(\LL)$ 
by the formula (52) to the thermodynamic limit, it was possible 
to obtain [9, 17]  Mayer coefficients representations 
in thermodynamic limit as tree sums. For short, the thermodynamic 
limit of Mayer coefficients $b_n(\LL)$ will be called 
the {\bf limiting Mayer coefficient} and denoted $b_n$. 
These representations are such:
\begin{equation}
b_n =
\frac{1}{n!}\sum_{t\in T\!R(n)}\left|TI(t)\right|I(t).
\label{32'}
\end{equation}
From the definition by the formula (52) of Mayer coefficients 
$b_n(\LL)$ representations and the definition by the formula (53)
of representations of limiting Mayer coefficients $b_n$ it follows: 
at all $n \ge 2$ the set of trees $TR(n)$ is the set of all
trees that are graphs-labels labeling both the integrands 
of the integrals, included in the tree sums representing Mayer 
coefficients $b_n(\LL)$, and the integrands of the integrals, included
in the tree sums representing limiting Mayer coefficients $b_n$.

The number of the trees in the set $TI(t)$ is 
completely defined by the tree $t$ according to the formula
\begin{equation}
\left|TI\,(t)\right|=
(n - 1)!\,\Bigl(\prod_{i=1}^{H(t)-1}n(t,i)!\Bigr)^{-1}
\Bigl(\prod_{i=1}^{n(t,H(t)-1)}(d(t,i) - 1)!
\Bigr)^{-1}.
\label{33}
\end{equation}
Here $H(t)$ is height [9, 17, 5] of the tree $t$;
$n(t, i)$ is the number of vertices of the tree $t$ located
at height $i$;
$d(t, i)$ is degree of the $i$-th vertex from the set
of all vertices of the tree $t$ located at the height $H(t) - 1$.

{\bf Lemma 3.} {\it For $n \ge 2$ the representation of Mayer 
coefficient $b_n(\LL)$ by the tree sum according to the formulas 
{\rm (52)} and {\rm (48)} and the representation of 
limiting Mayer coefficient $b_n$ by the tree sum according to the 
formulas {\rm (53)} and {\rm (47)} are base linear 
combinations of order $n$ with coefficients of negligible complexity.}

{\bf Proof.} The sum on the right-hand side of equality (52) and
the sum on the right-hand side of equality (53) have
the following properties: 1) the set of trees $TR\,(n)$ is a subset 
of the set $T_n$; 2) the integrals included in the first sum are 
defined by formula (48), and the integrals included in 
the second sum are defined by formula (47); 3) the coefficient 
of each of these integrals is the number of trees that are maximally 
isomorphic to the tree $t$ labeling the integrand of this integral;
this number is defined by the tree $t$ according to formula 
(54). Hence, by Definition 23, it follows that these sums are 
tree sums. By Theorem 5, these tree sums are base linear combinations 
of order $n$.

From the definition of the coefficients of these tree sums
by formula (54) it follows that the complexity 
of the calculation of these coefficients is negligible. Therefore, 
these tree sums are base linear combinations of the order $n$ 
with coefficients of the negligible complexity. Lemma is proven. 
$\blacktriangleright$

The number of trees in the set $TR(n)$ is calculated by the formula
\begin{multline}
\left|TR\,(n)\right|=1 + \left(2^{n - 2} - 1\right)+{}\\
+\sum_{H=3}^{n - 1} \,
\sum_{{\bf n} \in {\bf N}(H,\,\, n - 1)}\;
\frac{(n(H - 1) + n(H) - 1)!}{n(H)!(n(H-1)-1)!}
\prod_{i=2}^{H-1}\{[n(i-1)]^{n(i)}\}.
\label{34}
\end{multline}
Here ${\bf N}(H, k) = \{(n(1), n(2), \ldots, n(H))\}$ is the set
of $H$ -dimensional vectors whose components are natural numbers,
and the vectors themselves satisfy the condition:
$\sum\limits_{i = 1}^H n(i) = k$.

The results of the calculations by formula (55) are shown in
Table 1. This table lists the cardinalities of the sets $TR(n)$ 
for all $n$ satisfying the inequalities $2 \le n \le 10$. 

Recall that the set $TR(n)$ is the set of completed graphs-labels of
the integrands of all integrals included in a base linear combination 
that is a representation of the thermodynamic limit $b_n$ of Mayer 
coefficients $b_n(\LL)$ as a tree sum according to the formulas (53) 
and (47). The set $TR(n)$ is also the set of completed graphs-labels 
of the integrands of all integrals included in the linear combination 
representing Mayer coefficient $b_n(\LL)$ as a tree sum 
by formulas (52) and (48) for any volume $\LL$ that is a connected, 
bounded and Lebesgue measurable set contained in the space $\R$. 
All of these representations are base linear combinations 
of the same length equal to the cardinality of the set $TR(n)$, and 
differ only in their associated sets. Therefore, on all these 
representations the complexity criterion $Cr_1$ takes the same value 
equal to the cardinality of the set $TR(n)$.

Let us now compare the complexity of the representations of Mayer 
coefficients $b_n(\LL)$ according to the formulas (52) and 
(48) with the complexity of Ree-Hoover representations 
of the virial coefficients by the criterion $Cr_1$ in the case when
thermodynamic equilibrium one-component system of classical particles 
with pairwise interaction [24, 49] is enclosed in a bounded volume 
$\LL$, which is a connected, bounded and Lebesgue measurable set 
contained in the space $\R$. In this case, it is assumed that the 
pairwise interaction satisfies stability and regularity conditions, 
and the pair potential $\Phi({\bf r})$ is Lebesgue measurable function. 
Under these conditions, Ree-Hoover representation of the virial
coefficient $B_n(\LL)$ is defined for all $n \ge 2$ and is a
base linear combination of order $n$ with coefficients of
negligible complexity and with an associated set 
${\LL}^n \subset (\R)^n$. In this case, by definition 17
the considered representation of Mayer coefficient $b_n(\LL)$
and Ree-Hoover representation of the virial coefficient $B_n(\LL)$ 
are comparable for any $n \ge 2$ and for any $\LL$ satisfying 
the above conditions.

In the simplest case, when $n = 2$, both Mayer coefficient 
$b_2(\LL)$, and the virial coefficient $B_2(\LL)$ are represented 
by the same integral and their representations differ only in sign. 
There is nothing to simplify here.

Further, from Table 1 it is clear that for $n = 7, 8, 9, 10 $,
the representation of Mayer coefficient $b_n(\LL)$ by formula 
(52) contains a smaller number of summable integrals than 
Ree-Hoover representation of the virial coefficient $B_n(\LL)$. 
Therefore, for these values of $n$ according to the criterion $Cr_1$
the Ree-Hoover representation of the virial coefficient $B_n(\LL)$ is
considerably more complicated than representation of Mayer 
coefficient $b_n(\LL)$ as a tree sum according to formulas (52)
and (48).

Now let's see what result is obtained according to the criterion
$Cr_2$.

From the definition of the set $\XX_{ad}(t) = \{\{u, v\}\}$
of the admissible edges of the tree $t$ it follows that
for any $n > 2$, the tree sum defined by formulas (52) and
(48) satisfies the condition: in this sum only one integral,
labeled with the star [25, 28], all edges of which are incident
to its root, is a complete base integral; while everyone else
the integrals in this sum are incomplete base integrals.
Hence, by Definition 22 and Lemma 3, it follows that for any $n > 2$
representation of the Mayer coefficient $b_n(\LL)$ by the tree sum
according to formulas (52) and (48) is an incomplete base linear
combination of order $n$ with coefficients of the negligible 
complexity.

On the other hand, by Corollary 5, Ree-Hoover representation
of the virial coefficient $B_n(\LL)$ is a complete base linear 
combination of order of $n$ with coefficients of the negligible 
complexity.

From the above, by Corollary 6 it follows that for the values
$n = 7, 8, 9, 10$ Ree-Hoover representation of virial coefficient
$B_n(\LL)$ is considerably more complicated by the criterion $Cr_2$
than represention of Mayer coefficient $b_n(\LL)$ by the tree sum
defined according to formulas (52) and (48).

Note that for $n = 8, 9, 10 $, the number of integrals in the sum
representing according to Ree-Hoover method, the virial
coefficient $B_n(\LL)$ greatly exceeds the number of integrals 
in the sum representing Mayer coefficient $b_n(\LL)$ 
by formulas (52) and (48). Therefore, by Corollary 6, 
for these values of $n$, the representation of Mayer 
coefficient $b_n(\LL)$ by formulas (52) and (48) is
considerably simpler than the representation of the virial coefficient
$B_n(\LL)$ by Ree-Hoover method.

However, for $n = 3, 4, 5, 6$ the comparing representations
do not satisfy the conditions of Corollary 6. Hence, for these values
of $n$ this corollary cannot be applied for such comparison. At that 
for these values of $n$ according to the criteria $Cr_1$ and $Cr_2$ 
representation of Mayer coefficient $b_n(\LL)$ by formulas (52) and 
(48) is more complicated than the representation of the virial 
coefficient $B_n(\LL)$ by Ree-Hoover method.

7. {\bf Tree sums that are representations  of the coefficients $a_n$ 
of the expansion of the ratio of the activity $z$ to the density 
$\varrho(z)$ in a series in degrees of activity $z$} 

Another example of a represention of power series coefficients
in the form of tree sums is the representation of the coefficients
$a_n$ of the expansion of the ratio of the activity $z$
[23, 24, 44, 49] to the density $\varrho(z)$ in a series in degrees
of activity $z$:
\begin{equation}
z/\varrho(z) =1-
\sum\limits_{n=2}^{\infty}n a_n z^{n-1}.
\label{35}
\end{equation}
This expansion was considered by Lieb [41] and Penrose [45].

Penrose proposed two methods for finding the coefficients $a_n$:
either in a very complicated way using the Kirkwood-Salzburg equations;
or in a simpler way, proceeding from the relations
\begin{equation}
nb_n=\sum_{q=1}^{n -1}(q+1)a_{q+1}(n-q)b_{n-q}
\label{36}
\end{equation}
between these coefficients and Mayer coefficients $b_n$.

In [4, 31, 9, 17], it was proposed to represent the coefficients 
$a_n$ as a sum of integrals whose integrands are labeled with trees.
For this purpose, it was defined the set $T(n, 0)$ consisting
of all trees belonging to the set $T_n$ and satisfying the conditions:

a) any layer of a tree, with the exception of the zero and,
perhaps, the last, consists of at least two vertices;

 b) except for the zero layer, a tree has no layer, in which only
the highest vertex has a degree,  greater than one.

This made it possible to obtain [4, 31, 9, 17] free from asymptotic
catastrophe representations of the coefficients $a_n$
as the sum of all integrals whose integrands are labeled with trees
from the set $T(n, 0)$:
\begin{equation}
a_n = \frac{1}{n!}\sum_{t\in T(n, 0)}\,I(t),
\label{37}
\end{equation}
where $I(t)$ is the integral defined by formula (47).

Subsequently, these representations were simplified [9, 17].
For this purpose, the set $TR(n, 0) = TR(n) \cap T(n, 0)$
 was defined [9, 17].

From the definition of the maximal isomorphism relation of rooted
labeled trees and the definitions of the sets $T(n, 0)$ and $TR(n, 0)$
it follows that the set $T(n, 0)$ decomposes into classes $TI(t)$
of maximally isomorphic trees, where $t$ is the tree that is a label
of a class $TI(t) \subset T(n, 0)$ and belongs to the set TR(n, 0).
And the set $TR(n, 0)$ consists of all trees $t$ that are labels
of the included in the set $T(n, 0)$ classes $TI(t)$ of maximally
isomorphic trees.

Using the representation of the coefficients $a_n$ by formula
(58),
the concept maximal isomorphism of labeled rooted trees, decomposition
of the set $T(n, 0)$ into classes of maximally isomorphic trees and
properties of maximally isomorphic trees, the author proposed simpler
representations of the coefficients $a_n$ free from asymptotic
catastrophe:
\begin{equation}
a_n = \frac{1}{n!}\sum_{t\in T\!R(n,0)}\,\left|TI(t)\right|I(t).
\label{38}
\end{equation}
Here, as in formula (58), $I(t)$ is the integral defined by
formula (47);
$\left|TI(t) \right|$ is the defined by formula (54) number
of trees in the set $TI(t)$, labeled with the tree $t$.

The number of trees in the set $TR(n, 0)$ is calculated by the formula
\begin{multline}
\left|TR\,(n,0)\right|=1+\MO{\sum\nolimits^{\LT{\prime}
}_{\LT{\scriptstyle 2}}}_{\bf n}\quad
\left(\frac{[n(2)+n(1)-1]!}{[n(1)-1]!\,n(2)!}- 1\right)+{}\\
{}\\
+\sum_{H=3}^N
\MO{\sum\nolimits^{\LT{\prime}}
_{\LT{\scriptstyle H}}}_{\bf n}\quad
\left(\frac{[n(H)+n(H-1)-1]!}{[n(H-1) - 1]!\,n(H)!} -1\right)
\prod_{i=2}^{H-1}\left([n(i-1)]^{n(i)} -1\right),
\label{39}
\end{multline}
where $N = \lceil (n - 1) / 2 \rceil$ is the smallest of those
integers that
is at least $(n - 1) / 2$, and the symbol
$\sum\limits_{\bf n}{\vphantom{\left(sum\right)'}}'_{\scriptscriptstyle H}$
\rule[-3ex]{0em}{3ex}
in formula (60) means summation over all $H$-dimensional vectors
$(n_1, n_2, \ldots, n_{{}_H})$ whose components are natural
numbers, and the vectors themselves satisfy the conditions:

\ \ \ \ a)\quad $n_i \ge 2$, \quad $i=1,2,\ldots,H-1$; \qquad
b)\quad $n_{{}_H} \ge 1$; \qquad
c)\quad~$\sum\limits_{i=1}^H n(i) = n-1$.

{\bf Lemma 4.} {\it Let the potential of the paired 
interaction $\Phi({\bf r})$ is a measurable function, and 
the pair interaction satisfies the conditions of stability and 
regularity. Then the representation of the coefficient $a_n$ 
by the tree sum according to the formulas {\rm (59)} and 
{\rm (47)} for $n > 3$ is a base linear combination of order $n$ 
with coefficients of negligible complexity.}

{\bf Proof.} The sum on the right-hand side of equality (59) has
the following properties: 1)~the trees set $TR\,(n, 0)$ is
a subset of the set $T_n$;
2) the integrals included in this sum are defined by formula
(47);
3) the coefficient for each of these integrals is the number of trees,
maximally isomorphic to the tree $t$ labeling the integrand of this 
integral; this number is defined by the tree $t$ by formula (54). 
Hence, by Definition 23 it follows that this sum is a tree sum.

By Theorem 5, this tree sum is a base linear combination
of order $n$.

From the definition of the coefficients of this tree sum
by formula (54) it follows that the complexity of the calculation
of these coefficients is negligible. Therefore this tree sum is a base
linear combination of order $n$ with coefficients of the negligible
complexity. The lemma is proved. $\blacktriangleright$

{\bf Remark 10.} From the definition [4, 31, 9, 17] of the set
$\XX_{ad}(t) = \{\{u, v \}\}$ of admissible edges of a tree $t$
it follows that for any $n > 3$, the tree sum defined by formulas
(59) and (47) satisfies the condition: in this sum only
one integral, whose integrand is labeled with the star, all edges 
of which are incident to its root, is a complete base integral; and 
everyone else the integrals in this sum are incomplete base integrals.
Hence, by Definition 22 and Lemma 4, it follows that for any $n > 3$
representation of the coefficient $a_n$ by the tree sum according
to formulas (59) and (47) is an incomplete base linear
combination of order $n$ with coefficients of the negligible complexity.
$\blacksquare$

From representation (53) of Mayer coefficients $b_n$ and
from representation (59) of coefficients $a_n$ it is obvious 
that $b_2 = a_2$. The indicated representations of these coefficients
coincide and have the same complexity.

And from the definitions of the sets $TR(n)$ and $TR(n, 0)$ for $n > 2$
it follows that the set $TR(n, 0)$ is a proper subset of the set 
$TR(n)$. This has two corollaries:

1. for any $n > 2$, the length of the base linear combination, which 
is a tree sum representing Mayer coefficient $b_n$ by formulas
(53) and (47), is more the length of the base linear
combination, which is the tree sum representing the coefficient $a_n$
by formulas (59) and (47).

2. for any $n > 1$, each integral included in the sum representing
by formulas (59) and (47) the coefficient $a_n$ is also
included in the sum representing by formulas (53) and (47)
the Mayer coefficient $b_n$.

The definition of the set of trees $TR(n)$ implies that
the set $TR(3)$ consists of two trees, which are the graphs $G$
and $G_1$, introduced in examples 2 and 3, respectively. Further,
from the definition of the trees set $TR(n, 0)$ it follows that
the set $TR(3,0)$ consists of one tree, which is the graph $G_1$.
From the results obtained in example 3, it is clear that
the base linear combination, which is the tree sum representing Mayer
coefficient $b_3$ by formulas (53) and (47), is negligibly
more complicated than a base linear combination, which is the tree sum
representing coefficient $a_3$ by formulas (59) and (47).

For $n > 3$, the set $TR(n)$ contains at least one tree, which
does not belong to the set $T(n, 0)$ and has a non-empty set
of admissible edges. Such trees include, in particular, all trees 
from the set $TR(n)$ of height $H > 1$ that are not a chain and have 
such layer of vertices, in which only the highest vertex has degree
greater than one. Obviously, the integrals, whose integrands are 
labeled with such a trees, have a positive value of the criterion 
$N_1$ of the complexity of their
estimations. They are included in the base linear combination, which
is the tree sum representing Mayer coefficient $b_n$ by formulas
(53) and (47), and are not included in the base linear
combination, which is the tree sum representing a coefficient $a_n$
by formulas (59) and (47).

Thus, in the situation under consideration, all conditions
of Corollary 4 are satisfied. From this, by Corollary 4, it follows
that according to the criterion $Cr_3$ for $n > 3$ the base linear
combination, which is the tree sum, representing the Mayer
coefficient $b_n$ by the formulas (53) and (47), is
considerably more complicated than the base linear
combination, which is the tree sum representing the coefficient $a_n$
by formulas (59) and (47).

Table 3 shows the $Cr_3$ criterion values calculated
for $n = 3, 4, 5, 6$ for the base linear combinations that are the
representations of Mayer coefficients $b_n$ in the form
of tree sums by formulas (53) and (47), and for base 
linear combinations that are the representations of the coefficients 
$a_n$ in the form of tree sums by formulas (59) and (47).

These values are a numerical confirmation of the obtained by a
theoretical way of comparative estimations of the complication
of these base linear combinations.

Hence it follows that for estimation the coefficients $a_n$
the direct method, based on their representation in the form
of tree sums by formulas (59) and (47), is simpler and
more rational than the method proposed by Penrose for estimation
the coefficients $a_n$ proceeding from relations (57),
between these coefficients and the Mayer coefficients  $b_n$.
Relations (57) are more expedient to use to represent
coefficients $b_n$ in terms of coefficients $a_n$, in order then 
to apply these representations both for estimating Mayer 
coefficients $b_n$, and to estimate the virial coefficients $B_n$.

These conclusions are also numerically confirmed by the $Cr_1$ 
criterion values calculated for $n = \overline{2, 10}$ for base 
linear combinations, which are representations of limiting Mayer 
coefficients $b_n$ in the form of tree sums according to the formulas 
(\ref{32'}) and (\ref{30}), and for base linear combinations, which 
are representations of the coefficients $a_n$ in the form of tree sums 
according to the formulas (\ref{38}) and (\ref{30}).

The value of the $Cr_1$ criterion for the base linear combination, 
which is the representation of the coefficient $a_n$ by the tree sum 
is equal to the number of integrals in this tree sum. From 
the representation of the coefficient $a_n$ as a tree sum according 
to the formula (\ref{38}) it follows that the number of integrals 
in this tree sum is equal to the number of trees in the set 
$TR(n, 0)$, which is calculated by the formula (60).

The results of calculating the cardinality of the set $TR(n, 0)$ 
for $n = \overline{2, 10}$ are given in Table 1. The data in this 
table support the conclusions already drawn. According to these data, 
for $n = \overline{4, 10}$, the number of integrals in the sum 
representing limiting Mayer coefficient $b_n$ by formula (53) exceeds 
the number of integrals in the representation of the coefficient $a_n$ 
by formula (59) by more than $2$ times.
Hence, according to the simplest criterion, i.e. the length of 
the base linear combination, the conclusion follows:
for $n = \overline{4, 10}$ such a representation of the coefficient
$a_n$ is several times simpler than Mayer representation coefficient 
$b_n$ as a tree sum according to to formulas (53) and (47).

8. {\bf Representations of virial coefficients by polynomials in tree sums. 
An algorithm for computing estimates of virial coefficients using 
these representations and the complexity of calculations at the stages
of the algorithm}

Another example of successful application of the frame sums method
is the representations of virial coefficients obtained by this method.
Within the framework of this method, two ways of representing virial
coefficients have been developed.

The first is as follows: each virial coefficient
is represented as a polynomial in tree sums. As examples of
this way of representing virial coefficients can be given
representations of the virial coefficients free of the asymptotic
catastrophe of two types:
1) in the form of polynomials in tree sums representing Mayer
coefficients $b_n$, and 
2) in the form of polynomials in tree sums representing
the coefficients $a_n$.

Representations of the virial coefficients in the form of polynomials
in tree sums representing Mayer coefficients $b_n$ can be obtained
by using the results obtained by Mayer [23, 42, 43, 44].
In [44] is given a representation (in the form of polynomials in Mayer
coefficients $b_n$) of the quantities $\beta_{\mu}$,
by which the virial coefficients are expressed according
to the formula
\begin{equation}
B_n(\LL) = -{\frac{n - 1}{n}}\beta_{n-1}(\LL), \qquad n > 1.
\label{40}
\end{equation}
Let us present this representation, somewhat simplifying the notation
and at the same time correcting noticed a typo. For this purpose,
we introduce the notation

${\bf M}(n) = \{{\bf m}\}$ is the set of $(n - 1)$-dimensional 
vectors ${\bf m} = (m_1, m_2, \ldots, m_{n-1})$ whose components are 
whole non-negative numbers satisfying the condition:
\begin{equation}
\sum_{j=1}^{n - 1}jm_j = n - 1.
\label{41}
\end{equation}

For each vector ${\bf m} \in {\bf M}(n)$ define the {\bf vector norm},
denoting it $||{\bf m}||$ and setting
\begin{equation}
||{\bf m}|| = \sum_{i = 1}^{n - 1} m_i.
\label{42}
\end{equation}
In this notation, the quantity $\beta_{\mu}$ is represented as follows:
\begin{equation}
\beta_{\mu}(\LL) = -{\frac{1}{{\mu}!}}
\sum_{{\bf m} \in {\bf M}({\mu} + 1)}
({\mu} + ||{\bf m}|| - 1)!
\prod_{j = 1}^{\mu}\frac{1}{m_j!}[-(j + 1)b_{j + 1}(\LL)]^{m_j}.
\label{43}
\end{equation}

Formulas (61) and (64) imply the representations of the virial
coefficients as polynomials in Mayer coefficients $b_n$:
\begin{equation}
B_n(\LL) = 
\frac{n - 1}{n!}\sum_{{\bf m} \in {\bf M}(n)}(n + ||{\bf m}|| - 2)!
\prod_{j = 1}^{n-1}\frac{1}{m_j!}[-(j + 1)b_{j + 1}(\LL)]^{m_j}.
\label{44}
\end{equation}

The thermodynamic limit $B_n$ of virial coefficients $B_n(\LL)$ 
can be represented in a similar way in the form of polynomials 
in limiting Mayer coefficients $b_n$:
\begin{equation}
B_n = 
\frac{n - 1}{n!}\sum_{{\bf m} \in {\bf M}(n)}(n + ||{\bf m}|| - 2)!
\prod_{j = 1}^{n-1}\frac{1}{m_j!}[-(j + 1)b_{j + 1})]^{m_j}.
\label{44'}
\end{equation}

Formulas (65) and (66) will be called {\bf Mayer 
formula}.

For short, the thermodynamic limit of virial coefficients $B_n(\LL)$ 
will be called the {\bf limiting virial coefficient} and denoted $B_n$. 

Let Mayer coefficients $b_n(\LL)$ in formula (65) be defined 
by their representations in the form of tree sums according to formulas
(52) and (48). Then formulas (65), (52) and 
(48) are representations of the virial coefficient $B_n(\LL)$ 
as polynomials in tree sums representing Mayer coefficients 
$b_2(\LL), b_3(\LL), \ldots, b_n(\LL)$. Such the representation 
of the virial coefficient $B_n(\LL)$ will be called
its {\bf representation by Mayer formula and formulas (52) and 
(48)}. Similarly, the representation of the limiting 
virial coefficient $B_n$ in the form of a polynomial in tree sums 
representing limiting Mayer coefficients $b_2, b_3, \ldots, b_n$ 
we will call its {\bf representation by the Mayer formula and 
formulas (53) and (47)}.

Further, for the sake of brevity, we will omit the $\LL$ argument 
of the virial coefficients $B_n$ where it will not cause difficulties
for the reader to understand.

 Obviously, the procedure for calculating the estimate a limiting  
virial coefficient $B_n$ on base of its representation by Mayer 
formula and by formulas (53) and (47) has the same 
complexity as the evaluation procedure virial coefficient $B_n(\LL)$ 
on base of its representation by Mayer formula and formulas 
(52) and (48).
 
If the procedure of the calculation of the estimate
of a virial coefficient $B_n(\LL)$ is based on its representation
by Mayer formula and by formulas (52) and (48), then,
for brevity, the complexity of this procedure we will call
{\bf the complexity of representation of the virial coefficient
by Mayer formula and formulas (52) and (48).}

The question of interest is: what is the complexity of calculation of
the estimates of virial coefficients using these representations?
To answer this question, you need to clearly define the process
of the calculation of these estimates.
This article suggests the following scheme of this process:

{\bf Stage 1.} The calculation of the estimates of Mayer coefficients
included in the representation of a given virial coefficient
according  to Mayer formula.

{\bf Stage 2.} The calculation of the estimate of a given virial
coefficient. The calculation is performed according to Mayer formula,
in which instead of Mayer coefficients, the calculated
estimates of these coefficients are substituted.

To estimate the complexity of these the calculations using Mayer
formula, we present this formula in a slightly different, more
convenient form for solving this problem.

For this purpose, we introduce the notation:
\begin{equation}
Q_n({\bf x}; {\bf y}; {\bf m}) =
\prod_{j = 1}^{n-1} \frac{1}{m_j!}(y_j x_j)^{m_j}.
\label{45}
\end{equation}
Let
\begin{equation}
x_i = -b_{i + 1}(\LL), \quad i = 1, 2, \ldots , n - 1; \quad
{\bf b}(\LL) = \{b_2(\LL), b_3(\LL), \ldots, b_n(\LL)\};
\label{46}
\end{equation}
\begin{equation}
y_i = i + 1, \quad i = 1, 2, \ldots , n - 1.
\label{47}
\end{equation}

In this notations, Mayer formula (65)
takes the form
\begin{equation}
B_n(\LL) =
\frac{n - 1}{n!}\sum_{{\bf m} \in {\bf M}(n)}
(n + ||{\bf m}|| - 2)! Q_n({\bf x}; {\bf y}; {\bf m}).
\label{48}
\end{equation}

Condition (62) implies that the norm of any vector
$m \in {\bf M}(n)$ satisfies the inequality
\begin{equation}
||{\bf m}|| \le n - 1.
\label{49}
\end{equation}

{\bf Remark 11} [39]. From the definition of the function
$Q_n({\bf x}; {\bf y}; {\bf m})$ by formula (67) it follows that
in the case when the values of the components
of the vector $\bf y$ are calculated by formulas (69), and
the values of the components of the vector $\bf x$ are given,
to calculate the value of the function
$Q_n({\bf x}; {\bf y}; {\bf m})$ it is required to perform no more
than $5||{\bf m}||$ arithmetic operations.

Also in the case when the values of the components
of the vector $\bf y$ are calculated according to the formula
\begin{equation}
y_i = -i - 1, \quad i = 1, 2, \ldots , n - 1, \quad
\label{47'}
\end{equation}
and the values of the components of the vector $\bf x$ are given,
to calculate the value of the function
$Q_n({\bf x}; {\bf y}; {\bf m})$ it is required to perform no more
than $5||{\bf m}||$ arithmetic operations.

In the case when all the components of the vector $\bf y$ are equal
to the number $1$, and the values of the components
of the vector $\bf x$ are given, to calculate the value
of the function $Q_n({\bf x}; {\bf y}; {\bf m})$ for given values
of vectors $\bf x$ and $\bf y$ it is required to perform no more
than $3||{\bf m}||$ arithmetic operations. $\blacksquare$

{\bf Remark 12} [39]. From the definition of the sum
$\sum\limits_{{\bf m} \in {\bf M}(n)}$
\rule[-3ex]{0em}{3ex}
it follows that the number of terms in this sum is equal to the number
of all unordered expansions of the number $n - 1$ into a sum
of natural terms. Following [26, 29], we denote this number
by $p(n - 1)$.

 The value of $p(n)$ grows with the growth of $n$
rather slow. Its values are given in the book [26, 29]
(see Table 4.2). So, at $n = 9$ this value takes on the value $30$,
and at $n = 10$ this value is $42$. $\blacksquare$

From Remark 12, from formula (67) and from inequality (71)
it follows that for $n \le 10$ to calculate the sum
$$
 \sum_{{\bf m} \in {\bf M}(n)} (n + ||{\bf m}|| - 2)!
Q_n({\bf x}; {\bf y}; {\bf m}),
$$
where $\bf x$ and $\bf y$ are defined by formulas (68) and
(69) accordingly, it takes less than $2430$ arithmetic
operations. From this estimate and Mayer formula, it follows that
to calculate the estimate of the virial coefficient $B_n$ according
to Mayer formula by use of the known estimates of Mayer
coefficients $b_n$ for $n \le 10$ require perform less than $2440$
arithmetic operations.

This is a negligible number of arithmetic operations compared
with the number of operations required to obtain an estimate of even
the first virial coefficients such as $B_4$, $B_5$, $B_6$ (and as
Mayer coefficients $b_4$, $b_5$, $b_6$) by known methods. Indeed, 
in the procedure for calculating estimates of these coefficients 
by Monte Carlo method, about $10^{10}$ and more statistical tests are 
performed. This implies the following

{\bf Remark 13.} In the case when the process of the calculation of
the estimate of a virial coefficient is based on the representation
of this coefficient by Mayer formula and formulas (52) and
(48), the complexity of this process is negligibly exceeds
the complexity of all the calculations performed at the first stage
of this process.
This makes it possible to use the criterion of the complexity
of all the calculations performed at the first stage, as a criterion
of the complexity of the representation of this virial coefficient
by Mayer formula and formulas (52) and (48).
$\blacksquare$

Of course, the complexity of the procedure of the calculation of
the estimates of Mayer coefficients depends on their representations.
For brevity, the complexity of the procedure of the calculation of
the estimates of all Mayer coefficients from the set
$\{b_2(\LL), b_3(\LL), \ldots, b_n(\LL)\}$
with the help of given representations of all these coefficients
we will call {\bf the complexity of the given set of the representations
of Mayer coefficients}.

Based on Remark 13, we will hold that a criterion of the complexity
of the representation of a virial coefficient $B_n(\LL)$ by Mayer 
formula and formulas (52) and (48) is a criterion 
of the complexity of the set of the representations of Mayer 
coefficients $b_2(\LL), b_3(\LL), \ldots, b_n(\LL)$. Similarly, 
we will hold that the complexity criterion of the representation 
of the limiting virial coefficient $B_n$ by Mayer formula 
and formulas (53) and (47) is the complexity criterion 
of the set, consisting of representations  of the limiting Mayer 
coefficients $b_2, b_3, \ldots, b_n$.

In the cases considered below,  Mayer coefficients are
represented by formulas of the form (52) and (48).
By Lemma 3, these representations are base linear combinations
with coefficients of the negligible complexity.

9. {\bf Base set of base linear combinations and comparative criteria 
for the complexity of estimating base sets}

In order to estimate the complexity of the set of base linear 
combinations representing Mayer coefficients 
$b_2, b_3, \ldots, b_n$, it is necessary to introduce criteria 
for the complexity of evaluating a finite set of base linear 
combinations with coefficients of negligible complexity. 
For this purpose, we introduce the following notation:

${\mathfrak L} = \{L\}$ is a finite set of base linear combinations 
with coefficients of negligible complexity;

${\mathfrak U}({\mathfrak L}) = \{U(L) : L \in {\mathfrak L}\}$ is 
the totality of all sets associated with a base linear combinations 
belonging to the set ${\mathfrak L}$.

D\,e\,f\,i\,n\,i\,t\,i\,o\,n 24. The totality 
${\mathfrak U}({\mathfrak L})$ is called {\bf the sets totality,
associated with the set ${\mathfrak L}$ of base linear combinations}. 
$\blacksquare$

D\,e\,f\,i\,n\,i\,t\,i\,o\,n 25. The totality of sets
${\mathfrak U}({\mathfrak L})$ is called {\bf ordered} if there exists 
a connected, bounded and Lebesgue measurable set $\LL \subset \R$ 
such that for any linear combination 
$L \in {\mathfrak L}$ its the associated set $U(L)$ can be represented
as: $U(L) = \LL^k$, where $k$ is order of the linear combination $L$. 
In this case, the set $\LL$ is called {\bf conjugate to the set
${\mathfrak L}$}. $\blacksquare$

D\,e\,f\,i\,n\,i\,t\,i\,o\,n 26. A linear combinations set
${\mathfrak L}$ is called a {\bf base set} if it satisfies 
one of the following two conditions:

1) Each base linear combination of order $k$ belonging to it
belongs to the set ${\mf L}(k, (\R)^{k - 1})$; in this case the space
$\R$ is called {\bf conjugate to the set ${\mathfrak L}$}.

2) Each base linear combination of order $k$ belonging to it
belongs to the set ${\mf L}(k)$, and the population of sets
${\mathfrak U}({\mathfrak L})$ is ordered. $\blacksquare$

D\,e\,f\,i\,n\,i\,t\,i\,o\,n 27. The largest of the numbers serving
as order of one of the base linear combinations included
to the base set ${\mathfrak L}$ is called {\bf order} of this set.
$\blacksquare$

D\,e\,f\,i\,n\,i\,t\,i\,o\,n 28. The base sets ${\mathfrak L}_1$ 
and ${\mathfrak L}_2 $ are called {\bf comparable},
if they both have the same order $n$ and if they both satisfy one 
of the following two conditions:

1) any base linear combination of order $k$ belonging to
at least one of these two base sets, belongs to the set 
${\mf L}(k, (\R)^{k - 1})$, where $k \le n$, $n$ is the order 
of these base sets;

2) each of these two base sets has a conjugate set, and these two 
conjugate sets coincide with each other. $\blacksquare$

In what follows, we will consider only such sets of base linear 
combinations that are base sets.

In the article are proposed three criteria of the complexity
of estimation of a base set of linear combinations. Each
of these criteria is generated by one of the above the criteria
of the complexity of base linear combinations. The criterion
generated by the $Cr_i$ criterion, where $i = 1, 2, 3$, we denote
$Cr'_i$.

We define the complexity criterion $Cr'_i({\mathfrak L})$ on all base
sets consisting of such base linear combinations on which 
the criterion of complexity $Cr_i$ is defined.

On each such base set ${\mathfrak L} = \{L\}$, let's define the value 
of the criterion $Cr'_i({\mathfrak L})$, putting
\begin{equation}
Cr_i'({\mathfrak L}) = \sum_{L \in {\mathfrak L}}Cr_i(L), \quad
i = 1, 2,3.
\label{50}
\end{equation}

Since the criteria $Cr_1$ and $Cr_2$ are defined on all base linear 
combinations, the criteria $Cr'_1$ and $Cr'_2$, according to their 
definition by the formula (73), are defined on all base sets.
And since the criterion $Cr_3$ is defined only on base linear 
combinations of base improper convergent integrals, then
the criterion $Cr'_3$ according to its definition by the formula (73) 
is defined on all base sets consisting only of base linear 
combinations of base improper convergent integrals.

So, we have defined the complexity criteria $Cr'_i$, $Cr'_2$ and
$Cr'_3$. At this definition  the domain of the complexity criterion
$Cr'_i$ (where  $i = 1,  2, 3$)  is the totality of all finite subsets
of the set of all base linear combinations at which the complexity
criterion $Cr_i$ defined.

It was noted above that the value of each of the criteria $Cr_1$, 
$Cr_2$ and $Cr_3$ on a linear combination included in its definition
domain, depends only on the set of graphs serving as labels 
of the integrands of the integrals, which are included in this linear 
combination, and does not depend on the associated set of this 
linear combination. Hence and from the definition of the criteria 
$Cr'_1$, $Cr'_2$ and $Cr'_3$ by the formula (73) it follows that 
the value of each of the criteria $Cr'_1$, $Cr'_2$ and $Cr'_3$ 
on a base set included in its definition domain depends only 
on the set of graphs serving as labels of the integrands of integrals 
included in the linear combinations that belong to this set, 
and this value does not depend on the conjugate set of this base set.

D\,e\,f\,i\,n\,i\,t\,i\,o\,n 29. Let the criterion $Cr'_i$, where $i$ 
can take the values $i = 1, 2, 3$, is defined on comparable base 
sets ${\mathfrak L}$ and ${\mathfrak L}_1$ of linear combinations.

We will hold that {\bf by the criterion $Cr'_i$,} the
base set ${\mathfrak L}_1$ is {\bf considerably more complicated 
than the base set ${\mathfrak L}$}, if
$Cr'_i({\mathfrak L}_1) > Cr_i({\mathfrak L})$.
If \mbox{$Cr_i({\mathfrak L}_1) = Cr_i({\mathfrak L})$}, then we will 
hold  that, according to the criterion $Cr'_i$, the complexity of one 
of these two base sets is {\bf equal or negligibly different} from 
complexity another of them, and say that according to the criterion 
$Cr'_i$, the complexity of one of them {\bf is approximately equal to} 
the complexity of the other. 
   If it is known that the base set ${\mathfrak L}_1$ is more  
complicated than the base set ${\mathfrak L}$, and
$Cr_i({\mathfrak L}_1) = Cr_i({\mathfrak L})$, then we will hold that 
by the criterion $Cr'_i$, the set ${\mathfrak L}_1$ {\bf is negligibly
more complicated} then the set ${\mathfrak L}$. $\blacksquare$

Let $L_0$ be a base linear combination with the coefficients 
of negligible complexity, which belongs to the domain 
of the complexity criterion $Cr_i$. Let us put in correspondence 
to the linear  combination $L_0$ the base set 
${\mathfrak L}_0 = \{L_0\}$, consisting of one linear
combinations $L_0$. Obviously, the base linear combination $L_0$ and
the set ${\mathfrak L}_0$ have the same computational complexity.

The set $ {\mathfrak L}_0$, by its definition, belongs to the domain 
of definition of the complexity criterion $Cr'_i$. Therefore, 
the value of the complexity criterion $Cr'_i$ is defined for it.
According to the definition of the criterion $Cr'_i$ 
by the formula (73), the following equality holds:

\begin{equation}
Cr'_i({\mathfrak L_0}) = Cr_i(L_0).
\label{51}
\end{equation}

D\,e\,f\,i\,n\,i\,t\,i\,o\,n 30. A base set ${\mathfrak L}$ and 
a base linear combination $L_0$ are called {\bf comparable} if they 
both have the same order $n$, and if any base linear combination 
$L \in {\mathfrak L}$ of order $n$ is comparable to the linear 
combination $L_0$. $\blacksquare$

For any $i = 1, 2, 3$ this definition, together with equality (74), 
makes it possible to introduce a definition that makes it possible 
to compare the complexity of any basic linear combination $L_0$, 
on which the criterion $Cr_i$ is defined, with the complexity 
of the  base set ${\mathfrak L'} = \{L\}$, which is comparable 
to the base linear combination $L_0$ and on which the criterion 
$Cr'_i$ has been defined.

D\,e\,f\,i\,n\,i\,t\,i\,o\,n 31. Let $L$ be a base linear 
combination, on which a criterion $Cr_i$ is defined, and
${\mathfrak L}$ be a linear combinations base set, comparable with 
the base linear combination $L$.
We will hold that {\bf according to the criterion $Cr'_i$
the base linear combination $L$ is considerably more
complicated than the base linear combinations base set}
${\mathfrak L}$, if $Cr_i(L) > Cr_i'({\mathfrak L})$. If
$Cr_i(L) < Cr_i'({\mathfrak L})$, then we will hold that
{\bf according to the criterion $Cr'_i$ the base linear combination
$L$ is considerably simpler than the base linear combinations base set}
${\mathfrak L}$.

In the case when $Cr_i(L) = Cr_i'({\mathfrak L})$, we will hold
that {\bf according to the criterion $Cr'_i$ the complexity
of the base linear combination $L$ is approximately equal
to the complexity of the base linear combinations base set}
${\mathfrak L}$, $\blacksquare$

Let us denote by ${\mathfrak L}_{TR}(n, \LL) = \{L\}$ the base set 
of tree sums, each of which is the representation of a coefficient 
from the set of coefficients 
${\bf b}_{1, n - 1}(\LL) = \{b_2(\LL), b_3(\LL), \ldots, b_n(\LL)\}$ 
according to formulas (52) and (48). Following the above,
we hold that a complexity criterion of the base set
${\mathfrak L}_{TR}(n, \LL)$ of tree sums is a complexity criterion 
of the virial coefficient $B_n(\LL)$ representation according 
to Mayer formula (65) and formulas (52) and (48).

{\bf Lemma 5.} {\it Let the potential $\Phi({\bf r})$ of a pairwise 
 interaction  be a measurable function, and the pairwise interaction 
satisfies the conditions of stability and regularity. 
And let the set $\LL$ be a connected, bounded and Lebesgue measurable 
set contained in the space $\R$. Then the set 
${\mathfrak L}_{TR}(n, \LL)$ is a base set of base linear 
combinations. This set is of order $n$, and the set $\LL$ is 
the conjugate set of this base set.}

{\bf Proof.} From the definition of the tree sums set
${\mathfrak L}_{TR}(n, \LL)$ it follows that any tree sum belonging 
to this set is a representation of some Mayer coefficient 
$b_k(\LL)$ belonging to the Mayer coefficients set
${\bf b}_{1, n - 1}(\LL) = \{b_2(\LL), b_3(\LL), \ldots, b_n(\LL)\}$. 
From the definition of the set ${\mathfrak L}_{TR}(n, \LL)$ by Lemma 3
it follows that this tree sum is a base linear combination of order $k$ 
with coefficients of negligible complexity. Thus, the set 
${\mathfrak L}_{TR}(n, \LL)$ is a finite set of all base linear 
combinations that are definded  by the formulas (52) and (48) 
and are representations Mayer coefficients belonging to the set 
${\bf b}_{1, n - 1}(\LL)$. In this case, the representation of Mayer 
coefficient $b_k(\LL) \in {\bf b}_{1, n - 1}(\LL)$ is the base linear 
combination of order $k$ from the set ${\mathfrak L}_{TR}(n, \LL)$.

From the definition by the formulas (52) and (48) of the base linear 
combinations belonging to the set ${\mathfrak L}_{TR}(n, \LL)$  
it follows that the set $\LL^k$ is associated to the base linear 
combination of order $k$ from the set ${\mathfrak L}_{TR}(n, \LL)$. 
From the conditions of Lemma 5 it follows that for any natural 
number $k$ the associated set $\LL^k$ is a connected, bounded and 
Lebesgue measurable set [21] contained in the space $(\R)^k$.

From this, first, it follows that for any $k = 2, 3, \ldots, n$ 
the base linear combination of order $k$ from the set 
${\mathfrak L}_{TR}(n, \LL)$ belongs to set ${\mf L}(k)$ 
by the definition of this set. Second, from this, by Definition 25, 
it follows that the totality of all sets associated to base linear 
combinations belonging to the set ${\mathfrak L}_{TR}(n, \LL)$, 
is ordered, and the set $\LL$ is conjugate to the set 
${\mathfrak L}_{TR}(n, \LL)$.

From the results obtained, by Definition 26, it follows that the set
${\ mathfrak L}_{TR}(n, \LL)$ is a base set of base linear 
combinations.

Any Mayer coefficient $b_k(\LL)$ from Mayer coefficients set 
${\bf b})_{1, n - 1}(\LL)$ is represented by the base linear 
combination of order $k$ from the base set 
${\mathfrak L}_{TR}(n, \LL)$, and 
this base set contains only base linear combinations that are 
representations of Mayer coefficients belonging to the set 
${\bf b}_{1, n - 1}(\LL)$. Therefore, no base linear combination 
of order more than $n$ belongs to the base set 
${\mathfrak L}_{TR}(n, \LL)$.
On the other hand, this base set contains a base linear combination 
of order $n$, which is the representation of Mayer coefficient 
$b_n(\LL)$ belonging to the set ${\bf b}_{1, n - 1}(\LL)$.
Hence, the number $n$ is the largest of the numbers that serve as 
the order of one of the base linear combinations included to 
the base set ${\mathfrak L}_{TR}(n, \LL)$. From here
by definition 27 it follows that the number $n$ is  order of 
the base set ${\mathfrak L}_{TR}(n, \LL)$. Lemma 5 is completely 
proven. $\blacktriangleright$

{\bf Example 4}. Let us consider the set ${\mathfrak L}_{TR}(n, \LL)$ 
of all tree sums that according to the formulas (52) and (48) are 
representations of  Mayer coefficients, belonging to the set 
\linebreak
${\bf b}_{1, n - 1}(\LL) = \{b_2(\LL), b_3(\LL), \ldots, b_n(\LL)\}$.
Moreover, we will assume that the conditions of Lemma 5 are satisfied.
By Lemma 5, this set ${\mathfrak L}_{TR}(n, \LL)$ is a base set of 
base linear combinations with coefficients of negligible complexity 
and has order $n$, and the set $\LL$ is the conjugate set of this 
base set. The set ${\mathfrak L}_{TR}(n, \LL)$ contains only one base 
linear combination of order $n$. Its associated set is the set 
$\LL^n$. By Definition 17, this linear combination of order $n$ 
is comparable to Ree-Hoover representation of the virial coefficient 
$B_n(\LL)$. This statement is based on the analysis of Ree-Hoover 
representation set out in Example 1, where it is shown that
this representation of the coefficient $B_n(\LL)$ is a base linear 
combination with coefficients of negligible complexity and has order 
$n$, and the set $\LL^n$ is the associated set of this base linear 
combination. From this statement, by Definition 30, it follows 
that the base set ${\mathfrak L}_{TR}(n, \LL)$ is comparable to 
Ree-Hoover representation of the virial coefficient $B_n(\LL)$. 
Since the criteria $Cr_1$ and $Cr_2$ are defined on this Ree-Hoover 
representation, and the criteria $Cr'_1$ and $Cr'_2$ are defined 
on the base set ${\mathfrak L}_{TR}(n, \LL)$, the complexity of 
the representation of the virial coefficient $B_n(\LL)$ 
by the formulas (65), (52) and (48) was been compared with 
the complexity of Ree-Hoover representation of this coefficient 
at the stated below values of $n$. Since the values of the criteria 
$Cr'_1$ and $ Cr'_2$ do not depend on the set $\LL$ conjugate 
to a base set, then in examples 4 and 5 the symbol $\LL$ only denotes 
that the set $\LL$ conjugate to a base set is a connected, bounded 
and measurable by Lebesgue set contained in the space $\R$.

Table 4 shows the calculated values of the criterion
$Cr'_1({\mathfrak L}_{TR}(n, \LL))$ for \mbox{$n = \overline{2,10}$}.
where ${\mathfrak L}_{TR}(n, \LL)$ is representation of the virial
coefficient $B_n(\LL)$ according to Mayer formula (65) and
formulas (52) and (48). In particular,
$Cr'_1({\mathfrak L}_{TR}(8, \LL)) = 857$,
$Cr'_1({\mathfrak L}_{TR}(9, \LL)) = 3709$,
$Cr'_1({\mathfrak L}_{TR}(10, \LL)) = 17756$.
Comparing these values with the values of the complexity criterion
$Cr_1$ of Ree-Hoover representations given in Table 4, we see that
the values of the criterion $Cr'_1({\mathfrak L}_{TR}(n, \LL))$ for
$n = 8, 9, 10$ are less than the values of the complexity criterion
$Cr_1(L_{RH}(n, \LL))$ (see tables notations) for corresponding 
Ree-Hoover representations. Therefore, by Definition 31, at these 
values of $n$, the representation of the virial coefficient $B_n(\LL)$
according to formulas (65), (52) and (48) are considerably simpler 
than Ree-Hoover representation of this coefficient at any bounded 
volume $\LL \subset \R$.  $\blacktriangleright$

{\bf Example 5} Let us compare, according to the criterion $Cr'_2$,
the complexity of Ree-Hoover representations of the virial 
coefficients $B_3(\LL)$, $B_4(\LL)$, $B_5(\LL)$, $B_6(\LL)$ and 
$B_7(\LL)$ with the complexity of their representations in the form 
of a polynomial in tree sums by formulas (65), (52) and (48). 

Table 5 shows, in particular, the following results:
\begin{multline}
Cr'_2({\mathfrak L}_{TR}(3, \LL)) = 6, \quad 
Cr'_2({\mathfrak L}_{TR}(4, \LL)) = 28, \quad  
Cr'_2({\mathfrak L}_{TR}(5, \LL)) = 121, \\
Cr'_2({\mathfrak L}_{TR}(6, \LL)) = 524, \quad 
Cr'_2({\mathfrak L}_{TR}(7, \LL)) = 2406, \\
Cr_2(L_{RH}(3)) = 3, \quad Cr_2(L_{RH}(4)) = 12, \quad 
Cr_2(L_{RH}(5)) = 50, \\  Cr_2(L_{RH}(6)) = 345,\quad 
Cr_2(L_{RH}(7)) = 3591.
\end{multline}

Table 5 shows, in particular, that  the inequality 
$Cr'_2({\mathfrak L}_{TR}(n) > Cr_2(L_{RH}(n))$ holds for 
\mbox{$n = 3, 4, 5, 6$.} 
From this, by Definition 31, it follows that for values 
$n = 3, 4, 5, 6$ the representation
of the virial coefficient $B_n(\LL)$ by the formulas (65),
(52) and (48) is considerably more complicated than
Ree-Hoover representation of this coefficient for any bounded volume 
$\LL \subset \R$. And for $n = 7$ the inquality 
$Cr'_2({\mathfrak L}_{TR}(7) < Cr_2( L_{RH}(7))$ holds. From this
inequality, by Definition 31, it follows that the representation 
of the virial coefficient $B_7(\LL)$ by the formulas (65), 
(52) and (48) is considerably simpler than Ree-Hoover 
representation of this coefficient for any bounded volume 
$\LL \subset \R$.   $\blacktriangleright$

10. {\bf Two examples of representations of the thermodynamic limits 
of virial coefficients in the form of polynomials in tree sums and 
the application of the introduced criteria to their comparison 
in terms of complexity}

Let us now turn to representations of limiting virial coefficients 
$B_n$ in the form of polynomials 
in tree sums representing the coefficients $a_n$ by formulas 
(59) and (47). These representations of limiting virial 
coefficients for $n > 1$ have the form [9, 11, 17, 36, 39]:
\begin{equation}
B_n = \sum_{{\bf m} \in {\bf M}(n + 1)}||{\bf m}||!\,
e_{\scriptscriptstyle ||{\bf m}||}
\prod_{j=1}^n(m_j!)^{-1}[\tau_j]^{m_j}, \quad n \ge 2,
\label{53}
\end{equation}
where coefficients $e_{\mu}$ and $\tau_{\mu}$ are defined
by the formulas
\begin{equation}
e_1 = \tau_1 = 1;\quad
e_{\mu} = \mu^{-1}
\sum_{{\bf m} \in {\bf M}(\mu)}||{\bf m}||!
\prod_{j=1}^{\mu - 1}(m_j!)^{-1}[(j + 1)a_{j + 1}]^{m_j},\quad \mu \ge 2;
\label{54}
\end{equation}
\begin{eqnarray}
\tau_{\mu} = (\mu - 1)!
\sum_{{\bf m} \in {\bf M}(\mu)}\left[(\mu-||{\bf m}||)!\,\right]^{-1}
\prod_{j = 1}^{\mu - 1}(m_j!)^{-1}\{-(j + 1)a_{j + 1}\}^{m_j}.
\quad \mu \ge 2.
\label{55}
\end{eqnarray}
According to these formulas, a limiting virial coefficient $B_n$ is
represented as a polynomial in tree sums representing coefficients
$a_n$.

Of interest is the question: what is complexity of the calculation of
the estimate of a limiting virial coefficient $B_n$ using its 
representation by the formulas (76), (77) and (78)?

To estimate complexity of these calculations, first of all
we represent the limiting virial coefficient $B_n$ and the quantities 
$e_m$ and $\tau_m$ in a form more convenient for this purpose.

Namely, using the function $Q_n({\bf x}; {\bf y}; {\bf m})$ introduced
by formula (67), transform the representations of the quantities
$B_n$, $e_m$ and $\tau_m$ by formulas, respectively (76),
(77) and (78) as follows:
\begin{equation}
e_1 = 1;\quad
e_{\mu} =
\mu^{-1}\sum_{{\bf m} \in {\bf M}(\mu)}
||{\bf m}||!Q_m({\bf x}; {\bf y}; {\bf m}),\quad \mu \ge 2,
\label{56}
\end{equation}
where
\begin{equation}
x_j = a_{j + 1},\quad y_j = j + 1, \quad 1 \le j < \mu;
\label{57}
\end{equation}
\begin{eqnarray}
\tau_1 = 1; \quad \tau_{\mu} = (\mu - 1)!
\sum_{{\bf m} \in {\bf M}(\mu)}
\left\{[\mu - ||{\bf m}||]!\,\right\}^{-1}
Q_m({\bf x}; -{\bf y}; {\bf m}),   \quad \mu \ge 2,
\label{58}
\end{eqnarray}
where the vectors ${\bf y}$ and ${\bf x}$ are defined by formulas
(80),
and the vector $-{\bf y}$ is defined by the formula
\begin{equation}
-{\bf y} = (-y_1, -y_2, \ldots, -y_{\mu - 1});
\label{59}
\end{equation}
\begin{equation}
B_n = \sum_{{\bf m} \in {\bf M}(n + 1)}
||{\bf m}||!\,{\displaystyle e}_{\scriptscriptstyle ||{\bf m}||}
Q_{n + 1}({\bf x}; {\bf y}; {\bf m}), \quad n \ge 2,
\label{60}
\end{equation}
 where the values $e_j$ for $j = \overline{1.n}$ are defined
by the formulas (79),
\begin{equation}
x_j = \tau_j, \quad y_j = 1 \quad \mbox{\rm for}\quad j =
\overline{1,n},
\label{61}
\end{equation}
and the quantities $\tau_j$ are defined by formulas (81), where
the vectors ${\bf y}$ and ${\bf x}$ are defined by formulas (80), and
the vector $-{\bf y}$ defined by formula (82).

In these transformed representations, the limiting virial coefficient 
$B_n$ also, as in the representations by formulas (76), (77)
and (78), is presented as a polynomial in the tree sums
representing the coefficients $a_n$.

Further, in order to answer the question posed, you need to clearly
define the process of the calculation of the estimate of the limiting 
virial coefficient $B_n$. This article suggests the following scheme
of this process:

{\bf Stage 1.} A calculation of estimates of the values 
of the coefficients $ak$ for all $k = \overline{2, n}$. 
The estimate of the value of the coefficient $a_k$ is denoted 
by $a'_k$, \, $k = \overline {2, n}$.

{\bf Stage 2.} A calculation of estimates of the values of all 
quantities from the set ${\bf e}_n = \{e_2, e_3, \ldots, e_n \}$. 
The estimate of the value of $e_k$ is denoted by $e'_k$. 
The calculation is performed according to the formulas (\ref{56}) 
and (\ref{57}), into  which, instead of the coefficients $a_k$, 
where $k = \overline{2, n}$, are substituted the their estimates $a'_k$
that were calculated at stage 1, and instead of the quantity $e_k$, 
is substituted the estimate $e'_k$ of the value of this quantity. 

{\bf Stage 3.} A calculation of estimates for the values 
of all quantities from the set 
${\bm\tau_n} = \{\tau_2, \tau_3, \ldots, \tau_n\}$.
The calculation is performed according to the formula (\ref{58}) and 
(\ref{57}), into which, instead of the coefficients $a_k$, 
where $k = \overline{2, n}$, are substituted the their estimates $a'_k$
that were calculated at stage 1, and instead of the quantity$\tau_k$, 
is substituted the estimate $\tau'_k$ of the value of this quantity. 

{\bf Stage 4.} A calculation of the estimate of the value of the 
given limiting virial coefficient. The estimate of the value of this 
coefficient will be denoted by $B'_n$. The calculation is made 
according to the formula (\ref{60}), into which instead of this 
coefficient the its value estimate $B'_n$ is substituted, and instead 
of the quantities $e_k$ and $\tau_k$, are substituted the estimates 
of the values of these quantities respectively $e'_k$ and $\tau'_k$ 
calculated at stages 2 and 3. 

Our immediate goal is to find an upper bound of the number
of arithmetic operations required for the computations performed
in stages 2--4. Let's introduce the notation:
\begin{equation*}
{\bf e}'_n = (e'_1, e'_2, \ldots, e'_n), \quad
\mbox{$\bm \tau'_n$} = (\tau'_1, \tau'_2, \ldots, \tau'_n), \quad
{\bf a}'_n = \{a'_1, a'_2, \ldots, a'_n\}, \quad n \ge 2;
\end{equation*}
$E_1(\mu, {\bf m}\,\mid {\bf a_{\mu}})$ is an upper bound
of the number of arithmetic operations, which at a given value
of $\mu \ge 2$ and at a given vector ${\bf m} \in {\bf M}(\mu)$
are required to calculate the estimate of the value of the product
$||{\bf m}||!Q_{\mu}({\bf x}; {\bf y}; {\bf m})$, where the
$(\mu - 1)$-dimensional vectors $\bf x$ and $\bf y$ are defined
by the formulas (80), in which instead of the coefficients $a_k$ 
the these coefficients values estimates calculated at the stage 1  
are substituted; {} \\
\mbox{$E_2(\mu, {\bf m} \mid {\bf a}_{\mu})$} is an upper bound of
the number of arithmetic operations that at a given value of
$\mu \ge 2$ and a given vector ${\bf m} \in {\bf M}(\mu)$ are required
to calculate the estimate of the value of the product
$\mu!\left\{[\mu - ||{\bf m}||]! \, \right\}^{- 1}
Q_{\mu}({\bf x}; -{\bf y}; {\bf m})$, where the $(\mu - 1)$-dimensional vectors
${\bf x}$ and ${\bf y}$ are defined by formulas (80), in which instead
of the coefficients $a_k$ the these coefficients values estimates 
calculated at the stage 1 are substituted, and the vector ${-\bf y}$ 
is defined by formula (82);
\begin{equation}
\alpha(n, {\bf m}\mid {\bf e}_n, \mbox{$\bm\tau_n$}) =
||{\bf m}||!\,{\displaystyle e}_{\scriptscriptstyle ||{\bf m}||}
Q_{n + 1}({\bf x}; {\bf y}; {\bf m}), \quad {\bf m} \in {\bf M}(n + 1),
\label{62}
\end{equation}
where the $n$-dimensional vectors ${\bf y}$ and $\bf x$ are defined
by formulas (84), in which instead of the coefficients $a_k$ the
these coefficients values estimates calculated at the stage 1 are 
substituted;

$E_3(n, {\bf m} \mid {\bf e}_n, \mbox{$\bm \tau_n$})$ is an upper
bound of the number of arithmetic operations, which at a given vector
${\bf m} \in {\bf M}(n + 1)$ are required to calculate the estimate 
of the value of the product
\mbox{$\alpha(n, {\bf m} \mid {\bf e}_n, \mbox{$\bm \tau_n$})$}, where
the $n$-dimensional vectors ${\bf y}$ and $\bf x$ are defined
by the formulas (84), in which instead of the quantities $\tau_k$
the these quantities values estimates calculated at the stage 3 are 
substituted,
and instead of the quantitie 
${\displaystyle e}_{\scriptscriptstyle ||{\bf m}||}$
the this quantitie value estimate calculated at the stage 2 is 
substituted;

$E(e_{\mu} \mid {\bf a}_{\mu})$ is upper estimate of the number 
of arithmetic operations required at the stage 2 to calculate 
the estimate of the value of the quantity $e_{\mu}$ under all 
estimates, which belong to the set 
${\bf a}'_{\mu} = \{a'_1, a'_2, \ldots, a'_{\mu}\}$ and are calculated 
at the stage 1;

$E(\tau_{\mu} \mid {\bf a}_{\mu})$ is an upper bound of the number
of arithmetic operations required at the stage 3 to calculate the 
estimate of the value of quantity $\tau_{\mu}$ under all estimates,
which belong to the set 
${\bf a}'_{\mu} = \{a'_1, a'_2, \ldots, a'_{\mu}\}$ and are calculated 
at the stage 1;

$E({\bf e}_n \mid {\bf a}_n)$ is an upper bound of the number
of arithmetic operations, required at the stage 2 to calculate the
estimates of the values of all quantities from the set 
${\bf e}_n = \{e_1, e_2, \ldots, e_n\}$ under all estimates, which 
belong to the set ${\bf a}'_{\mu} = \{a'_1, a'_2, \ldots, a'_{\mu}\}$ 
and are calculated at the stage 1;

$E(\mbox{$\bm \tau_n$} \mid{\bf a}_n)$ is an upper bound of
the number of arithmetic operations required at the stage 3 
to calculate the estimates of  the values of all quantities from 
the set $\mbox{$\bm \tau_n$} = \{\tau_1, \tau_2, \ldots, \tau_n\}$
under all estimates, which belong to the set 
${\bf a}'_{\mu} = \{a'_1, a'_2, \ldots, a'_{\mu}\}$ and are calculated 
at the stage 1;

$E(B_n \mid \, {\bf e}_n, \mbox{$\bm \tau_n$})$ is an upper estimate 
of the number of arithmetic operations required at stage 4 
to calculate the estimate of the limiting virial coefficient $B_n$ 
under the estimates of the values of all quantities from 
the population 
${\bf e}_n = \{e_1, e_2, \ldots, e_n \}$ and of the values of all 
quantities from the set 
$\mbox{$\bm\tau_n$} = \{\tau_1, \tau_2, \ldots, \tau_n\}$ obtained 
as results of the calculations at the stages 1, 2 and 3;

$E(B_n \mid \, {\bf a}_n)$  is an upper estimate of the number 
of arithmetic operations required at stage 4 to calculate 
the estimate of the limiting virial coefficient $B_n$  under 
the estimates obtained as results of the calculations at the stages 1,
2 and 3, that is under the estimates of the values of all 
coefficients from the set ${\bf a}_n = \{a_1, a_2, \ ldots, a_n \}$, 
under the estimates of the values of all quantities from 
the population ${\bf e}_n = \{e_1, e_2, \ ldots, e_n \}$ and under 
the estimates of the values of all quantities from the set 
$\mbox{$\bm\tau_n$} = \{\tau_1, \tau_2, \ldots, \tau_n\}$.

Let us find an upper bound for the number of arithmetic operations 
required at stage 2 to calculate the estimates of the values of all 
quantities from the set ${\bf e}_n = \{e_1, e_2, \ldots, e_n\}$
under all estimates, which belong to the set 
${\bf a}'_n = \{a'_1, a'_2, \ldots, a'_n\}$ and have been 
calculated at the stage 1.

From the definition of the vectors set ${\bf M}(\mu)$ it follows
that for any ${\mu} \ge 2$ every vector ${\bf m} \in {\bf M}(\mu)$
satisfies  the inequality
\begin{equation}
||{\bf m}|| \le \mu - 1.
\label{64}
\end{equation}
From the definition of the estimate
$E_1(\mu, {\bf m} \mid {\bf a}_n)$, the definition of the function
$Q_n({\bf x}; {\bf y}; {\bf m})$ by formula (67),
inequality (86), and Remark 11\, it follows that for any
${\mu} \ge 2$ and any vector ${\bf m} \in {\bf M}(\mu)$
the inequality
\begin{equation}
E_1(\mu, {\bf m} \mid {\bf a}_{\mu}\,) \le 7(\mu - 1)
\label{65}
\end{equation}
holds.

From the definition of $e_{\mu}$ by formula (79),
inequality (87), Remark 12 and definitions of the estimates
$E(e_{\mu} \mid {\bf a}_{\mu})$ and
$E_1(\mu, {\bf m} \mid {\bf a}_n)$ implies the estimate
\begin{equation}
E(e_{\mu} \mid {\bf a}_{\mu}) \le
\sum_{{\bf m} \in {\bf M}(\mu)}E_1(\mu, {\bf m}|\,{\bf a}_{\mu}) \le
7p(\mu -1)(\mu - 1).
\label{66}
\end{equation}

Using inequality (88) and the monotonic increase of the function
$p(n)$, from the definitions of estimates
$E(e_{\mu} \mid {\bf a}_{\mu})$ and
$E({\bf e}_n \mid {\bf a}_n)$ we obtain the inequality
\begin{equation}
E({\bf e}_n \mid {\bf a}_n) \le
\sum_{\mu = 2}^n E(e_{\mu} \mid {\bf a}_{\mu}) \le
7p(n - 1)\sum_{\mu = 2}^n (\mu - 1) = 7p(n - 1)n(n - 1)/2.
\label{67}
\end{equation}

Let us find an upper bound for the number of arithmetic operations 
required at stage 3 to calculate the estimates of the values of all 
quantities from the set 
$\mbox{$\bm \tau_n$} = \{\tau_1, \tau_2, \ldots, \tau_n\}$
under all estimates, which belong to the set 
${\bf a}'_n = \{a'_1, a'_2, \ldots, a'_n\}$ and have been 
calculated at the stage 1.

From the definition of the estimate 
$E_2(\mu, {\bf m} \mid {\bf a}_{\mu})$, from the definition  
of the function $Q_n({\bf x}; {\bf y}; {\bf m})$ 
by formula (67), from inequality (86) and Remark 11 it
follows that for any ${\mu} \ge 2$ and any vector 
${\bf m} \in {\bf M}(\mu)$ the inequality
\begin{equation}
E_2(\mu, {\bf m}|\,{\bf a}_{\mu}) \le 7(\mu - 1)
\label{68}
\end{equation}
holds.

From the definition of the quantity $\tau_{\mu}$ 
by formula (81), from inequality (90), from Remark 12 
and the definitions of estimates $E(\tau_{\mu} \mid {\bf a}_{\mu})$ 
and $E_2(\mu, {\bf m} \mid {\bf a}_{\mu})$ the estimate 
\begin{equation}
E(\tau_{\mu} \mid {\bf a}_{\mu}) \le
\sum_{{\bf m} \in {\bf M}(\mu)}E_2(\mu, {\bf m}|\,{\bf a}_{\mu}) \le
7p(\mu -1)(\mu - 1).
\label{69}
\end{equation}
follows.

Using the inequality (91) and the monotonic increase 
of the function $p(n)$, from the definitions of the estimates
$E(\tau_{\mu} \mid {\bf a}_{\mu})$ and
$E(\mbox{$\bm \tau_n$} \mid {\bf a}_n)$ we obtain the inequality
\begin{equation}
E(\mbox{$\bm \tau_n$} \mid {\bf a}_n) \le
\sum_{\mu = 1}^n E(\tau_{\mu} \mid {\bf a}_{\mu}) \le
7p(n - 1)\sum_{\mu = 2}^n (\mu - 1) =
7p(n - 1)n(n - 1)/2.
\label{70}
\end{equation}

Let us find an upper bound for the number of arithmetic operations
required to calculate the estimate of the limiting virial coefficient
$B_n$ under all estimates, which belong to the set 
${\bf e}_n = \{e_1, e_2, \ldots, e_n\}$ and have been calculated 
at the stage 2, and under all estimates, which belong to the set 
$\mbox{$\bm \tau_n$} = \{\tau_1, \tau_2, \ldots, \tau_n\}$ and have 
been calculated at the stage 3.

From inequality (86), the definition of the product
$\alpha(n, {\bf m} \mid {\bf e}_n, \mbox{$\bm \tau_n$})$
by formula (85), the definition of the estimate
$E_3(n, {\bf m} \mid {\bf e}_n, \mbox{$\bm \tau_n$})$,
the definition of the function $Q_n({\bf x}; {\bf y}; {\bf m})$
by formula (67) and Remark 11 it follows that for any $n \ge 2$
and any vector ${\bf m} \in {\bf M}(n + 1)$ the inequality
\begin{equation}
E_3(n, {\bf m} \mid {\bf e}_n, \mbox{$\bm \tau_n$}) \le 5n
\label{71}
\end{equation}
holds.

Definition by formula (83) of the limiting virial coefficient $B_n$ 
and definition by formula (85) of the product
\mbox{$\alpha(n, {\bf m} \mid {\bf e}_n, \mbox {$\bm \tau_n$})$}
implies that the coefficient $B_n$ can be represented
by the sum
\begin{equation}
B_n =
\sum_{{\bf m} \in {\bf M}(n + 1)}
\alpha(n, {\bf m} \mid {\bf e}_n, \mbox{$\bm \tau_n$}).
\label{72}
\end{equation}

Hence, using the definitions of the estimates
$E_3(n, {\bf m} \mid {\bf e}_n, \mbox{$\bm \tau_n$})$ and
$E(B_n \mid {\bf e}_n, \mbox{$\bm \tau_n$})$, we obtain the inequality
\begin{equation}
E(B_n \mid {\bf e}_n, \mbox{$\bm \tau_n$})) \le
\sum_{{\bf m} \in {\bf M}(n + 1)}
E_3(n, {\bf m} \mid {\bf e}_n, \mbox{$\bm \tau_n$}).
\label{73}
\end{equation}

Hence, by Remark 12 and inequality (93), the estimate follows
\begin{equation}
E(B_n \mid {\bf e}_n, \mbox{$\bm \tau_n$})) \le 5np(n).
\label{74}
\end{equation}

From the proposed scheme of the computation process for the estimate
of the virial coefficient $B_n$ it follows that the sole purpose
of all calculations at stages 2, 3 and 4 of this scheme is to estimate
this coefficient by the estimates of the coefficients
$a_1, a_2, \ldots, a_n$ calculated at stage 1.
The number of all arithmetic operations required to achieve this goal
is the sum of all arithmetic operations that should be performed
on these stages. Hence, applying the definitions of estimates
$E({\bf e}_n \mid {\bf a}_n)$, $E(\mbox{$\bm \tau_n$} \mid {\bf a}_n)$,
\mbox{$E(B_n \mid {\bf e}_n, \mbox {$\bm \tau_n$})$} and
$E(B_n \mid {\bf a}_n)$,
we get the estimate
\begin{equation}
E(B_n \mid {\bf a}_n) \le E({\bf e}_n \mid {\bf a}_n) +
E(\mbox{$\bm \tau_n$} \mid {\bf a}_n) +
E(B_n \mid {\bf e}_n, \mbox{$\bm \tau_n$})).
\label{75}
\end{equation}
The inequalities (97), (89), (92), and (96) imply the estimate 
\begin{multline}
E(B_n \mid {\bf a}_n) \le 7p(n - 1)n(n - 1)/2 + 7p(n - 1)n(n - 1)/2 +
p(n)5n ={}\\
7p(n - 1)n(n - 1) + 5np(n).
\label{76}
\end{multline}

In particular,  from formula (98) and Remark 12 it follows that
for $n \le 10$ it takes less than $21000$ of arithmetic operations
to compute the estimate of the limiting virial coefficient $B_n$
by the  estimates of the coefficients
$a_2, a_3, \ldots, a_n$ computed at stage 1.

This is a negligible number of arithmetic operations compared
to the number of operations necessary to obtain an estimate of any
of the coefficients $a_4, a_5, \ldots$. Indeed, it takes
about $10^{10}$ and more statistical trials to compute  estimates
of these coefficients by the Monte Carlo method. This implies

{\bf Remark 14.} For $n \ge 4$ the main difficulty of the calculation 
procedure of the estimate of a limiting virial coefficient by means 
of its representation as the polynomial in the coefficients $a_n$ 
according to formulas (76), (77), (78), (59) and (47)  consists 
in complexity of the estimation procedure of all coefficients 
from the set $\{a_2, a_3, \ldots, a_n\}$. Moreover, complexity 
of the calculation procedure of the estimate of the limiting virial 
coefficient $B_n$ negligibly exceeds the complexity of the calculation
procedure of the estimates of all coefficients $a_m$ from this set. 
Hence, the criterion of complexity of representation of this set 
is a criterion for the complexity of the given representation 
of the limiting virial coefficient $B_n$. $\blacksquare$

Let us introduce the notation:

${\mathfrak L}_{TR}(n, 0) = \{L\}$ is the set of all tree sums,
each of which by the formulas (59) and (47) represents
coefficient from the set of coefficients 
${\bf a}_{1, n - 1} = \{a_2, a_3, \ldots, a_n\}$,
where $n \ge 2$;

${\mathfrak L}_{TR}(n) = \{L\}$ is the set of all tree sums, each 
of which by the formulas (53) and (47) represents 
the limiting Mayer coefficient from the set of coefficients
${\bf b}_{1, n - 1} = \{b_2, b_3, \ldots, b_n\}$, where $n \ge 2$.

{\bf Lemma 6.} {\it Let a pair interaction potential $\Phi({\bf r})$ 
be a measurable function, and the pair interaction satisfies 
the conditions of stability and regularity.
Then the set ${\mathfrak L}_{TR}(n)$ is a base set of base linear 
combinations. This set has order $n$, and for each 
$k \in \{2, 3, \ldots, n\}$ the base linear combination of order $k$ 
belonging to this base set belongs to the set ${\mf L}(k, (\R)^{k - 1})$}.

{\bf Proof.} From the definition of the tree sums set
${\mathfrak L}_{TR}(n)$ it follows that every tree sum belonging 
to this set is a representation of some limiting Mayer coefficient 
$b_k \in {\bf b}_{1, n -1}$, where $1 < k \le n$. 
 By Lemma 3, this tree sum is a base linear 
combination of order $k$ with coefficients of negligible complexity. 
Thus, the set ${\mathfrak L}_{TR}(n)$ is a finite set of all base
linear combinations that by the formulas (53) and (47) are 
representations of the limiting Mayer coefficients belonging 
to the set ${\bf b}_{1, n - 1}$. At that, the representation
of the limiting Mayer coefficient $b_k \in {\bf b}_{1, n - 1}$ 
is a base linear combination of order $k$ from the set
${\mathfrak L}_{TR}(n)$.

From the definition of this base linear combination of order $k$ 
by the formulas (53) and (47) it follows that the space 
$(\R)^{k - 1}$ is the integration domain of all integrals 
included in this linear combination. Therefore, this linear 
combination of order $k$ belongs to the set 
${\mf L}(k, (\R)^{k - 1})$ by the definition of this set.
So, the set ${\mathfrak L}_{TR}(n)$ is a base linear combinations
finite set, in which each base linear combination of order $k$ 
belonging to it belongs to the set ${\mf L}(k, (\R)^{k - 1})$. 
This means that this set is the base set of base linear combinations 
by definition 26.

Any Mayer coefficient $b_k$ from the set of Mayer coefficients 
${\bf b}_{1, n - 1}$ is represented by a base linear combination 
of order $k$ from the base set ${\mathfrak L}_{TR}(n)$, and this 
base set contains only base linear combinations that are 
representations of Mayer coefficients belonging to the set 
${\bf b}_{1, n - 1}$.
Therefore, no base linear combination of order more than $n$ 
belongs to the base set ${\mathfrak L}_{TR}(n)$. On the other hand, 
this base set contains a base linear combination of order $n$, 
which is a representation of Mayer coefficient $b_n$ belonging 
to the set ${\bf b}_{1, n - 1}$. Hence, the number $n$ is 
the largest of the numbers serving as the order of one of the base 
linear combinations included to the base set ${\mathfrak L}_{TR}(n)$. 
Hence, by Definition 27, it follows that 
the number $n$ is the order of the base set ${\mathfrak L}_{TR}(n)$. 
Lemma 6 completely proven. $\blacktriangleright$
                  
{\bf Lemma 7.} {\it Let a pair interaction potential $\Phi({\bf r})$ 
be a measurable function, and the pair interaction satisfies 
the conditions of stability and regularity. Then the set 
${\mathfrak L}_{TR}(n, 0)$ is a base set of base linear combinations. 
This set is of order $n$, and each its base linear combination 
of order $k$ belongs to the set ${\mf L}(k, (\R)^{k - 1})$}.

{\bf Proof.} From the definition of the set of tree sums
${\mathfrak L}_{TR}(n, 0)$ it follows that any tree sum belonging 
to this set is a representation by the formulas (59) and (47) 
of a certain coefficient $a_k$ from the coefficients set
${\bf a}_{1, n - 1} = \{a_2, a_3, \ldots, a_n\}$, where $1 < k \le n$.
By Lemma 4, this tree sum is a base linear combination of order $k$ 
with coefficients of negligible complexity. 
Thus, the set ${\mathfrak L}_{TR}(n, 0)$ is a finite set of all base
linear combinations that by the formulas (59) and (47) are 
representations of the coefficients belonging to the set 
${\bf a}_{1, n - 1}$. At that, the representation of the coefficient 
$a_k \in {\bf a}_{1, n - 1}$ is a base linear combination of order $k$ 
from the set ${\mathfrak L}_{TR}(n)$.

From the definition of this base linear combination of order $k$ 
by the formulas (59) and (47) it follows that the space 
$(\R)^{k - 1}$ is the integration domain of all integrals 
included in this linear combination. Therefore, this linear 
combination of order $k$ belongs to the set 
${\mf L}(k, (\R)^{k - 1})$ by the definition of this set.
So, the set ${\mathfrak L}_{TR}(n, 0)$ is a base linear combinations 
finite set, in which each base linear combination of order $k$ 
belonging to it belongs to the set ${\mf L}(k, (\R)^{k - 1})$. 
This means that this set is the base set of base linear combinations 
by definition 26.

Any coefficient $a_k$ from the coefficients set ${\bf a}_{1, n - 1}$ 
is represented by a base linear combination of order $k$ from 
the base set ${\mathfrak L}_{TR}(n, 0)$, and this 
base set contains only base linear combinations that are 
representations of the coefficients belonging to the set 
${\bf a}_{1, n - 1}$. Therefore, no base linear combination 
of order more than $n$ belongs to the base set 
${\mathfrak L}_{TR}(n, 0)$. On the other hand, this base set contains 
a base linear combination of order $n$, which is a representation 
of the coefficient $a_n$ belonging to the set ${\bf a}_{1, n - 1}$. 
Hence, the number $n$ is the largest of the numbers serving as 
order of one of the base linear combinations included to the base 
set ${\mathfrak L}_{TR}(n, 0)$. Hence, by Definition 27, it follows 
that the number $n$ is the order of the base collection 
${\mathfrak L}_{TR}(n, 0)$. Lemma 7 completely proven. 
$\blacktriangleright$

By Definition 28, Lemma 6 and Lemma 7 imply

{\bf Corollary 7} {\it Base sets ${\mathfrak L}_{TR}(n)$ and
${\mathfrak L}_{TR}(n, 0)$ are comparable.}
                                               
For any $k > 1$, the set ${\mf L}(k, (\R)^{k-1})$ is a subset 
of the set $D(Cr_3)$ defined by the formula (43). The set 
$D(Cr_3)$ is the definitional domain of the complexity criterion 
$Cr_3$. From here by Lemmas 6 and 7 it follows that for any $n > 1$ 
the sets ${\mathfrak L}_{TR}(n, 0)$ and ${\mathfrak L}_{TR}(n)$ 
are base sets containing only such base linear combinations that 
belong to the set $D(Cr_3)$. The set $D(Cr_3)$ is contained
in the set $D(Cr_1)$ that is defined by the formula (39) 
and is the definitional domain of the complexity criteria $Cr_1$ and 
$Cr_2$. This means that three complexity criteria are defined 
on the set ${\mf L}(k, (\R)^{k-1})$: $Cr_1$, $Cr_2$ and $Cr_3$. 
Hence it follows that for any $n > 1$ the sets 
${\mathfrak L}_{TR}(n, 0)$ and ${\mathfrak L}_{TR}(n)$ are base sets 
containing only such base linear combinations on that three complexity 
criteria are defined: $Cr_1$, $Cr_2$ and $Cr_3$. Therefore, these sets 
belong to the definitional domain of complexity criteria: 
$Cr'_1$, $Cr'_2$ and $Cr'_3$, defined by the formula 
(73). This makes it possible to compare by these criteria  
the complexity of the finite set ${\mathfrak L}_{TR}(n, 0)$ 
of the tree sums, which are the representations of the coefficients 
$a_2, a_3, \ldots, a_n$, with the complexity of the finite set 
${\mathfrak L}_{TR}(n)$ of tree sums, which are the representations 
of the limiting Mayer coefficients $b_2, b_3, \ldots, b_n$.

As an example, for \mbox{$n = \overline{2,10}$}, the criterion 
$Cr'_1({\mathfrak L})$ values were calculated for the sets 
of the tree sums of the form ${\mathfrak L}_{TR}(n, 0) = \{L\}$ and 
for the set ${\mathfrak L}_{TR}(n)$ of the tree sums.
The results are shown in Table 4. Further, for $n = \overline{2,6}$,
the criteria $Cr_2'({\mathfrak L})$ and $Cr'_3({\mathfrak L})$ values 
were calculated for the set ${\mathfrak L}_{TR}(n, 0)$ of the tree 
sums and for the set ${\mathfrak L}_{TR}(n)$ of the tree sums. 
The results are shown in Tables 5 and 6, respectively.    
                                     
Comparison the values of criteria $Cr'_1$, $Cr'_2$ and $Cr'_3$ 
on the sets of tree sums of the form ${\mathfrak L}_{TR}(n, 0)$ with 
their values on the sets of tree sums of the form 
${\mathfrak L}_{TR}(n)$ confirms the conclusion immediately following 
from the above results: for $n > 3$ the base set 
${\mathfrak L}_{TR}(n, 0)$ is considerably simpler than the comparable 
base set ${\mathfrak L}_{TR}(n)$. Hence, for $n > 3$ any function 
of negligible complexity of the base set ${\mathfrak L}_{TR}(n, 0)$ 
is considerably simpler than any function of negligible complexity 
of the comparable base set ${\mathfrak L}_{TR}(n)$.

11. {\bf Representations of virial coefficients by frame sums that are 
not tree sums, and application of the introduced criteria to their 
comparison in terms of complexity with the tree sums representing
coefficients $b_n$ and $a_n$}

Using the frame sum method, you can get also representions of
power series coefficients that are not tree sums. So, by the method
of frame sums, the author obtained representations of virial
coefficients in the form:
\begin{equation}
B_n = -\frac{n-1}{n!}\sum_{{\bf C}\in{\mathfrak C}(n)}J({\bf C}).
\label{77}
\end{equation}
Here \mbox{${\mathfrak C}(n)$} is the set of ensembles of frame
cycles [14-16, 18-20, 37--39] of all doubly connected graphs with
the set of vertices $V_n = \{1, 2, \ldots, n\}$;
$\bf C$ is an ensemble of frame cycles from the set
${\mathfrak C}(n)$;
\begin{eqnarray}
J({\bf C}) =
\int_{(\R)^{n-1}}\prod_{\{u,v\}\in X(S({\bf C}))}f_{uv}
\prod_{\{\widetilde u,\widetilde v\}\in X_{ad}({\bf C})}
(1+f_{\widetilde u, \widetilde v})(d{\bf r})_{1,n-1},
\label{78}
\end{eqnarray}
Where
$S(\bf C)$ is the union of all cycles of the ensemble $\bf C$
[14, 15, 19, 37];
$X(S({\bf C}))$ is the set of all edges of the graph $S(\bf C)$
[14, 15, 19, 37];
$X_{ad}({\bf C})$ is the set of all admissible edges
[14, 15, 19, 37] of the ensemble ${\bf C}$;
$\{u, v\}$ is an edge incident to the vertices $u$ and $v$.

From the definition of integrals of the form $J({\bf C})$
by formula (100) it follows that in each of the integrals that are
terms of the sum on the right-hand side (99), the integrand is
the product of Mayer functions labeled with the edges of the cycles
included into the frame cycles ensemble that labels this integral,
and Boltzmann functions labeled with edges from the set
$X_{ad}({\bf C}) = \{\{u, v\}\}$.
We will call such a sum of integrals {\bf a frame sum.}

From the definition of the set $X_{ad}({\bf C})$ [14, 15, 19, 37]
follows that this set consists of pairwise distinct edges, and each
edge, contained in this set connects two non-adjacent vertices of
the graph $S({\bf C})$.

{\bf Theorem 6.} {\it If the potential of the pairwise 
interaction $\Phi({\bf r})$ is a measurable function and the pairwise
 interaction satisfies the conditions of stability and regularity,
then for any ensemble of frame cycles ${\bf C} \in {\mathfrak C}(n)$
the integral $J({\bf C})$ is a convergent improper base integral 
of order $n$, and the graph $S({\bf C})$ is a completed graph-label 
of the integrand of this integral.}

{\bf Proof.} First, we prove that the integrand of the integral
$J({\bf C})$ is a base product of order $n$.

For this purpose, we first of all prove that the sets of edges
$X(S({\bf C}))$ and $X_{ad}({\bf C})$ form a canonical pair of sets
${\bf X} = (X(S({\bf C})), X_{ad}({\bf C}))$ of order $n$. From
the definition of the edges set $X(S({\bf C}))$ it follows that this
set consists of pairwise different edges. As noted above, the set
$X_{ad}({\bf C})$ also consists of pairwise distinct edges, and each
edge contained in this set connects two non-adjacent vertices of
the graph $S({\bf C})$. Two conclusions follow from this:

1) disjoint sets $X(S({\bf C}))$ and $X_{ad}({\bf C})$
form an ordered pair ${\bf X} = (X(S({\bf C})), X_{ad}({\bf C}))$ of
sets;

2) the vertices of all edges from the set $X_{ad}({\bf C})$ belong
to the set of vertices of the graph $S({\bf C})$.

Since $\bf C$ is an ensemble of frame cycles from of the set
${\mathfrak C}(n)$, then, as is known [19],
the graph $S({\bf C})$ is a doubly connected graph with the set
vertices $V_n$.

Hence, the equality
\begin{equation}
V(X(S({\bf C}))) \cup V(X_{ad}({\bf C})) = V_n
\label{79}
\end{equation}
holds.
Here $V(X(S({\bf C})))$ is the set of all vertices of the graph
$S({\bf C})$, and $V(X_{ad}({\bf C}))$ is the set of vertices
of all admissible edges of the ensemble $\bf C$.
From equality (101) by Definition 5 it follows that the ordered
pair of sets ${\bf X} = (X(S({\bf C}), X_{ad}({\bf C}))$ is
a canonical pair of order $n$.

From the obtained results it follows that the graph $S({\bf C})$, to
which  the set $X_{ad}({\bf C})$ is putted in correspondence,
belongs to the set of graphs $\widetilde{\mathfrak G}_n$
by the definition of this set.

Hence, by Lemma 2, it follows that the Mayer and Boltzmann functions
product $\widetilde P_{\widetilde{\mathfrak G}_n}(S({\bf C}))$
labeled by this graph is a base product of order $n$ and is defined
by the formula
\begin{equation}
\widetilde P_{\widetilde{\mathfrak G}_n}(S({\bf C})) =
\prod_{\{i,j\} \in X(S({\bf C}))}\prod_{\{i',j'\} \in X_{ad}({\bf C})}
f_{ij}\widetilde f_{i'j'}.
\label{80}
\end{equation}

Hence, by Theorem 2, it also follows that the graph $S({\bf C})$ is
a completed graph-label of the product
 $\widetilde P_{\widetilde{\mathfrak G}_n}(S({\bf C}))$.

Comparison of formulas (100) and (102) implies that
the integrand of the integral $J({\bf C})$ is identical to the
functions base product
$\widetilde P_{\widetilde{\mathfrak G}_n}(S({\bf C}))$.
Therefore, this integrand is a functions base product of order $n$, it
is labeled with the graph $S({\bf C})$, and the graph $S({\bf C})$
is a completed graph-label of the integrand of the integral
$J({\bf C})$. Hence, by theorem 3, it follows that the improper 
integral $J({\bf C})$ is an improper convergent base integral 
of order $n$. Theorem 6 is proved. $\blacktriangleright$

Theorem 6 implies the following

{\bf Corollary 8.} {\it The frame sum on the right-hand side
{\rm (99)} is, by Definition {\rm 11} and Remark {\rm 6}, a base
linear combination with coefficients of negligible complexity.}

This circumstance makes it possible to use the proposed in this
article criteria $Cr_1$, $Cr_2$ and $Cr_3$ for comparison 
in complexity of representations of the virial coefficients by frame 
sums with other base linear combinations with coefficients 
of negligible complexity.

This circumstance also makes it possible to use the criteria $Cr'_1$,
$Cr_2'$ and $Cr'_3$ proposed in this article for comparison 
in complexity of representations of limiting virial coefficients
by frame sums with representations of these coefficients 
by polynomials in base linear combinations with coefficients 
of negligible complexity.

From tables 1, 2, 3, 4, 5 and 6 the conclusions follow.

According to the criteria $Cr_1$, $Cr_2$ and $Cr_3$, the complexity 
of the representation of the limiting virial coefficient $B_3$ 
by the frame sum according to the formulas (99) and (100) 
differs negligibly from the complexity of the representation 
of the coefficient $a_3$ by the tree sum according to formulas 
(59) and (47).

According to the criteria $Cr_1$ and $Cr_2$, this representation 
of the limiting virial coefficient $B_3$ by the frame sum is 
considerably simpler than the representation of the limiting Mayer 
coefficient $b_3$ by tree sums according to formulas (53) 
and (47). But according to the $Cr_3$ criterion, these two
representations in their complexity differ negligibly from each other.

According to the criteria $Cr'_1$ and $Cr_2'$, the representation
of the limiting virial coefficient $B_3$ by the frame sum is 
considerably simpler than its representation by formula (66) 
in the form of the polynomial in tree sums, representing the limiting
coefficients $b_n$ by formulas (53) and (47); also according 
to the criteria $Cr'_1$ and $Cr_2'$,
this representation of the limiting virial coefficient $B_3$ 
by the frame sum is considerably simpler then its represention 
by formulas (76), (77) and (78) in the form
of the polynomial in tree sums representing the coefficients $a_n$
by formulas (59) and (47).
But according to the criterion $Cr'_3$, all these three
representations in their complexity differ negligibly from each other.

According to the criteria $Cr_1$, $Cr_2$ and $Cr_3$, 
the representation of the limiting virial coefficient $B_4$ 
by the frame sum according to the formulas (99) and (100) 
is considerably more complicated than the representation 
of the coefficient $a_4$ by the tree sum according to the formulas 
(59) and (47).

The complexity of the representation of the limiting virial 
coefficient $B_4$ by the frame sum according to the criterion $Cr_1$ 
negligibly differ from the complexity of the representation 
of the limiting Mayer coefficient $b_4$ by the tree sum according to
 formulas (53) and (47). However, according 
to the criteria $Cr_2$ and $Cr_3$, this representation of the limiting 
virial coefficient $B_4$ is considerably more complicated than 
the above representation of limiting Mayer coefficient $b_4$. 
Since the criteria $Cr_2$ and $Cr_3$ are more accurate, then, 
apparently, it should be assumed that the representation 
of the limiting virial coefficient $B_4$ by the frame sum considerably
more complicated than the above representation of limiting Mayer
coefficient $b_4$.

Further, according to the criteria $Cr'_1$ and $Cr_2'$ 
the representation of the limiting virial coefficient $B_4$ 
by the frame sum is considerably simpler then the representatiun 
of this coefficient by the formula (66) in the form 
of the polynomial in tree sums, representing the limiting Mayer
coefficients $b_n$ by the formulas (53) and (47). 
But according to the $Cr'_3$ criterion, the first of these two 
representations of the limiting virial coefficient $B_4$ is 
considerably more complicated than the second one.
Since the criterion $Cr'_3$ is more accurate than the criteria 
$Cr'_1$ and $Cr_2'$, then, apparently, it should be assumed that 
the given representation of the limiting virial coefficient $B_4$ 
by the frame sum is considerably more complicated than 
the representation of this limiting coefficient as a polynomial 
in tree sums representing coefficients $b_n$.

Finally, according to the criteria $Cr_1$, $Cr_2$ and $Cr_3$, 
the representation of the limiting virial coefficient $B_4$ by the 
frame sum according to the formulas (99) and (100) is 
considerably more complicated than its representation 
by the formulas (76), (77) and (78) as a polynomial 
in tree sums representing the coefficients $a_n$ by formulas 
(59) and (47).

{\bf Acknowledgment.}
The author considers it his pleasant duty to express his deep thanks
to prof. D.A. Kofke for a fast and efficient informational support and
for fast and detailed answers to questions of interest to the author,
 to prof. G.A. Martynov for effective information  support and
useful discussion, to Dr. R. Hellman for a fast and efficient
informational support,
to Dr. N. Cleesby for quick and effective informational and
organizational support and high appreciation of the author's works
and to PhD in Physics and Mathematics V.I.~Cebro for fast and
effective information and technical support and helpful advices.

\newpage

\begin{center}
\bf Complexity tables of representations of Mayer coefficients $b_n$
and coefficients $a_n$ by tree sums, representations of virial
coefficients by frame sums and Ree-Hoover representations of virial
coefficients
\end{center}

\begin{center} \bf Table 1 of complexity by the criterion $Cr_1$
\end{center}
\begin{center}
\begin{tabular}{rccccccccc}
$n$ & 2 & 3 & 4 & 5 & 6 & 7 & 8 & 9 & 10 \\
$Cr_1(L_{TR}(n))$ & 1 & 2 & 5 & 14 & 44 & 157 & 634 & 2852 & 14047 \\
$Cr_1(L_{TR}(n.0))$ & 1 & 1 & 2 & 5 & 15 & 55 & 239 & 1169 & 6213 \\
$Cr_1(L_F(n))$ & 1 & 1 & 5 & 57 & - & - & - & - & - \\
$Cr_1(L_{RH}(n))$ & 1 & 1 & 2 & 5 & 23 & 171 & 2606 & 81564 & 4 980 756
\end{tabular}
\end{center}

\begin{center} \bf Table 2 of complexity by the criterion $Cr_2$
\end{center}
\begin{center}
\begin{tabular}{rccccccccc}
$n$ & 2 & 3 & 4 & 5 & 6 & 7 & 8 & 9 & 10 \\
$Cr_2(L_{TR}(n))$ & 1 & 5 &  22 & 93 & 403 & 1882 & 9671 & 54370 & 329325 \\ 
$Cr_2(L_{TR}(n,0))$ & 1 & 3 & 11 & 42 & 172 & 804 & 4330 & 25930 & 166666 \\
$Cr_2(L_F(n))$ & 1 & 3 & 26 & - & - \\
$Cr_2(L_{RH}(n))$ & 1 & 3 & 12 & 50 & 345 & 3591 & 72968 & 2936304 & 224134020 \\
\end{tabular}
\end{center}

\begin{center} \bf Table 3 of complexity by the criterion $Cr_3$
\end{center}
\begin{center}
\begin{tabular}{rccccccccc}
$n$ & 2 & 3 & 4 & 5 & 6 & 7 & 8 & 9 & 10 \\
$Cr_3(L_{TR}(n))$ & 0 & 1 & 7 & 37 & 183 & 940 & 5233 & 31554 & 202902 \\
$Cr_3(L_{TR}(n,0))$ & 0 & 1 & 5 & 22 & 97 & 474 & 2657 & 16578 & 110749 \\
$Cr_3(L_F(n))$ & 0 & 1 & 11 & - & - \\
\end{tabular}
\end{center}

The tables use the following designations:

$n$ is index of Mayer (virial) coefficient;

$L_{TR}(n)$ is the representation of Mayer coefficient $b_n(\LL)$
by tree sum, defined according to formulas (52) and 
(48), and the representation of the limiting Mayer coefficient 
$b_n$ by tree sum, defined according to formulas (53) and 
(47);

$\LL \subseteq \R$ is the volume containing a particle system;

$L_{TR}(n.0)$ is the representation of the coefficient $a_n$ by
tree sum, defined according to formulas (59) and (47);

$L_F(n)$ is representation of the limiting virial coefficient $B_n$
by the frame sum according to formulas (99) and (100);

$L_{RH}(n)$ is Ree-Hoover representation of the virial coefficient
$B_n(\LL)$.

\pagebreak

\begin {center}
\bf Complexity tables of representations of virial coefficients:
1)~representations by means of the Mayer coefficients $b_n$,
presented by tree sums; 2) representations by means
of the coefficients $a_n$, represented by tree sums;
3) representations by frame sums; 4) Ree-Hoover representation;
\end{center}

\begin{center} \bf Table 4 of complexity by the criterion $Cr'_1$
\end{center}
\begin{center}
\begin{tabular}{rccccccccc}
$n$ & 2 & 3 & 4 & 5 & 6 & 7 & 8 & 9 & 10 \\
$Cr'_1({\mathfrak L}_{TR}(n))$ & 1 & 3 & 8 & 22 & 66 & 223 & 857 & 3709 & 17756 \\
$Cr'_1({\mathfrak L}_{TR}(n.0))$ & 1 & 2 & 4 & 9 & 24 & 79 & 318 & 1487 & 7700 \\
$Cr'_1(L_F(n))$ & 1 & 1 & 5 & 57 & - & - & - & - & - \\
$Cr'_1(L_{RH}(n))$ & 1 & 1 & 2 & 5 & 23 & 171 & 2606 & 81564 & 4 980 756
\end{tabular}
\end{center}

\begin{center} \bf Table 5 of complexity by the criterion $Cr_2'$
\end{center}
\begin{center}
\begin{tabular}{rccccccccc}
$n$ & 2 & 3 & 4 & 5 & 6 & 7 & 8 & 9 & 10 \\
$Cr_2'({\mathfrak L}_{TR}(n))$ & 1 & 6 & 28 & 121 & 524 & 2406 & 12077 & 66447 & 395772 \\
$Cr_2'({\mathfrak L}_{TR}(n,0))$ & 1 & 4 & 15 & 57 & 229 & 1033 & 5363 & 31293 & 197959 \\
$Cr_2'(L_F(n))$ & 1 & 3 & 26 & - & - & \\
$Cr_2'(L_{RH}(n))$ & 1 & 3 & 12 & 50 & 345 & 3591 & 72968 & 2936304 & 224134020 \\
\end{tabular}
\end{center}

\begin{center} \bf Table 6 of complexity by the criterion $Cr'_3$
\end{center}
\begin{center}
\begin{tabular}{rccccccccc}
$n$ & 2 & 3 & 4 & 5 & 6 & 7 & 8 & 9 & 10\\
$Cr'_3({\mathfrak L}_{TR}(n))$ & 0 & 1 & 8 & 45 & 228 & 1168 & 6401 & 37955 & 240857 \\
$Cr'_3({\mathfrak L}_{TR}(n,0))$ & 0 & 1 & 6 & 28 & 125 & 599 & 3256 & 19834 & 130583 \\
$Cr'_3(L_F(n))$ & 0 & 1 & 11 & - & - \\
\end{tabular}
\end{center}

The tables use the following designations:

$n$ is index of virial coefficient;

${\mathfrak L}_{TR}(n)$ is the representation of the virial 
coefficient $B_n(\LL)$ by Mayer formula (\ref{44}) as a polynomial 
in all tree sums being representations of Mayer coefficients 
$b_2(\LL), b_3(\LL), \ldots, b_n(\LL)$
by formulas (52) and (48), and the representation 
of the limiting virial coefficient $B_n$ by Mayer formula (\ref{44}) 
as a polynomial in all tree sums that are representations of 
the limiting Mayer coefficients $b_2, b_3, \ldots, b_n$
by formulas (53) and (47);

${\mathfrak L}_{TR}(n.0)$ is representation of the limiting virial 
coefficient $B_n$ by formulas (\ref{56})--(\ref{61}) as a polynomial
in all tree sums that are representations of the coefficients
$a_2, a_3, \ldots, a_n$ by formulas (\ref{38}) and (47);

${\mathfrak L}_F(n)$ is frame sum representation of the limiting
virial coefficient $B_n$ according to formulas (\ref{77}) and 
(\ref{78});

$L_{RH}(n)$ is representation of the virial coefficient $B_n(\LL)$
by Ree-Hoover  method;

{\bf Note.} In Tables 1 and 4, the values of lengths of the base
linear combinations that are  Ree-Hoover representations
of the virial coefficients $B_n$, were borrowed from the article [27].
Criterion values $Cr_2$ for  Ree-Hoover representations
of virial coefficients $B_n$ were calculated based
on the definition [46, 47, 48] of these representations and using
length values of base linear combinations given in [27].

\pagebreak

\renewcommand{\refname}
\end{document}